\documentclass[aps, prd, showpacs, showkeys, preprintnumbers, bibnotes, floatfix, longbibliography, notitlepage, nofootinbib, superscriptaddress, superscriptgaddress]{revtex4-2}

\pdfoutput=1
\usepackage[colorlinks=true,breaklinks=true]{hyperref}
\usepackage{amsmath}
\usepackage{amsfonts}
\usepackage{amssymb}
\usepackage{mathrsfs}
\usepackage{graphicx}
\usepackage{color}
\usepackage[utf8]{inputenc}
\usepackage{amsfonts}
\usepackage{graphicx}
\usepackage{booktabs}
\usepackage{multirow}
\usepackage{makecell}
\usepackage{siunitx}
\usepackage{ragged2e}
\usepackage{rotating}
\usepackage{float}
\usepackage[dvipsnames]{xcolor}
\definecolor{darkred}{rgb}{0.5,0,0}
\definecolor{darkblue}{rgb}{0,0,0.5}
\definecolor{firebrick}{rgb}{0.75,0.125,0.125}
\definecolor{darkgreen}{rgb}{0,0.5,0}
\hypersetup{urlcolor=darkblue,
        citecolor=darkgreen,
        linkcolor=firebrick}

\usepackage{orcidlink}
\usepackage{lipsum}
\usepackage[normalem]{ulem}
\usepackage{enumitem}
\usepackage[capitalise]{cleveref}
\usepackage{pifont}
\usepackage{subfigure}
\usepackage{physics}
\usepackage{comment}
\usepackage{mwe}
\usepackage{setspace}

\long\def\exclude#1{}

\newcommand{\ie}{{\it i.e.}}

\newcommand{\eg}{{\it e.g.}}

\newcommand{\eq}{Eq.}

\newcommand{\fig}{Fig.}

\newcommand{\Refe}{Ref.}
\newcommand{\Refes}{Refs.}
\newcommand{\equ}[1]{\eq~(\ref{equ:#1})}
\newcommand{\figu}[1]{\fig~\ref{fig:#1}}

\newcommand{\ship}{\textsc{SHiP}}

\graphicspath{{figures/}}

\begin{document}

\begin{center}
{\Large\bfseries Notice to readers (November 2026)}
\end{center}
\vspace{0.3cm}
\onehalfspacing
{\large

After this paper appeared on the arXiv, we found that the signature it proposed for the Majorana nature of neutrinos does not arise. We explain why in G.~Barenboim, M.~Bustamante, and Q.~Liu, ``The renormalization-group running of neutrino mixing is undetectable,'' \href{https://arxiv.org/abs/2610.28744}{arXiv:2610.28744}, which supersedes the present analysis.

This paper proposed that the Majorana nature of neutrinos could be established by combining a measurement of a GeV-scale heavy neutral lepton (HNL) at a fixed-target experiment, such as SHiP, with the flavor composition of high-energy astrophysical neutrinos measured at IceCube. A Majorana HNL of mass $M$ induces a finite one-loop correction, $\delta m$, to the light-neutrino mass matrix. We argued that detection at momentum transfer $Q > M$ would ``resolve'' the HNL and see a mixing that differs, by the effect of $\delta m$, from the mixing seen in oscillation experiments at lower $Q$, and that the disagreement between the two would reveal the Majorana nature.

We have since found that this premise is incorrect. The correction $\delta m$ is a self-energy of the light neutrino, evaluated on its mass shell, $p^2 = m_i^2$. It is therefore part of the physical mass matrix that fixes the mixing in every process, at every momentum transfer: the mixing measured in oscillation experiments already includes it. The momentum transfer of a charged-current process is carried by the $W$ boson and the charged lepton, not by the neutrino, which stays on its mass shell. A hard collision therefore does not ``resolve'' the HNL and cannot undo $\delta m$; it does not change the flavor content of the incoming on-shell neutrino. Consequently, the flavor composition of astrophysical neutrinos at Earth is the standard one, independent of $Q$. No comparison of flavor measurements at different momentum transfers can isolate $\delta m$ or reveal the Majorana nature. Independently, short-baseline data from NuTeV and NOMAD already exclude the implied flavor change at zero baseline for the benchmarks considered here. The implied $\nu_\mu \to \nu_e$ transition probabilities exceed the NuTeV limits by factors of 7 to 67.

We present this argument in full, together with three other cases in which a momentum-transfer--dependent mixing was proposed, in arXiv:2610.28744. We note, in passing, that the expression for $\delta m$ used here, \equ{deltamnu_LEFT}, was itself incorrect; the corrected expression is given there. This does not change our conclusion: even with the correct $\delta m$, the proposed signature does not exist.

The original paper follows, unchanged. It also remains available as v1.

}
\clearpage
\singlespacing
\title{Are neutrinos Majorana? Fixed-target and high-energy astrophysical searches decide}

\author{Gabriela Barenboim}
\email{gabriela.barenboim@uv.es}
\affiliation{Departament de F\'isica Te\'orica and IFIC, Universitat de Val\`encia-CSIC, E-46100, Burjassot, Spain}

\author{Mauricio Bustamante}
\email{mbustamante@nbi.ku.dk}
\affiliation{Niels Bohr International Academy, Niels Bohr Institute,\\University of Copenhagen, 2100 Copenhagen, Denmark}

\author{Qinrui Liu}
\email{qinrui\_liu@sfu.ca}
\affiliation{Department of Physics, Simon Fraser University, Burnaby, BC V5A 1S6, Canada}
\affiliation{Arthur B. McDonald Canadian Astroparticle Physics Research Institute, Kingston ON K7L 3N6, Canada}

\date{June 16, 2026}

\begin{abstract}
Determining whether the neutrino is a Dirac or Majorana fermion remains a fundamental open question. Conventional searches rely on neutrinoless double beta decay, but this electron-only channel suffers from blind spots. We propose a new, complementary probe to overcome this limitation. A heavy neutral lepton (HNL) triggers a high-energy shift in how the active neutrino flavors ($\nu_e$, $\nu_\mu$, $\nu_\tau$) mix---but only if the neutrinos are Majorana. For GeV-scale HNLs, the upcoming beam-dump experiment SHiP can discover the HNL and measure how it mixes with the active flavors. Separately, the scattering of TeV--PeV astrophysical neutrinos can resolve the HNL, revealing a shift in the proportions of each flavor arriving at Earth that could be detected by neutrino telescopes, regardless of the unknown flavor composition at the astrophysical neutrino sources. Because this flavor shift is most sensitive to the muon and tau sectors, it bypasses the blind spots of neutrinoless double beta decay. A correlated signal at SHiP and next-generation neutrino telescopes would prove that neutrinos are Majorana; its absence would point to them being Dirac.
\end{abstract}

\maketitle

\section{Introduction}
\label{sec:intro}

Whether the neutrino is a Majorana or Dirac fermion is one of the deepest open questions in particle physics. The distinction is not merely taxonomic: a Majorana neutrino is its own antiparticle~\cite{Majorana:1937vz}, carries no conserved lepton number, and arises naturally from the seesaw mechanism, providing a compelling explanation for the smallness of neutrino masses. A Dirac neutrino requires a new conserved quantum number (lepton number) and leaves the origin of the neutrino masses unexplained. While any origin of neutrino mass extends the Standard Model (SM)~\cite{Mohapatra:2006gs}, establishing its Majorana nature would  prove lepton-number violation (LNV), illuminating the high-energy scales of the seesaw mechanism~\cite{King:2025eqv}.

The most well-established experimental approach to deciding whether neutrinos are Majorana is the search for neutrinoless double beta decay ($0\nu\beta\beta$)~\cite{Furry:1939qr, Agostini:2022zub}, a proposed nuclear transition in which two neutrons decay into two protons and two electrons without emitting any antineutrinos---a process possible only if neutrinos are Majorana.  However, relying exclusively on $0\nu\beta\beta$ imposes two structural limitations: it is sensitive only to the electron sector, thereby failing to probe LNV potentially manifesting in other flavors, and it can be rendered unobservable by the destructive interference of unknown Majorana CP-violation phases~\cite{Pascoli:2002xq}. Therefore, an independent, complementary probe is strongly motivated.

\begin{figure}[t!]
 \centering
 \includegraphics[width=\columnwidth]{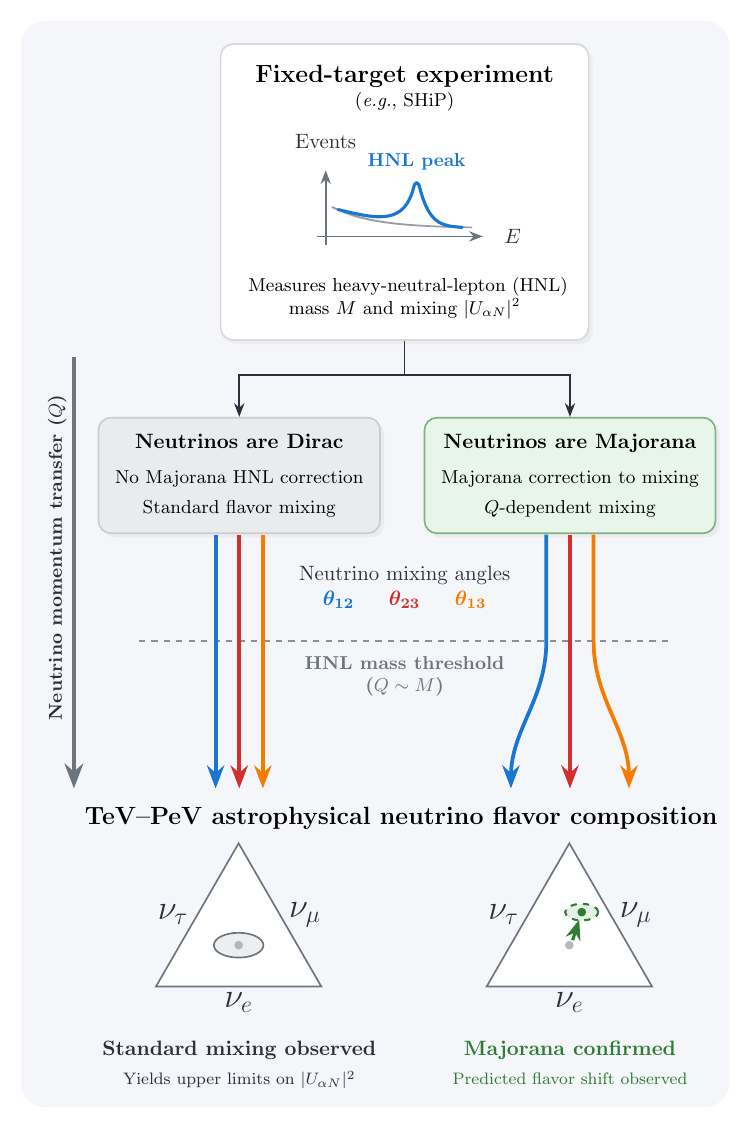}
 \vspace{-23pt}
 \caption{\textbf{Determining the nature of the neutrino.} Following the discovery of a heavy neutral lepton (HNL) in a fixed-target experiment, Majorana neutrinos---unlike Dirac ones---induce a specific high-energy correction to the neutrino mixing angles. Observing a predicted flavor shift in TeV--PeV astrophysical neutrinos confirms this correction.}
 \label{fig:schematic}
 \vspace*{-0.5cm}
\end{figure}

We propose such a probe and demonstrate its feasibility grounded on realistic experimental capabilities. We consider a minimally extended SM containing a Majorana heavy neutral lepton (HNL), $N$. We target primarily HNLs with GeV-scale masses, corresponding to low-scale realizations of the seesaw mechanism (though we also explore heavier HNLs). The HNL couples feebly to the active lepton flavors, \ie, $|U_{e N}|^2, |U_{\mu N}|^2, |U_{\tau N}|^2 \ll 1$, since larger couplings are excluded by collider bounds.  This well-motivated region of HNL parameter space allows us to combine two disparate experimental settings: fixed-target, beam-dump experiments, like the proposed Search for Hidden Particles (\ship)~\cite{SHiP:2018xqz}, that are optimized to detect GeV-scale feebly interacting particles; and high-energy neutrino telescopes, like the IceCube Neutrino Observatory, that are sensitive to TeV--PeV neutrinos of astrophysical origin. Unlike $0\nu\beta\beta$, our approach is sensitive to all three flavor sectors.

Figure~\ref{fig:schematic} illustrates our proposal. First, \ship\ discovers the HNL and measures its mass, $M$, and its couplings, $|U_{\alpha N}|^2$ ($\alpha = e, \mu, \tau$). If the HNL is Majorana, it induces a correction to the masses of the active neutrinos~\cite{Grimus:2002nk, AristizabalSierra:2011mn}, which, in turn, modifies the active-neutrino mixing parameters ($\theta_{12}$, $\theta_{23}$, $\theta_{13}$, $\delta_{\rm CP}$) relative to their standard values. The effect is visible to TeV--PeV astrophysical neutrinos: at detection, they scatter with transferred momenta $Q \gtrsim M \gtrsim \text{GeV}$, making the HNL corrections resolvable. As a result, the flavor composition of these neutrinos---\ie, the proportions of $\nu_e$, $\nu_\mu$, and $\nu_\tau$ that reach Earth---is shifted from its standard expectations. This shift serves as our primary observable. Because it is sensitive to the muon and tau sectors ($|U_{\mu N}|^2$, $|U_{\tau N}|^2$), it bypasses the electron exclusivity of $0\nu\beta\beta$. Most importantly, the shift vanishes if the HNL is Dirac, providing the arbiter of its Majorana nature that we sought.

Extracting this signature, however, requires unprecedented experimental precision. The astrophysical flavor shifts are subtle---of 5--15\%---placing them below the current sensitivity of IceCube~\cite{Abbasi:2025fjc}. Nevertheless, we show that they will be resolvable by multi-detector networks combining existing and future neutrino telescopes~\cite{Song:2020nfh}. 

This paper is organized as follows. Section~\ref{sec:threshold} derives the HNL-induced correction to the neutrino masses and mixing angles. Section~\ref{sec:SHiP} defines the joint SHiP-astrophysical discovery window. Section~\ref{sec:astro_production} introduces high-energy astrophysical neutrinos. Section~\ref{sec:astro_flavor_hnl_correction} introduces HNL corrections into their flavor composition. Section~\ref{sec:synergies} establishes the complementarity with $0\nu\beta\beta$ and explains how null new-physics results at the Large Hadron Collider (LHC) sharpen the interpretation. Section~\ref{sec:limits_and_discovery} shows the limits on and discovery prospects of HNLs using the flavor composition.  Section~\ref{sec:conclusions} presents our conclusions. This paper builds on and is self-contained with respect to the observational framework of \Refe~\cite{Bustamante:2026zst} (see also \Refe~\cite{Bustamante:2026aur}).


\section{Majorana corrections to neutrino mass and mixing}
\label{sec:threshold}

\textit{We derive the corrections to the light-neutrino masses induced by a Majorana heavy neutral lepton.  From this, we extract analytical expressions to understand the resulting corrections to the active-neutrino mixing parameters, mapping how they are determined at high momenta by the active-sterile couplings and Majorana phases.}


\subsection{Introducing the HNL}

We extend the SM by a single Majorana HNL, $N$, with mass $M$, coupled to the active neutrino flavors, $\nu_\alpha$, through Yukawa interactions, 
\begin{equation}
 \mathcal{L} \supset Y_\alpha \, \overline{L_\alpha} \tilde{H} N + \frac{1}{2} M \overline{N^c} N + \mathrm{h.c.}\,,
\end{equation}
where $Y_\alpha$ is the Yukawa coupling constant, $L_\alpha$ is the left-handed lepton doublet, $H$ is the Higgs doublet, and $\tilde{H} = i\sigma_2 H^*$ its conjugate. After electroweak symmetry breaking, the mixing between $N$ and $\nu_\alpha$ is
\begin{equation}
 |U_{\alpha N}|^2 = \frac{|Y_\alpha|^2 v^2}{M^2} \;,
\end{equation}
where $v = 246$~GeV is the Higgs vacuum expectation value. (This minimal low-scale seesaw setup assumes $N$ is the only new relevant degree of freedom below the electroweak scale, ensuring the loop-induced threshold correction derived below provides the leading modification to the effective mass matrix.) These mixing elements are the quantities to be measured by SHiP~\cite{SHiP:2018xqz}. We work in the limit $|U_{\alpha N}|^2 \ll 1$, consistent with current constraints and with the SHiP sensitivity range (Figs.~\ref{fig:discovery_window}, \ref{fig:discovery_window_var_angles_UmuN}).

Throughout this section, we consider the mass range $M \sim 0.3$--$6$~GeV, which, as explained in Sec.~\ref{sec:astro_flavor_hnl_correction}, is contained within the transferred momentum range probed by TeV--PeV astrophysical neutrinos. Since this range lies well below the electroweak scale, $M_W \simeq 80.4$~GeV, the appropriate framework for computing the threshold correction is the Low Energy Effective Field Theory (LEFT) obtained after integrating out the heavy SM gauge bosons ($W^\pm$, $Z$) and the top quark at $Q = M_W$. We discuss this matching explicitly next.


\subsection{Electroweak threshold matching}
\label{subsec:EW_matching}

In the SM, extended with massive Majorana neutrinos, the Weinberg operator~\cite{Weinberg:1979sa}, $\mathcal{O}_5 = (L^T C^{-1} L)(HH)/\Lambda$ (where $C$ is the charge-conjugation operator and $\Lambda$ is the scale of new physics), generates the neutrino masses and mixing after electroweak symmetry breaking.  The Weinberg operator runs in the full SM above $Q = M_W$ under the complete electroweak renormalization group.  At $Q = M_W$, the heavy SM fields ($W^\pm$, $Z$, top quark, and Higgs) are integrated out via finite one-loop threshold matching conditions~\cite{Jenkins:2013zja}.  Below $M_W$, the theory reduces to LEFT, in which the neutrino mass operator is a dimension-3 Majorana mass term $\mathcal{L}_\text{LEFT} \supset - \frac{1}{2}(m_\nu)_{\alpha\beta}\bar\nu_\alpha^c\nu_\beta$ running under QCD and QED only.  Since neutrinos are $SU(3)_c \times U(1)_{\rm em}$ singlets, their mass matrix does not run in LEFT at one loop. 

Thus, the effective neutrino mass matrix is
\begin{equation}
 (m_\nu)_{\alpha\beta}
 =
 (m_\nu^{(0)})_{\alpha\beta}
 +
 \delta(m_\nu)_{\alpha\beta} \;,
 \label{equ:mass_matrix_full}
\end{equation}
where the tree-level term is the standard seesaw contribution generated by $N$,
\begin{equation}
 (m_\nu^{(0)})_{\alpha\beta}
 =
 - \frac{v^2}{2M} Y_\alpha Y_\beta \;,
\end{equation}
and the threshold correction, $\delta(m_\nu)_{\alpha\beta}$, is the one-loop radiative correction to the Type-I seesaw mass matrix.

The dominant contribution to $\delta(m_\nu)_{\alpha\beta}$ from integrating out $N$ at the scale $Q = M \ll M_W$ comes entirely from the finite matching condition at $Q = M$ in LEFT. The relevant diagram is a one-loop neutrino self-energy with $N$ and a virtual gauge or Higgs boson in the loop. After performing the matching~\cite{Grimus:2002nk, AristizabalSierra:2011mn} (equating the full and effective-theory amplitudes at the threshold $Q = M$), the threshold correction takes the form
\begin{widetext}
\begin{equation}
 \delta(m_\nu)_{\alpha\beta} =
 -\frac{Y_\alpha Y_\beta M}{16\pi^2}
 \left[
 \frac{3M_W^2}{M^2 - M_W^2}\ln\frac{M^2}{M_W^2}
 + \frac{M_Z^2}{M^2 - M_Z^2}\ln\frac{M^2}{M_Z^2}
 - \frac{M_h^2}{M^2 - M_h^2}\ln\frac{M^2}{M_h^2}
 \right]\,.
 \label{equ:deltamnu_LEFT}
\end{equation}
\end{widetext}
\textit{This expression exists only if neutrinos are Majorana; it vanishes identically if they are Dirac.}

(For clarity, we focus on the case of a single HNL.  The extension to $n > 1$ HNLs, each $N_k$ with mass $M_k$, is straightforward: their corrections, $\delta(m_\nu)_{\alpha\beta}^{(k)}$, simply add, \ie, $\delta(m_\nu)_{\alpha\beta} = \sum_{k=1}^n \delta(m_\nu)_{\alpha\beta}^{(k)}$. Each contribution has its own flavor fingerprint via $U_{\alpha N_k}$.)

\medskip

\textbf{\textit{Light HNLs.---}}In the limit $M \ll M_W$, where SHiP is sensitive, \equ{deltamnu_LEFT} reduces to
\begin{align}
 \left. \delta(m_\nu)_{\alpha\beta} \right\vert_{M \ll M_W}
 &\simeq
 -\frac{\varepsilon(M)}{v^2}\, Y_\alpha Y_\beta M \nonumber \\
 &=
 -\varepsilon(M)\frac{M^3}{v^4}\,U_{\alpha N} U_{\beta N}\,,
 \label{equ:deltamnu_approx}
\end{align}
where the loop function is
\begin{equation}
 \varepsilon(M) = \frac{v^2}{16\pi^2}
 \left( 3\ln\frac{M_W^2}{M^2} + \ln\frac{M_Z^2}{M^2}
 - \ln\frac{M_h^2}{M^2} \right)\,,
 \label{equ:epsilon}
\end{equation}
and $M_W \simeq 80.4$~GeV, $M_Z \simeq 91.2$~GeV, $M_h \simeq 125.1$~GeV.  The coefficient $\varepsilon(M)$ is expressed entirely in terms of known SM masses, with no dependence on UV couplings, a consequence of working in LEFT below $M_W$. For $M = 1$~GeV, $\varepsilon \simeq 0.16\,v^2$; for $M = 5$~GeV, $\varepsilon \simeq 0.10\,v^2$, so the function varies slowly across the SHiP mass range and the dominant dependence on $M$ of the mass correction comes from the $M^3/v^4$ prefactor in \equ{deltamnu_approx}.  This is the approximate expression that applies in the mass range pertinent to the main narrative of our work, where \ship\ and astrophysical flavor measurements physically discriminate Majorana from Dirac neutrinos (Sec.~\ref{sec:limits_and_discovery-discovery}).

\medskip

\textbf{\textit{Heavy HNLs.---}}In the limit $M \gg M_h$, \equ{deltamnu_LEFT} reduces instead to
\begin{equation}
 \left. \delta(m_\nu)_{\alpha\beta} \right\vert_{M \gg M_h}
 \simeq
 -\varepsilon^\prime(M)\frac{M}{v^4}\,U_{\alpha N} U_{\beta N}\,,
 \label{equ:deltamnu_approx_heavy}
\end{equation}
where now
\begin{align}
 \varepsilon^\prime(M) 
 &= 
 \frac{v^2}{16\pi^2}
 \left( 3M_W^2\ln\frac{M_W^2}{M^2} + M_Z^2\ln\frac{M_Z^2}{M^2} \right. \nonumber \\
 & \quad \left.~- M_h^2\ln\frac{M_h^2}{M^2} \right)\,.
 \label{equ:epsilon_prime}
\end{align}
This is the approximate expression applicable when we place upper limits on the mixing elements $|U_{\alpha N}|^2$, using high-energy astrophysical neutrinos, for HNL masses above the SHiP range (Sec.~\ref{sec:limits_and_discovery-limits}).  

For heavy HNLs, the evolution of the mass correction with $M$ is more pronounced than for light HNLs: for $M = 150$~GeV, $\varepsilon^\prime/\text{GeV}^2 \simeq -170\,v^2$; for $M = 1$~TeV, $\varepsilon^\prime/\text{GeV}^2 \simeq -460\,v^2$.  Further, because now the correction has a sign opposite to the tree-level term of the effective neutrino mass [\equ{mass_matrix_full}], since $M \gg M_W, M_Z, M_h$ in \equ{epsilon_prime}, it can partially cancel the tree-level contribution. This allows the Yukawa couplings $Y_\alpha$---and, by extension, the couplings $|U_{\alpha N}|^2$---to be larger than the naive seesaw bound would normally allow, while still producing the small active neutrino masses.

\medskip

We use the above approximate analytical expressions for the mass threshold correction exclusively to develop physical insight, but we implement the full expression, \equ{deltamnu_LEFT}, when producing numerical results.

\begin{figure*}[t!]
 \centering
 \includegraphics[width=\textwidth]{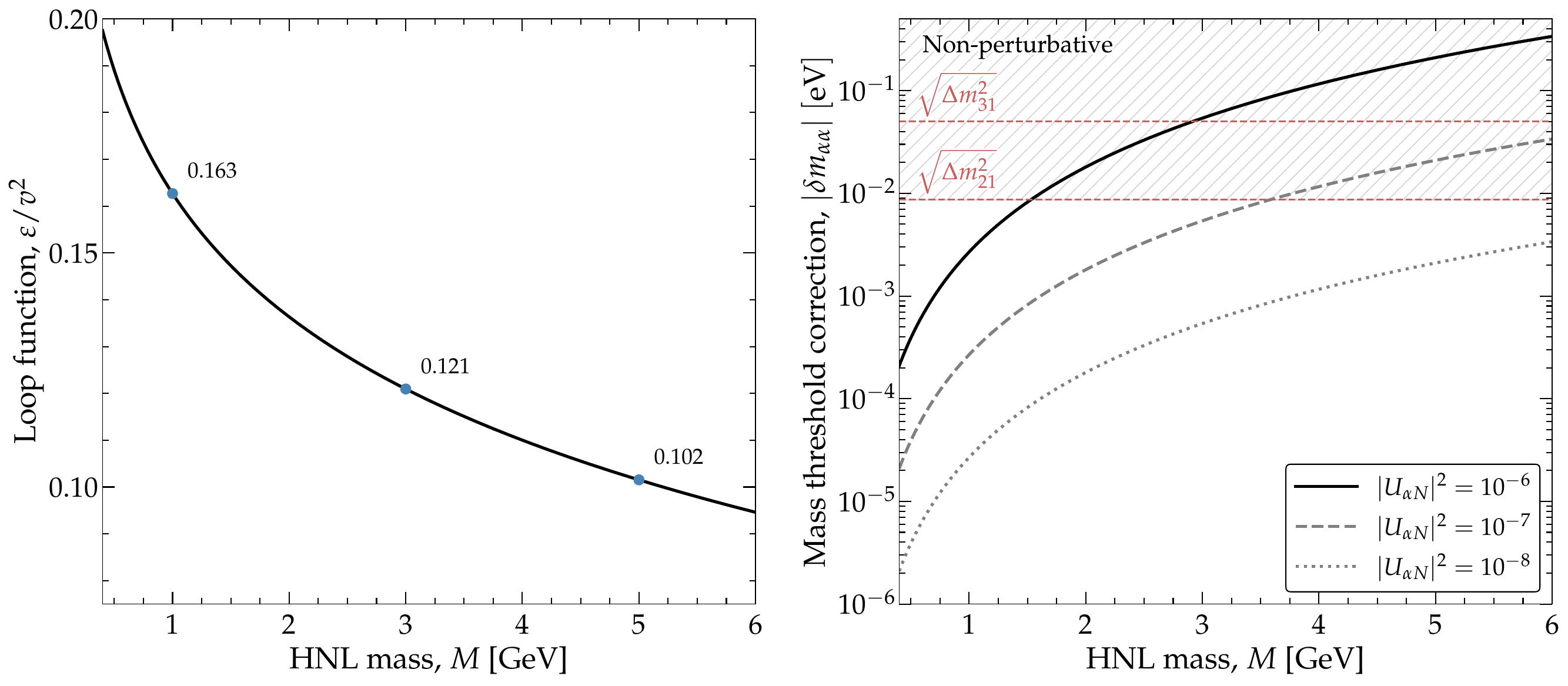}
 \caption{\textbf{Majorana mass threshold correction.} \textit{Left:} The loop function $\varepsilon(M)/v^2$ as a function of the HNL mass $M$, computed from \equ{epsilon}. The function varies slowly across the SHiP mass range ($0.3$--$6$~GeV), confirming that the dominant $M$-dependence of the threshold correction comes from the explicit $M^3/v^4$ prefactor in \equ{deltamnu_approx}. Annotated values at $M = 1, 3, 5$~GeV for illustration. \textit{Right:} Absolute size of the approximate diagonal threshold correction for a light HNL, $|\delta m_{\alpha\alpha}| = \varepsilon(M)\,M^3/v^4\cdot|U_{\alpha N}|^2$, for three reference values of $|U_{\alpha N}|^2$. Horizontal red dashed lines mark the solar ($\Delta m_{21}$) and atmospheric ($\Delta m_{31}$) mass splittings; the correction must lie below these for the perturbative expansion of Sec.~\ref{sec:threshold} to be valid.
 }
 \label{fig:loop_function}
\end{figure*}


\subsection{Physical nature of the mass threshold correction}
\label{subsec:physical}

In LEFT, we work in the flavor basis where the charged-lepton mass matrix is diagonal, and the active neutrino flavor fields relate to the physical mass eigenstates, $\nu_i$, via $\nu_{\alpha} = \sum_i U_{\alpha i} \nu_{i}$. Inserting this into the LEFT Lagrangian ($\mathcal{L}_\text{LEFT}$ in Sec.~\ref{subsec:EW_matching}), we assume the unitary Pontecorvo-Maki-Nakagawa-Sakata (PMNS) matrix $U$ diagonalizes only the tree-level contribution, yielding $U^T m_\nu^{(0)} U = \mathrm{diag}(m_1, m_2, m_3)$. Inverting this expresses the tree-level mass matrix as $(m_\nu^{(0)})_{\alpha\beta} = [U^* \, \mathrm{diag}(m_1, m_2, m_3) \, U^\dagger]_{\alpha\beta}$. This isolates the threshold correction $\delta(m_\nu)_{\alpha\beta}$ as a calculable radiative perturbation, while allowing the dominant tree-level parameters to be constrained by standard oscillation data.

Because the full mass matrix, \equ{mass_matrix_full}, is no longer diagonalized by the PMNS matrix $U$, we must introduce a new effective mixing matrix, $\hat{U}$, defined such that $\hat{U}^T m_\nu \hat{U}$ is diagonal. This  realignment from $U$ to $\hat{U}$ at the mass boundary forces an abrupt shift in the elements of the mixing matrix and, as a consequence, in the physical neutrino mixing parameters that define them (more on this later).  As a result, the mass correction introduces a distinct kink in the momentum-dependence of the mixing parameters exactly at $Q = M$, separating the high-$Q$ regime from the effective low-$Q$ regime.

We draw a key distinction between the threshold correction computed here and the scheme-dependent kinks that appear in the renormalization-group (RG) running of the neutrino mixing parameters computed in the $\overline{\rm MS}$ renormalization scheme, such as those discussed in \Refe~\cite{Bustamante:2026aur}. Those kinks are artifacts of the mass-independent renormalization scheme: particles are treated as massless above their threshold and instantaneously integrated out below it via a step function, producing an abrupt discontinuity that is not physically observable and would be smoothed by proper kinematic decoupling in a more complete formalism.

In contrast, the threshold correction $\delta(m_\nu)_{\alpha\beta}$ in \equ{deltamnu_approx} is a \emph{finite matching condition}---a genuine physical shift in the light-neutrino mass matrix generated at $Q = M$ when the heavy Majorana state is integrated out. It is scheme-independent in the sense that it appears in any consistent regularization scheme as a finite, non-zero contribution to the low-energy effective operator, and it would persist even in a fully matched EFT calculation with smooth kinematic decoupling. Most importantly, it vanishes identically if $N$ is Dirac, because no Majorana mass matrix exists to receive the correction. Therefore, the threshold correction is an unambiguous probe of the Majorana nature of $N$, in a way that scheme-dependent running effects are not. 


\subsection{Corrections to the mixing parameters}
\label{subsec:corrections_mixing_angles}

\textbf{\textit{The PMNS matrix.---}}The PMNS mixing matrix, \ie, the light-neutrino mixing matrix that diagonalizes the tree-level effective mass matrix, $m_\nu^{(0)}$, is
\begin{equation}
 U_{\alpha i} = \tilde{U}_{\alpha i} \cdot
 \mathrm{diag}\!\left(e^{i\rho},\, e^{i\sigma},\, 1\right)\,.
 \label{equ:PMNS_Majorana}
\end{equation}
Here, $\tilde{U}_{\alpha i}$ is the Dirac PMNS matrix in its standard parametrization~\cite{ParticleDataGroup:2024cfk} consisting of the three mixing angles ($\theta_{12}$, $\theta_{23}$, $\theta_{13}$) and one Dirac CP-violation phase ($\delta_{\rm CP}$); and $\rho$ and $\sigma$ are Majorana CP-violation phases. The Majorana phases cancel in neutrino oscillation probabilities, which depend on $|U_{\alpha i}|^2 = |\tilde{U}_{\alpha i}|^2$, and are therefore unconstrained by oscillation experiments. They do enter the $0\nu\beta\beta$ amplitude through the effective mass $m_{\beta\beta}$ (Sec.~\ref{subsec:0vbb}) and, crucially for us, they enter the threshold correction through $U_{\alpha i}$ in the $\xi_i$ factors defined below.

Today, the mixing angles are known to within 1--3\% and the Dirac CP phase, to within 16\% [at 68\% confidence level (C.L.)]~\cite{deSalas:2020pgw, Capozzi:2021fjo, NuFIT6}.  Because $\theta_{13} \approx 8^\circ$ is small, it suppresses the effect of $\delta_{\rm CP}$ in oscillations.  In the coming decade~\cite{Song:2020nfh}, measurements by DUNE~\cite{DUNE:2020lwj}, Hyper-Kamiokande~\cite{Hyper-Kamiokande:2018ofw}, and JUNO are expected to improve the precision on the mixing angles to 0.5\% and on $\delta_{\rm CP}$ to 3\%.  Thus, in our approximate analytical approximations below, we treat them as known inputs.  (Later, when producing our numerical results, we do float them.) In contrast, the Majorana phases $\rho$ and $\sigma$, are unknown; they can have any value within their allowed range of $[0, 2\pi)$.

\medskip

\textbf{\textit{Approximate analytical corrections.---}} We define the intermediate quantities
\begin{equation}
 \xi_i \equiv \sum_{\alpha} U_{\alpha i}\, U_{\alpha N}
 = (U^T)_{i,N}\,,
 \qquad i = 1, 2, 3\;,
 \label{equ:xi}
\end{equation}
which encode the projection of the flavor couplings of the HNL onto the mass eigenstates, $\nu_i$. Substituting \equ{PMNS_Majorana} in \equ{xi}, the $\xi_i$ factors read
explicitly
\begin{align}
 \xi_1 &= e^{i\rho}\!\left(
  |U_{eN}|\tilde{U}_{e1}
  + |U_{\mu N}|\tilde{U}_{\mu 1}
  + |U_{\tau N}|\tilde{U}_{\tau 1}
 \right),\nonumber\\
 \xi_2 &= e^{i\sigma}\!\left(
  |U_{eN}|\tilde{U}_{e2}
  + |U_{\mu N}|\tilde{U}_{\mu 2}
  + |U_{\tau N}|\tilde{U}_{\tau 2}
 \right),
 \label{equ:xi_explicit}\\
 \xi_3 &=\phantom{e^{i\rho}\!\Big(}
  |U_{eN}|\tilde{U}_{e3}
  + |U_{\mu N}|\tilde{U}_{\mu 3}
  + |U_{\tau N}|\tilde{U}_{\tau 3}\;,\nonumber
\end{align}
where the $\tilde{U}_{\alpha i}$ are the elements of the Dirac PMNS matrix, which are complex and carry the Dirac phase $\delta_{\rm CP}$. The Majorana phases $\rho$ and $\sigma$ are isolated in the overall prefactors of $\xi_1$ and $\xi_2$. This structure means that the corrections to different mixing parameters have different dependencies on $\rho$ and $\sigma$, as made explicit below.

For light HNLs, after rotating to the mass basis, the perturbation takes the form
\begin{equation}
 A_{ij} \equiv (U^T \delta m_\nu \, U)_{ij}
 = -\varepsilon(M) \frac{M^3}{v^4}\, \xi_i \xi_j\,.
 \label{equ:Aij}
\end{equation}
First-order perturbation theory for a complex symmetric
matrix gives the correction to the mixing matrix as $\delta U_{\alpha i} = \sum_{j \neq i} \frac{A_{ij}}{\Delta m_{ij}}\, U_{\alpha j}$, \ie,
\begin{equation}
 \delta U_{\alpha i}
 = -\varepsilon(M) \frac{M^3}{v^4}
 \sum_{j \neq i} \frac{\xi_i \xi_j}{\Delta m_{ij}}\, U_{\alpha j}\,,
 \label{equ:deltaU}
\end{equation}
where $\Delta m_{ij} \equiv m_i - m_j$, and $m_i$, $m_j$ are the masses of the mass eigenstates ($i, j = 1, 2, 3$)

The corrections to the mixing angles follow from the standard relations~\cite{ParticleDataGroup:2024cfk} $\sin^2\theta_{13} = |U_{e3}|^2$,
$\sin^2\theta_{12} = |U_{e2}|^2/(1 - |U_{e3}|^2)$, and
$\sin^2\theta_{23} = |U_{\mu 3}|^2/(1 - |U_{e3}|^2)$.
To first order in $\delta m_\nu$, they are
\begin{align}
 \delta(\sin^2\theta_{13}) 
 &\!\simeq \!
 2\,\mathrm{Re}(U_{e3}^*\, \delta U_{e3}) \,,
 \label{equ:delta_s13} \\
 \delta(\sin^2\theta_{12}) 
 &\!\simeq \!
 2\,\mathrm{Re}(U_{e2}^*\, \delta U_{e2})
 + 2\sin^2\theta_{12}\,\delta(\sin^2\theta_{13}) \,,\!\!\!
 \label{equ:delta_s12} \\
 \delta(\sin^2\theta_{23}) 
 &\!\simeq \!
 2\,\mathrm{Re}(U_{\mu 3}^*\, \delta U_{\mu 3})
 + 2\sin^2\theta_{23}\,\delta(\sin^2\theta_{13})
 \label{equ:delta_s23} \,,\!\!\!\!
\end{align}
where
\begin{align}
 \mathrm{Re}(U_{e3}^*\, \delta U_{e3})
 &=
 -\frac{\varepsilon(M) M^3}{v^4} 
 \label{equ:delta_s13_explicit} \\
 &\quad \times \mathrm{Re}\!\left[
 \xi_3 \left(
 \frac{\xi_1 U_{e3}^* U_{e1}}{m_3 - m_1}
 + \frac{\xi_2 U_{e3}^* U_{e2}}{m_3 - m_2}
 \right)
 \right] \;,
 \nonumber \\
 \mathrm{Re}(U_{e2}^*\, \delta U_{e2})
 &=
 -\frac{\varepsilon(M) M^3}{v^4}
 \label{equ:delta_s12_explicit} \\
 &\quad \times \mathrm{Re}\!\left[
 \xi_2 \left(
 \frac{\xi_1 U_{e2}^* U_{e1}}{m_2 - m_1}
 + \frac{\xi_3 U_{e2}^* U_{e3}}{m_2 - m_3}
 \right)
 \right] \;,
 \nonumber \\
 \mathrm{Re}(U_{\mu 3}^*\, \delta U_{\mu 3})
 &=
 -\frac{\varepsilon(M) M^3}{v^4}
 \label{equ:delta_s23_explicit} \\
 &\quad \times \mathrm{Re}\!\left[
 \xi_3 \left(
 \frac{\xi_1 U_{\mu 3}^* U_{\mu 1}}{m_3 - m_1}
 + \frac{\xi_2 U_{\mu 3}^* U_{\mu 2}}{m_3 - m_2}
 \right)
 \right] \;.
 \nonumber
\end{align}

Because these corrections scale inversely with the neutrino mass differences ($\propto (m_i - m_j)^{-1}$), the change in the solar angle, $\delta(\sin^2\theta_{12})$, dominates all other corrections. This dominance is driven by $m_i - m_j = \Delta m_{ij}^2 / (m_i + m_j)$, which dictates that the $(m_2 - m_1)$ gap is the smallest due to its drastically smaller numerator ($\Delta m_{21}^2 \ll |\Delta m_{31}^2|$). While this enhancement is moderate under normal neutrino mass ordering (NO, where $m_1$ is the lightest mass), transitioning to inverted ordering (IO, where $m_3$ is lightest) forces the states $m_1$ and $m_2$ to be relatively heavy ($m_1 + m_2 \simeq 2\sqrt{|\Delta m_{31}^2|}$).  This inflates the denominator of the mass gap, causing $m_2 - m_1$ to shrink and the $\theta_{12}$ correction to dwarf all others.

\begin{figure*}[t!]
 \centering
 \includegraphics[width=\textwidth]{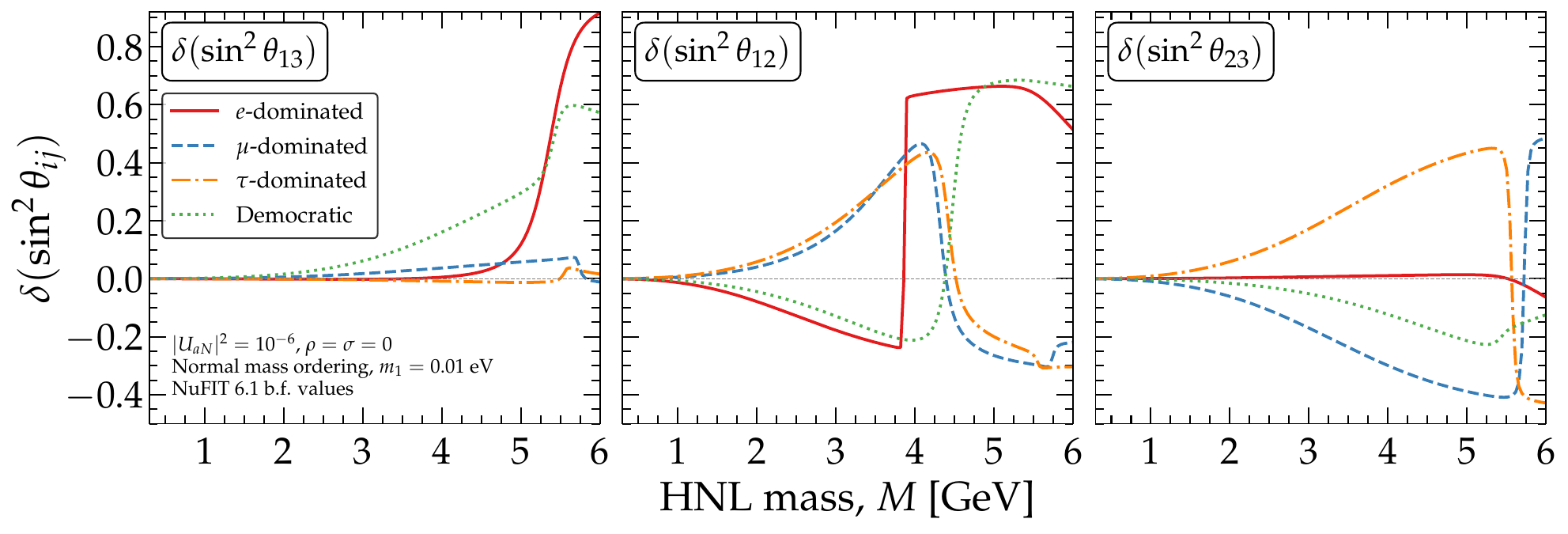}
 \caption{\textbf{Majorana mass threshold correction of the neutrino mixing angles.}  Threshold corrections $\delta(\sin^2\theta_{13})$ (\textit{left}), $\delta(\sin^2\theta_{12})$ (\textit{center}), and $\delta(\sin^2\theta_{23})$ (\textit{right}) as functions of the HNL mass, $M$, computed via full Autonne-Takagi diagonalization. Results are for four benchmark flavor scenarios: electron-philic (only $|U_{e N}|^2$ nonzero), muon-philic (only $|U_{\mu N}|^2$ nonzero), tau-philic (only $|U_{\tau N}|^2$ nonzero), and flavor-democratic (all $|U_{\alpha N}|^2$ nonzero and equal). In all cases, we fix the nonzero $|U_{\alpha N}|^2 = 10^{-6}$ and $\rho = \sigma = 0$ for illustration, and the standard mixing parameters to their best-fit values from NuFIT~6.1~\cite{NuFIT6} (assuming normal neutrino mass ordering, with Super-Kamiokande atmospheric data). The flavor fingerprint on the corrections is clearly visible: a muon- or tau-philic HNL drives $\delta(\sin^2\theta_{23})$, while an electron-philic HNL drives $\delta(\sin^2\theta_{13})$ and $\delta(\sin^2\theta_{12})$.}
 \label{fig:angle_corrections}
\end{figure*}

\medskip

\textbf{\textit{Evolution with HNL mass.---}}Figure~\ref{fig:angle_corrections} shows the evolution with $M$ of the corrections to the mixing angles, computed via full Autonne-Takagi diagonalization. Three distinct physical regimes and features emerge:
\begin{enumerate}
 \item 
  \textbf{\textit{Perturbative growth at low mass.---}}For $M \lesssim 3$~GeV, the corrections to the mixing angles grow smoothly. As predicted by the approximate LEFT threshold matching [\equ{deltamnu_approx}], this evolution is driven primarily by the $M^3$ scaling of $\delta(m_\nu)$. In this low-mass regime, the radiative shift remains a small perturbation compared to the tree-level mass splittings ($\Delta m^2_{21}$ and $\Delta m^2_{31}$), and the angles respond linearly to the perturbation.
 \item
  \textbf{\textit{The flavor fingerprint.---}}Figure~\ref{fig:angle_corrections} illustrates how specific HNL couplings map to specific mixing angles. An electron-philic HNL induces massive shifts in $\theta_{12}$ and $\theta_{13}$ while leaving $\theta_{23}$ nearly flat. Conversely, muon- and tau-philic HNLs strongly drive $\theta_{23}$ in opposite directions. This separation enforces the central complementarity of our work: astrophysical measurements are especially sensitive probes of the muon and tau sectors, bypassing the electron-only blind spots of $0\nu\beta\beta$.
 \item
  \textbf{\textit{Level crossings.---}}As $M$ approaches $4$--$6$~GeV, the $M^3$-enhanced radiative correction becomes large enough to rival the tree-level light neutrino mass scale ($m_i \sim 0.01$~eV). As the shifting mass eigenvalues become nearly degenerate, the mass basis undergoes rapid rotations to avoid level crossing and maintain the prescribed normal mass ordering chosen for \figu{angle_corrections}. These sudden realignments of the eigenvectors manifest as the sharp jumps in the mixing angles. The first-order perturbation breaks down, making exact  diagonalization necessary.
\end{enumerate}

\begin{figure}[t!]
 \centering
 \includegraphics[width=\columnwidth]{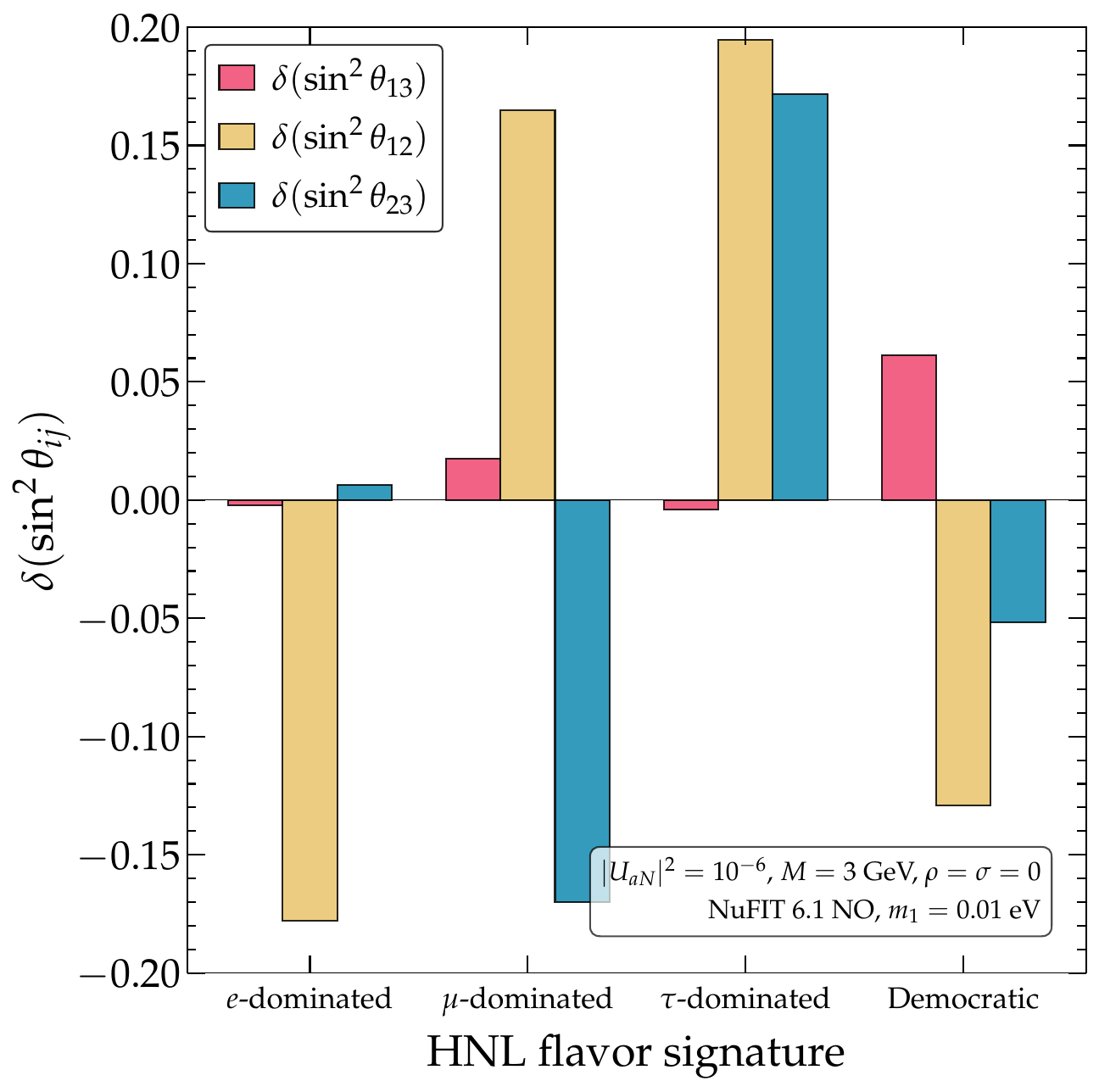}
 \caption{\textbf{Majorana mass threshold correction of the neutrino mixing angles.} The pattern of mixing-angle corrections, [$\delta(\sin^2\theta_{13}),\,\delta(\sin^2\theta_{12}),\,\delta(\sin^2\theta_{23})$], defines the unique flavor fingerprint for four benchmark HNL scenarios. Results are evaluated via exact numerical diagonalization at $M = 3$~GeV, $|U_{\alpha N}|^2 = 10^{-6}$, and $m_1 = 0.01$~eV, with Majorana phases $\rho = \sigma = 0$. Because these corrections---particularly the atmospheric-angle shift $\delta(\sin^2\theta_{23})$---are only weakly sensitive to the unknown Majorana phases, this distinctive flavor mapping is robust. Consequently, measuring the specific HNL flavor profile at \ship\ uniquely predicts the flavor shift in high-energy astrophysical neutrinos.}
 \label{fig:fingerprint}
\end{figure}

\textbf{\textit{The predictive flavor fingerprint.---}}Figure~\ref{fig:fingerprint} illustrates the signature that each HNL flavor profile imprints on the active neutrino mixing angles. An electron-philic HNL suppresses $\theta_{12}$ while leaving the atmospheric angle $\theta_{23}$ entirely untouched; a muon-philic HNL simultaneously enhances $\theta_{12}$ and suppresses $\theta_{23}$; and a tau-philic HNL sharply enhances both. Crucially, this distinctive flavor mapping is a robust prediction. As dictated by the structure of $\xi_3$ [\equ{xi_explicit}] and confirmed across the full parameter space by exact numerical diagonalization, the angle correction $\delta(\sin^2\theta_{23})$ is remarkably insensitive to the unknown Majorana phases $\rho$ and $\sigma$. Because this phase ambiguity is suppressed in the muon and tau sectors, the flavor couplings measured at \ship\ dictate the flavor shifts that neutrino telescopes must observe, providing a robust test of the Majorana nature of the HNL.

\medskip

Later, in Sec.~\ref{sec:astro_flavor_hnl_correction}, we show the effects that the corrections to the mixing parameters have on the flavor composition of high-energy astrophysical neutrinos.


\section{SHiP Sensitivity and the Joint Discovery Window}
\label{sec:SHiP}

\textit{We establish \ship\ as the fixed-target experiment to measure the HNL mass and flavor couplings, which are then used to predict the astrophysical neutrino flavor shift. We map the joint discovery window wherein an HNL can be simultaneously discovered by \ship\ and its induced flavor shift resolved by future neutrino telescopes.}


\subsection{SHiP as the natural fixed-target experiment}

The mass threshold correction derived in Sec.~\ref{sec:threshold} depends on two sets of parameters that must be measured in the laboratory: the HNL mass, $M$, and the three active-sterile mixing elements, $|U_{eN}|^2$, $|U_{\mu N}|^2$, and $|U_{\tau N}|^2$. The natural experiment to provide these inputs is SHiP~\cite{SHiP:2018xqz}.

SHiP is a beam-dump experiment at the CERN Super Proton Synchrotron (SPS) accelerator, which delivers 400-GeV protons on target.  SHiP is designed to search for feebly interacting particles with masses in the GeV range. Its sensitivity to HNLs arises from the production of an intense flux of charmed and bottom mesons in the target, which decay into HNLs via the active-sterile mixing $|U_{\alpha N}|^2$. The HNLs travel through a $\sim$50-m-long shielded decay vessel and produce detectable signals through their subsequent decays, $N \to \ell^\pm + W^{\mp*}$ or $N \to \nu + Z^*$. In particular, the \emph{flavor} of the final-state charged lepton tags the active-sterile mixing element responsible for the production vertex, making SHiP sensitive to $|U_{eN}|^2$, $|U_{\mu N}|^2$, and $|U_{\tau N}|^2$ \emph{separately}~\cite{SHiP:2018xqz}.

This flavor resolution is essential for the present analysis. As shown in Eqs.~\eqref{equ:delta_s13}--\eqref{equ:delta_s23_explicit}, the corrections to different active mixing angles depend on different combinations of the flavor couplings: $\delta(\sin^2\theta_{23})$ is governed primarily by $\xi_3 \propto |U_{\mu N}| + |U_{\tau N}|$ (for small $\theta_{13}$ and maximal $\theta_{23} \approx 45^\circ$) and is therefore sensitive to the muon and tau couplings; $\delta(\sin^2\theta_{12})$ depends on $\xi_1$ and $\xi_2$ and is sensitive to all three flavor couplings, with a strong dependence on the Majorana phases; and $\delta(\sin^2\theta_{13})$ involves products $\xi_3\xi_1$ and $\xi_3\xi_2$ and is more sensitive to $U_{eN}$. A flavor-blind measurement of the total mixing without resolving the individual $|U_{\alpha N}|^2$ would leave the flavor pattern---and hence the predicted size and direction of each angle correction---undetermined.

The SHiP sensitivity mass range, $M \sim 0.3$--$6$~GeV, is not merely compatible with the threshold-correction framework of Sec.~\ref{sec:threshold}: it is required by it. As shown in Sec.~\ref{sec:astro_flavor-exact_observables}, it is this mass range for which the threshold falls within the momentum-transfer window $Q \sim 1$--$10$~GeV probed by TeV--PeV astrophysical neutrinos.


\subsection{SHiP sensitivity to HNLs}
\label{subsec:sensitivity}

Figure~\ref{fig:discovery_window} (also \figu{discovery_window_var_angles_UmuN}) shows the SHiP sensitivity in the $M$--$|U_{\alpha N}|^2$ plane in the electron-, muon-, and tau-philic HNL cases, reproduced from \Refe~\cite{SHiP:2018xqz}. This is the parameter space accessible to SHiP at $90\%$~C.L.~after accumulating $2 \times 10^{20}$ protons on target.  SHiP will be able to improve on previous HNL searches---which come mainly from colliders---by orders of magnitude, especially in the weakly constrained case of a tau-philic HNL.

Below $M \sim$ 2 GeV, SHiP exhibits exceptional sensitivity to HNLs due to the immense number of charmed mesons produced at the SPS. In this light-mass regime, HNL production is overwhelmingly dominated by the decays of abundant $D$ and $D_s$ mesons. Because the $c\bar{c}$ production cross-section is large at the 400-GeV beam energy ($\sigma_{c\bar{c}} \approx 18\ \mu\text{b}$), the resulting vast parent flux allows SHiP to constrain exquisitely small $|U_{\alpha N}|^2$. However, this production channel is bounded by the kinematic limit of the $D_s$ meson ($M_{D_s} \approx 1.97$ GeV). For HNL masses exceeding this boundary, the charm channels close, and production must proceed exclusively via the decays of heavier bottom mesons. At these beam energies, the $b\bar{b}$ cross-section ($\sigma_{b\bar{b}} \sim 1$~nb) is suppressed by a factor of $\sim 10^4$ relative to charm production. This abrupt depletion of the source flux requires a correspondingly larger mixing element to maintain an observable event rate, manifesting as the cliff in the sensitivity boundary in \figu{discovery_window} exactly at the charm threshold.

Figure \ref{fig:discovery_window} also shows the variation of the mixing-angle correction $\delta(\sin^2 \theta_{23})$ within the SHiP sensitivity region.  Because $\delta(\sin^2\theta_{ij}) \propto M^3 |U_{\alpha N}|^2$ [\equ{deltamnu_approx}], larger $|U_{\alpha N}|^2$ yield larger angle corrections at fixed $M$. The figure shows that, \textit{regardless of the HNL flavor scenario}, SHiP will be able to test corrections in the unexplored range $\delta(\sin^2 \theta_{23}) \sim 10^{-6}$--$10^{-1}$, with the exact value depending on the HNL mass. This sensitivity extends to the corrections to $\theta_{12}$ and $\theta_{13}$, as we show next.

Figure~\ref{fig:discovery_window_var_angles_UmuN} shows, for the specific case of a muon-philic HNL, the variation of the different angle corrections, $\delta(\sin^2 \theta_{12})$, $\delta(\sin^2 \theta_{23})$, and $\delta(\sin^2 \theta_{13})$, within the SHiP sensitivity region. The correction to $\theta_{23}$ is dominant, while the corrections to $\theta_{12}$ and $\theta_{13}$ are suppressed, because $\xi_3 \propto |U_{\mu N}| + |U_{\tau N}| \gg \xi_1, \xi_2$; the same holds for a tau-philic HNL, show in \figu{discovery_window_var_angles_UtauN} in Appendix~\ref{app:var_angles_UeN_UtauN}.  Conversely, for the electron-philic case, shown in \figu{discovery_window_var_angles_UeN}, the corrections for $\theta_{12}$ and $\theta_{13}$ are larger than for $\theta_{23}$. 

The above observations on the SHiP sensitivity to HNLs prefigure two key consequences for the counterpart search of the HNL-induced flavor shifts of high-energy astrophysical neutrinos that we explore later (Sec.~\ref{sec:astro_flavor_hnl_correction}).  First, because the size of the angle corrections can be tiny, so can the predicted HNL-induced flavor shifts be, complicating their experimental identification.  Second, the hierarchy in the size of the corrections to the different angles---and in the variation of this hierarchy across different HNL flavor scenarios---impacts not only the size of the flavor shift, but also its direction. The size and direction of a predicted HNL-induced flavor shift determine whether neutrino telescopes will be able to detect it.


\subsection{The joint HNL discovery window}
\label{subsec:window}

\begin{figure*}[t]
 \centering
 \includegraphics[width=\textwidth]{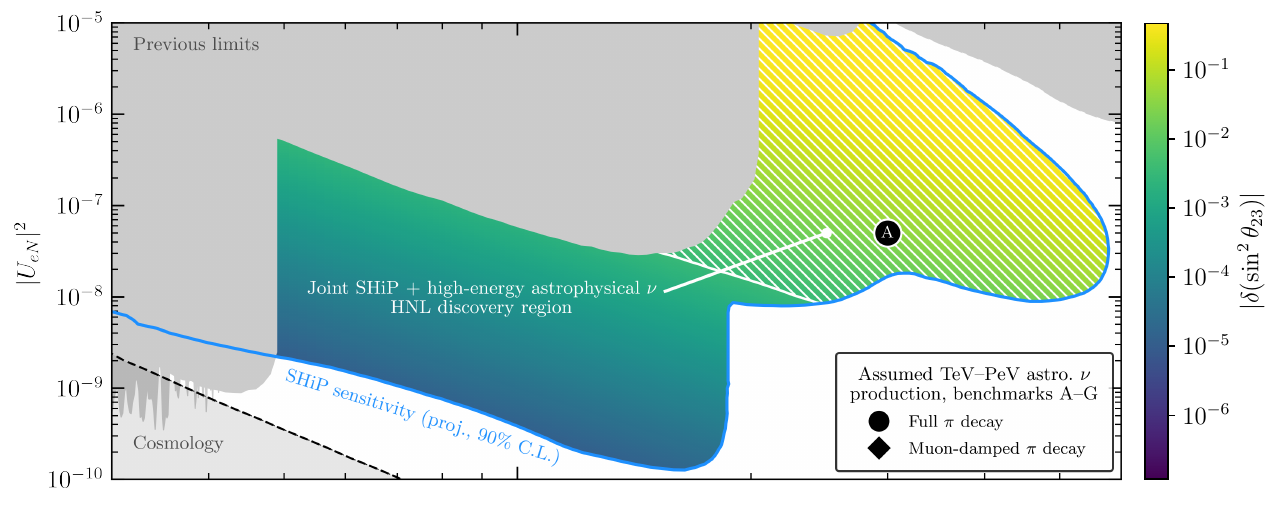}
 \\[-7pt]
 \includegraphics[width=\textwidth]{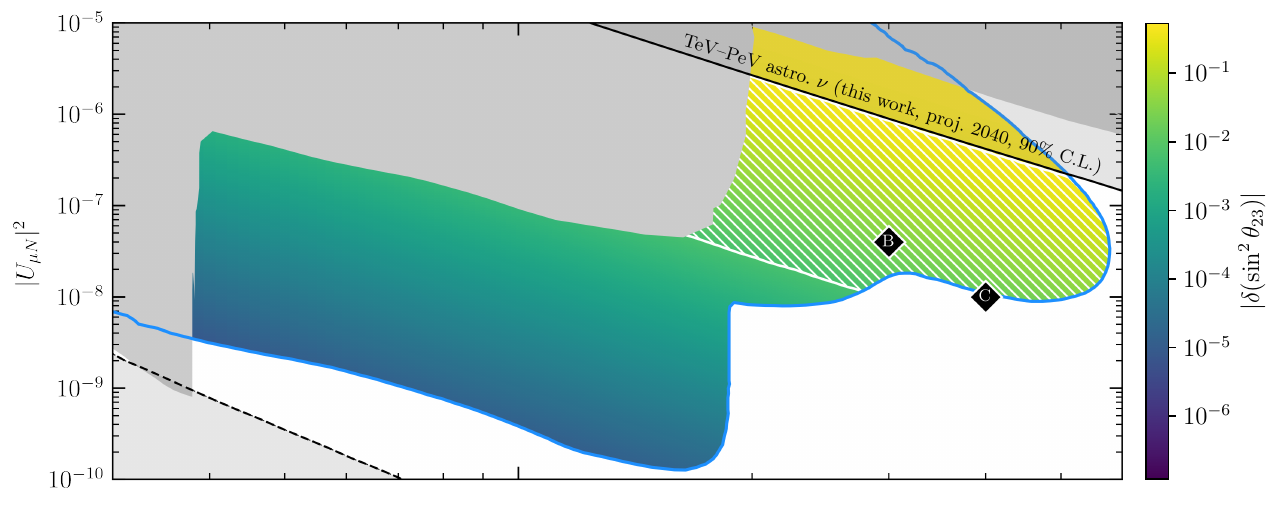}
 \\[-7pt]
 \includegraphics[width=\textwidth]{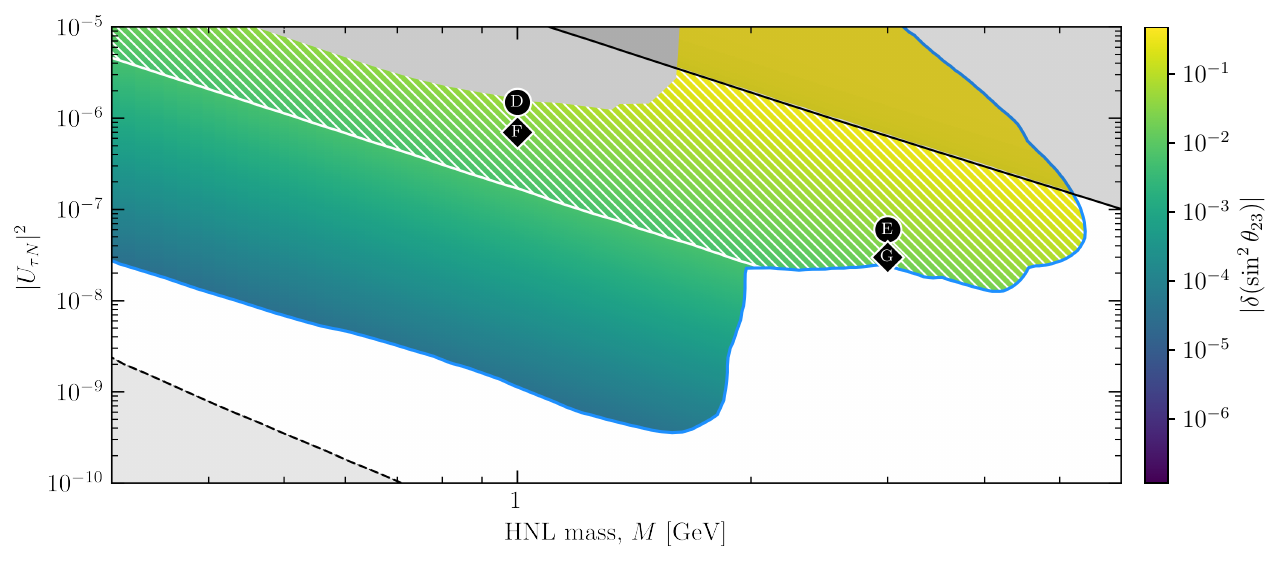}
 \caption{\textbf{Joint SHiP and high-energy astrophysical neutrino HNL discovery window.} The color fill shows the mass-threshold correction to the neutrino mixing angle $|\delta(\sin^2\theta_{23})|$, restricted to the interior of the SHiP 90\%~C.L. sensitivity~\cite{SHiP:2018xqz}. Previous limits are from colliders and cosmology, as summarized in \Refe~\cite{Fernandez-Martinez:2023phj}; we use \Refe~\cite{hnl_github} to import them.  The new projected limits from TeV--PeV astrophysical neutrinos are derived in Sec.~\ref{sec:limits_and_discovery-results}; we show them specifically assuming neutrino production via full pion decay. The joint discovery region is where SHiP-measured HNL parameters imply shifts to the flavor composition of astrophysical neutrinos that are allowed by present measurements and discoverable by future ones. Benchmark points A--G (Table~\ref{tab:benchmark_points}) are selected to illustrate joint HNL discovery prospects later. See Sec.~\ref{subsec:window} for details.}
\label{fig:discovery_window}
\end{figure*}

\begin{figure*}[t]
 \centering
 \includegraphics[width=\textwidth]{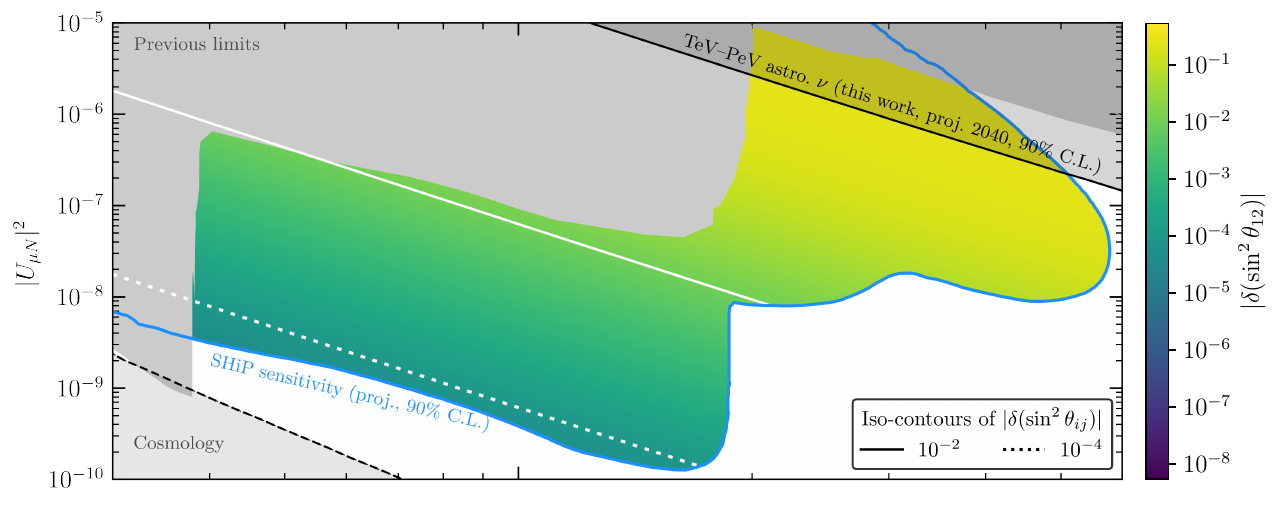}
 \\[-7pt]
 \includegraphics[width=\textwidth]{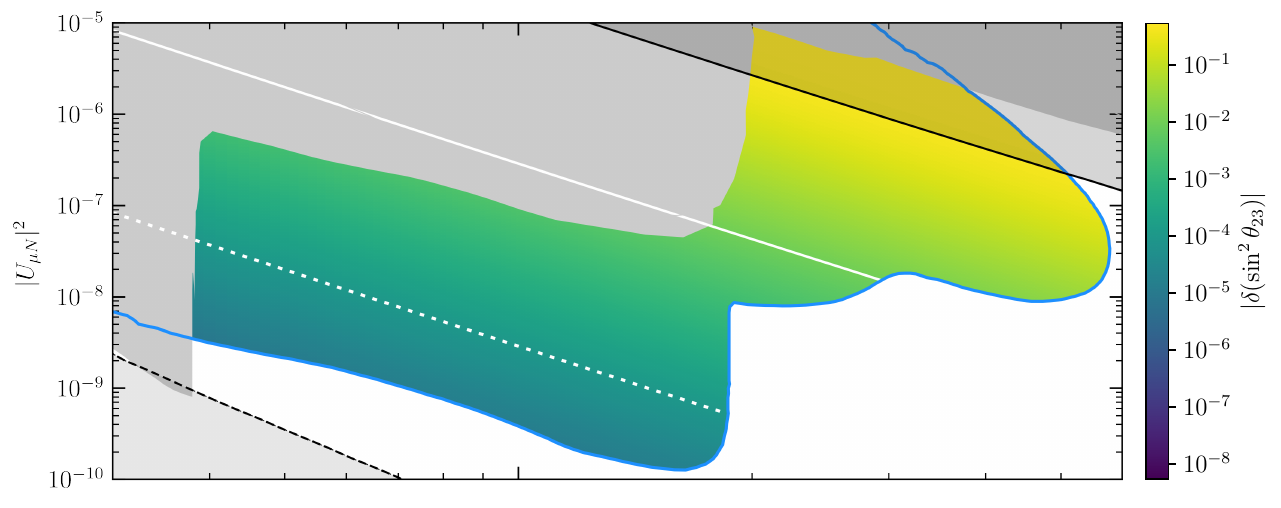}
 \\[-7pt]
 \includegraphics[width=\textwidth]{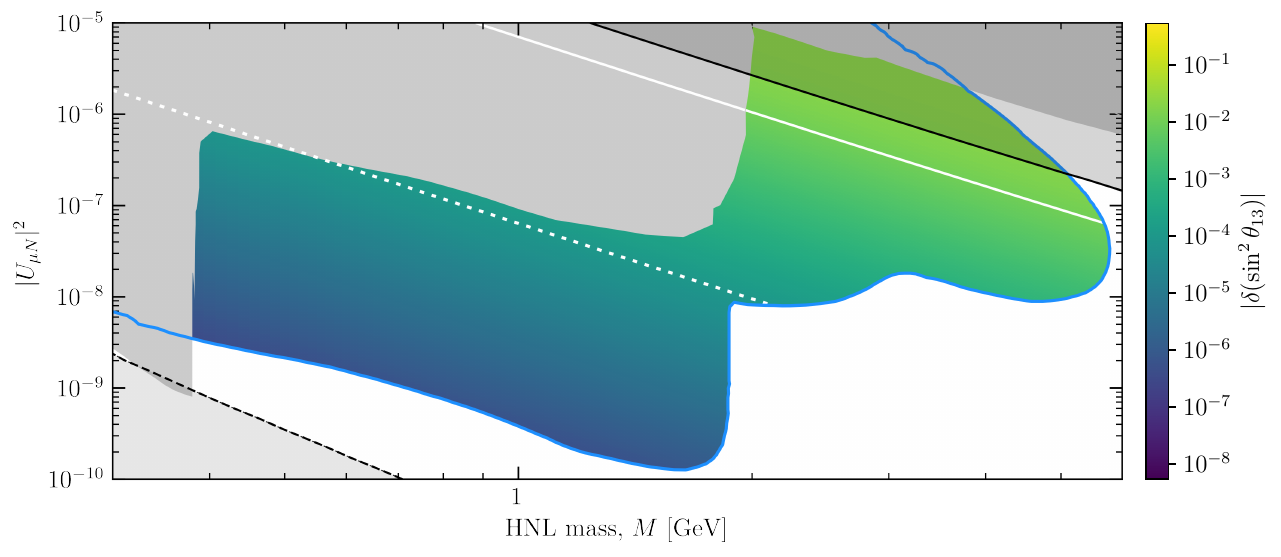}
 \caption{\textbf{Mass-threshold correction of the active neutrino mixing angles for a muon-philic HNL.} The corrections are $\delta(\sin^2 \theta_{12})$ (\textit{top}), $\delta(\sin^2 \theta_{23})$ (\textit{center}), and $\delta(\sin^2 \theta_{13})$ (\textit{bottom}). In the muon-philic scenario shown here, $|U_{\mu N}|$ is the only non-zero active-sterile mixing element. The SHiP sensitivity and limits are the same as in \figu{discovery_window}, and we retain NO and the best-fit values of the standard mixing parameters from NuFIT~6.1 (with Super-Kamiokande atmospheric data).  See Sec.~\ref{subsec:sensitivity} for details and Figs.~\ref{fig:discovery_window_var_angles_UeN} and \ref{fig:discovery_window_var_angles_UtauN} in Appendix \ref{app:var_angles_UeN_UtauN} for analogous results in the electron- and tau-philic HNL cases.}
\label{fig:discovery_window_var_angles_UmuN}
\end{figure*}

The \emph{joint HNL discovery window} of SHiP and high-energy neutrino telescopes is the region of $M$--$|U_{\alpha N}|^2$ space satisfying three conditions simultaneously:
\begin{enumerate}[label=(\roman*)]
 \item
  \textbf{SHiP:} The combination $(M,\, |U_{\alpha N}|^2)$ lies within the SHiP $90\%$~C.L.~sensitivity region shown in Figs.~\ref{fig:discovery_window} and \ref{fig:discovery_window_var_angles_UmuN}, so that the HNL can be discovered and its mass and mixing elements measured.
 \item 
  \textbf{Upper limits:} The combination $(M,\, |U_{\alpha N}|^2)$ lies below current experimental upper bounds from direct HNL searches~\cite{Bondarenko:2018ptm} and electroweak precision data~\cite{deBlas:2013qqa}, as summarized in Figs.~\ref{fig:discovery_window} and \ref{fig:discovery_window_var_angles_UmuN}.  (It also lies below the new projected upper limits that we derive later, in Sec.~\ref{sec:limits_and_discovery}, which are shown \figu{discovery_window}.)
 \item
  \textbf{Astrophysical:} The mixing-angle corrections, $|\delta(\sin^2\theta_{ij})|$, computed from the SHiP-measured $(M,\, |U_{\alpha N}|^2)$, yield a flavor composition of high-energy astrophysical neutrinos that is allowed by current measurements~\cite{Abbasi:2025fjc} and within the projected reach of combinations of neutrino telescopes, as quantified later in Sec.~\ref{sec:astro_flavor-flavor_regions}. In \figu{discovery_window}, we consider the breadth of flavor shifts spanned by assuming the likely cases of neutrino production via full or muon-damped pion decay (Sec.~\ref{sec:astro_production}).
\end{enumerate}

Figure~\ref{fig:discovery_window} shows the resulting joint discovery window for electron-, muon-, and tau-philic HNLs.  For electron- and muon-philic HNLs, the joint discovery window spans roughly $1.5 \lesssim M/\mathrm{GeV} \lesssim 6$, while for tau-philic HNLs it spans a wider range, $0.3 \lesssim M/\mathrm{GeV} \lesssim 5$, on account of the existing limits being weaker (Sec.~\ref{subsec:sensitivity}).  In all cases, the joint discovery window spans $|U_{\alpha N}|^2 \gtrsim 10^{-8}$, with the exact value depending on $M$.

The joint discovery window favors relatively high values of $M$, $|U_{\alpha N}|^2$, or both, given that the mass-threshold correction $\delta(m_\nu) \propto M^3 |U_{\alpha N}|^2$ [\equ{deltamnu_approx}]. Within the joint discovery region, the angle corrections are relatively large---in \figu{discovery_window}, $\delta(\sin^2 \theta_{23}) \gtrsim 10^{-3}$, and similarly large corrections to the other angles---leading to large, detectable shifts in the flavor composition of high-energy astrophysical neutrinos.  

The seven benchmark points A--G in \figu{discovery_window}, which we introduce in detail in Sec.~\ref{sec:astro_flavor-testable_scenarios}, are chosen to lie within the joint discovery windows of the different HNL flavor scenarios.  Later (Sec.~\ref{sec:limits_and_discovery}), we show how neutrino telescopes will be able to detect the mass-threshold corrections associated with these benchmarks, and to reconstruct $M$ and $|U_{\alpha N}|^2$ independently from SHiP.

\subsection{Dirac vs.\ Majorana at SHiP alone?}
\label{subsec:SHiP_dirac}

A natural question is whether SHiP itself can establish the Majorana nature of a discovered HNL, without invoking astrophysical neutrinos. SHiP can attempt this via the same-sign dilepton channel: a Majorana HNL produces $N \to \ell^+ + W^{-*}$ and $N \to \ell^- + W^{+*}$ with equal rates, yielding an LNV same-sign signal, whereas a Dirac HNL produces only lepton-number-conserving opposite-sign pairs. The ratio
\begin{equation}
 R_\text{LNV} \equiv
 \frac{N(\ell^+ \ell^+) + N(\ell^- \ell^-)}{N(\ell^+ \ell^-)}
 = \begin{cases} 1 & (\text{Majorana}) \\ 0 & (\text{Dirac}) \end{cases}
 \label{equ:RLNV}
\end{equation}
is the canonical SHiP discriminator. However, a statistically conclusive LNV signal requires $\mathcal{O}(10)$ same-sign events above background~\cite{SHiP:2018xqz}, which may not be achievable for couplings near the lower boundary of the SHiP sensitivity contour---precisely the region of most interest for the joint SHiP-astrophysics program we propose.

In that part of the window where the SHiP LNV measurement is inconclusive, the astrophysical neutrino measurement provides an independent and qualitatively different probe of the Majorana nature. Crucially, the two approaches have completely uncorrelated systematic uncertainties: the SHiP measurement depends on detector efficiency and backgrounds in the decay vessel, while the astrophysical measurement depends on source modeling and atmospheric neutrino and muon backgrounds. This makes their joint interpretation especially robust against experiment-specific systematics. This is the sense in which the two programs are not merely complementary in parameter space, but also in method.


\subsection{Robustness of flavor-coupling measurements}
\label{subsec:SHiP_robust}

Two sources of uncertainty in the SHiP measurement of the mixing elements $|U_{\alpha N}|$ deserve explicit discussion.

\medskip

\textbf{\textit{Tau-flavor coupling.---}}SHiP can measure $|U_{\tau N}|^2$ only through final states containing a tau lepton, which have lower reconstruction efficiency than electron or muon channels due to the missing energy carried away by $\nu_\tau$, and to the complexity of distinguishing hadronic tau decays from background. Consequently, SHiP's sensitivity to $|U_{\tau N}|^2$ is weaker by roughly an order of magnitude compared to the other flavors~\cite{SHiP:2018xqz}, as shown in \figu{discovery_window}. This matters because $\delta(\sin^2\theta_{23})$ is governed by $\xi_3 \propto |U_{\mu N}| + |U_{\tau N}|$; if $|U_{\tau N}|$ is dominant, the predicted correction may be underestimated when using SHiP inputs alone. We account for this by treating $|U_{\tau N}|^2$ as a nuisance parameter when projecting limits on or discovery potential of $|U_{e N}|^2$ and $|U_{\mu N}|^2$ from high-energy astrophysical neutrinos in Sec.~\ref{sec:limits_and_discovery}.

\medskip

\textbf{\textit{Majorana phase uncertainty.---}}Predicting the TeV--PeV astrophysical neutrino flavor composition at Earth based on SHiP-measured HNL parameters requires knowledge of the modified mixing parameters. However, these modifications depend, in general, on the Majorana CP-violation phases, $\rho$ and $\sigma$ (Sec.~\ref{subsec:corrections_mixing_angles}), which remain unconstrained by experiments like SHiP. We account for this by treating $\rho$ and $\sigma$ as nuisance parameters when projecting limits on or discovery potential of $|U_{\alpha N}|^2$ from high-energy astrophysical neutrinos in Sec.~\ref{sec:limits_and_discovery}.


\section{High-Energy Astrophysical Neutrinos}
\label{sec:astro_production}

\textit{We establish the experimental framework for the detection and flavor-composition measurement of the diffuse high-energy astrophysical neutrino flux. We detail the mechanism of optical-Cherenkov detection, the challenges of flavor identification, and the projected sensitivities of next-generation multi-detector networks. This establishes the observational baseline required to resolve the HNL-induced mass-threshold corrections defined in Sec.~\ref{sec:threshold}.}


\subsection{Neutrino production and the diffuse flux}
\label{sec:astro_production-prod}

High-energy astrophysical neutrinos, with TeV--PeV energies, are produced in cosmic accelerators where tens-of-PeV protons interact with ambient matter~\cite{Margolis:1977wt, Stecker:1978ah, Kelner:2006tc} or radiation~\cite{Stecker:1978ah, Mucke:1999yb, Kelner:2008ke, Hummer:2010vx}. This yields intermediate mesons---primarily, charged pions---that subsequently decay into neutrinos, \ie, $\pi^- \to \mu^- + \bar{\nu}_\mu$, followed by $\mu^- \to e^- + \bar{\nu}_e + \nu_\mu$, and the charge-conjugated processes. 

We focus exclusively on the isotropic diffuse flux of high-energy neutrinos---the integrated emission from all unresolved, predominantly extragalactic cosmic accelerators---since it is from this flux that flavor-composition measurements are inferred, as we explain later.  (Tens-of-TeV neutrinos from the Milky Way Plane~\cite{IceCube:2023ame} fall below the energy window of the flavor-composition measurements we adopt here.)  For this diffuse flux, the propagation baselines, between tens of Mpc and a few Gpc, exceed the oscillation length, guaranteeing the mass eigenstates arrive at Earth as a completely decohered ensemble of $\nu_1$, $\nu_2$, and $\nu_3$, which justifies our use of average flavor-transition probabilities in Sec.~\ref{sec:astro_flavor-std}. 


\subsection{The flavor composition at the sources}

Different high-energy neutrino production mechanisms (Sec.~\ref{sec:astro_production}) yield different proportions of $\nu_e$, $\nu_\mu$, and $\nu_\tau$ at their astrophysical sources. We denote the fraction of $\nu_\alpha + \bar{\nu}_\alpha$ in the total emitted flux ($\alpha = e, \mu, \tau$) as $f_{\alpha, {\rm S}} \in [0, 1]$, normalized such that $\sum_\alpha f_{\alpha, {\rm S}} = 1$.  We do not distinguish between $\nu_\alpha$ and $\bar{\nu}_\alpha$ because neither do high-energy neutrino telescopes like IceCube.

Neutrino production via full pion decay yields $(f_{e, {\rm S}}, f_{\mu, {\rm S}}, f_{\tau, {\rm S}}) = \left( \frac{1}{3}, \frac{2}{3}, 0 \right)$, which represents our nominal expectation. In astrophysical environments with strong magnetic fields, the intermediate muons may cool by synchrotron radiation prior to decaying, yielding $(0, 1, 0)_{\rm S}$~\cite{Kashti:2005qa, Kachelriess:2007tr, Lipari:2007su}. This \textit{muon-damped} scenario is expected to become dominant at the highest energies (see, \eg, \Refes~\cite{Winter:2013cla, Bustamante:2020bxp}), where the muon lifetime is sufficiently time-dilated to ensure severe energy losses prior to decay.  Henceforth, we adopt these two cases as our benchmarks. (A third mechanism is the beta-decay of neutrons~\cite{Anchordoqui:2003vc}, $n \to p + e^- + \bar{\nu}_e$, which yields $(1, 0, 0)_{\rm S}$. We exclude this scenario from our benchmarks because the $\bar{\nu}_e$ have lower energies than the neutrinos from pion decay.)

Astrophysical sources yield a negligible $\nu_\tau$ fraction due to the high kinematic threshold for tau-lepton production~\cite{Farzan:2021gbx}, ensuring $f_{\tau, \rm S} \approx 0$. Although exotic mechanisms like heavy dark matter decay~\cite{Feldstein:2013kka, Esmaili:2013gha, Bhattacharya:2014vwa}, leptoquark interactions~\cite{Barger:2013pla}, or sterile neutrinos~\cite{Arguelles:2019tum, Ahlers:2020miq} can produce substantial tau-neutrino fluxes, we focus strictly on the standard paradigm. Thus, the source flavor composition is inherently restricted to $\nu_e$ and $\nu_\mu$ mixtures, parameterized as $(f_{e, {\rm S}}, 1 - f_{e, {\rm S}}, 0)_{\rm S}$, which we consider when scanning the parameter space as part of our statistical analyses (Sec.~\ref{sec:limits_and_discovery}).  Further, in-source matter interactions are generally not expected to alter these ratios prior to escape~\cite{Mena:2006eq, Razzaque:2009kq, Sahu:2010ap, Varela:2014mma, Xiao:2015gea} (though see \Refe~\cite{Dev:2023znd}).

Although flavor composition is expected to vary with energy~\cite{Kashti:2005qa, Lipari:2007su}, measuring this dependence is, and will remain, difficult~\cite{Liu:2023flr}. Consequently, we adopt energy-independent flavor ratios that effectively represent  energy-averaged quantities. Further, lacking any current observational evidence for directional dependence~\cite{Telalovic:2023tcb, Telalovic:2025xor}, we assume the flavor ratios are isotropic across the neutrino sky.


\subsection{The flavor composition at the Earth}
\label{sec:astro_flavor-std}

The flavor composition is a versatile probe of astrophysics~\cite{Rachen:1998fd, Athar:2000yw, Crocker:2001zs, Barenboim:2003jm, Beacom:2003nh, Beacom:2004jb, Kashti:2005qa, Mena:2006eq, Kachelriess:2006ksy, Lipari:2007su, Esmaili:2009dz, Choubey:2009jq, Hummer:2010ai, Winter:2013cla, Palladino:2015zua, Bustamante:2015waa, Biehl:2016psj, Bustamante:2019sdb, Ackermann:2019ows, Bustamante:2020bxp, Song:2020nfh, AlvesBatista:2021eeu, Liu:2023lxz, Bhattacharya:2023mmp, Telalovic:2023tcb, Dev:2023znd} and fundamental physics~\cite{Beacom:2002vi, Barenboim:2003jm, Beacom:2003nh, Beacom:2003eu, Beacom:2003zg, Serpico:2005bs, Mena:2006eq, Lipari:2007su, Pakvasa:2007dc, Esmaili:2009dz, Choubey:2009jq, Esmaili:2009fk, Bhattacharya:2009tx, Bhattacharya:2010xj, Bustamante:2010nq, Mehta:2011qb, Baerwald:2012kc, Fu:2012zr, Pakvasa:2012db, Chatterjee:2013tza, Xu:2014via, Aeikens:2014yga, Arguelles:2015dca, Bustamante:2015waa, Pagliaroli:2015rca, Shoemaker:2015qul, deSalas:2016svi, Gonzalez-Garcia:2016gpq, Bustamante:2016ciw, Rasmussen:2017ert, Dey:2017ede, Bustamante:2018mzu, Farzan:2018pnk, Ahlers:2018yom, Brdar:2018tce, Palladino:2019pid, Ackermann:2019cxh, Arguelles:2019rbn, Ahlers:2020miq, Karmakar:2020yzn, Fiorillo:2020gsb, Song:2020nfh, AlvesBatista:2021eeu, Arguelles:2022tki, MammenAbraham:2022xoc, Telalovic:2023tcb, Liu:2024wmk, Telalovic:2025xor, Bustamante:2026aur, Bustamante:2026zst}.
Because the baselines traversed by high-energy astrophysical neutrinos are vastly larger than the neutrino oscillation coherence length, the mass eigenstates separate into independent wave packets. The transition probability for a neutrino produced with flavor $\alpha$ to be detected with flavor $\beta$ at Earth is given by the incoherent sum over the mass eigenstates $i = 1,2,3$:
\begin{equation}
 P_{\alpha\beta}^{(0)} = \sum_{i=1}^{3} |U_{\alpha i}|^2 |U_{\beta i}|^2 \,,
 \label{equ:P_standard}
\end{equation}
where $U$ is the standard low-energy PMNS matrix. The expected flavor fractions at Earth are then
\begin{equation}
 f_{\beta, \oplus}^{(0)} = \sum_{\alpha} f_{\alpha, {\rm S}} P_{\alpha\beta}^{(0)} 
 = \sum_{i=1}^3 W_i |U_{\beta i}|^2 \,,
 \label{equ:f_earth_standard}
\end{equation}
where the intermediate flavor weights $W_i \equiv \sum_\alpha f_{\alpha, {\rm S}} |U_{\alpha i}|^2$ (we use them later when introducing the HNL modifications of the flavor ratios).

Evaluating \equ{f_earth_standard} using the NuFIT~6.1~\cite{NuFIT6} global best-fit values of the standard mixing parameters (assuming normal mass ordering with Super-Kamiokande atmospheric data) establishes the standard-mixing expectations for our two benchmark scenarios. For neutrino production via full pion decay,  the expected flavor composition at Earth is near the flavor-democratic center of the flavor triangle, yielding $f_{\oplus}^{(0)} \approx (0.30, 0.36, 0.34)$. For muon-damped pion decay, $f_{\oplus}^{(0)} \approx (0.18, 0.45, 0.37)$.

\subsection{Measuring the flavor composition}
\label{sec:astro_production-flavor_measurement}

In optical-Cherenkov neutrino telescopes~\cite{Markov:1961tyz} such as IceCube~\cite{IceCube:2013low}, KM3NeT~\cite{KM3Net:2016zxf}, Baikal-GVD~\cite{Baikal-GVD:2025rhg}, and upcoming ones, TeV--PeV neutrinos are detected via deep inelastic scattering (DIS) on nucleons~\cite{CTEQ:1993hwr, Conrad:1997ne, Formaggio:2012cpf} within the detector medium (ice or water). The momentum transfer in these interactions peaks broadly at $Q \simeq 20$~GeV~\cite{Bustamante:2026zst} (Sec.~\ref{sec:astro_flavor-exact_observables}),  exceeding the GeV-scale HNL masses $M$ targeted in this work. These DIS interactions proceed via charged-current (CC) or neutral-current (NC) channels. The final-state charged particles shower and emit Cherenkov radiation, which is detected by an array of photomultiplier tubes. From the spatial and temporal distribution of the detected light, analyses infer the arrival time, energy, direction, and flavor of the interacting neutrino, to varying degrees of success.

High-energy neutrino telescopes classify neutrino-induced events into three primary topologies:
\begin{itemize}
 \item
  \textbf{Cascades:} Produced by CC $\nu_e$ and CC $\nu_\tau$ interactions, as well as NC interactions from all flavors.
 \item \textbf{Tracks:}
  Produced primarily by CC $\nu_\mu$ interactions, where the final-state muon traverses km-scale distances, and by CC $\nu_\tau$ interactions where the tau undergoes muonic decay ($\approx 18\%$ branching ratio).
 \item
  \textbf{Double cascades:} A unique signature of CC $\nu_\tau$ interactions occurring when a first shower from the DIS vertex and a second shower from the subsequent tau decay are resolved.
\end{itemize}
The indistinguishability between CC $\nu_e$ showers and the hadronic showers of NC or untagged CC $\nu_\tau$ events imposes a fundamental degeneracy between $\nu_e$ and $\nu_\tau$ that precludes event-by-event flavor identification. The flavor fractions at Earth, $f_{\alpha, \oplus}$, are instead extracted statistically from the relative numbers of events of each topology, which map imperfectly onto flavors.

Traditionally, these analyses use ``starting events'', where the neutrino interaction occurs within the instrumented volume and all event topologies can be distinguished, providing sensitivity to all flavors. However, these analyses are limited by the low event rate of astrophysical neutrinos (about 10 neutrinos per ${\rm km}^3$ per year over 60 TeV~\cite{IceCube:2020wum}).  To overcome this, flavor measurements can be augmented with through-going tracks---events where a $\nu_\mu$ interacts outside the instrumented volume, producing a muon that crosses the detector. While these events only constrain the $\nu_\mu$ fraction, they are more numerous.  By combining them with starting events, they  tighten the inferred flavor composition, as shown in \Refe~\cite{IceCube:2015gsk}.  So far, flavor measurements have been carried out only on IceCube data, either by the IceCube Collaboration~\cite{IceCube:2015rro, IceCube:2015gsk, IceCube:2018pgc, IceCube:2020fpi, Abbasi:2025fjc} or externally to it~\cite{Mena:2014sja, Palomares-Ruiz:2015mka, Palladino:2015zua}.


\subsection{Present and future flavor measurements}
\label{sec:astro_production-flavor_present_future}

For our present-day results, we adopt the flavor measurements based on the 11.4-year IceCube Medium Energy Starting Events (MESE) sample~\cite{Abbasi:2025fjc}, which report non-zero contributions from all three active flavors at the 68\% C.L. MESE events span from 1~TeV to 10~PeV and have high astrophysical purity~\cite{IceCube:2014rwe, IceCube:2025tgp}. However, even for this modern flavor measurement, the uncertainties on $f_{\alpha, \oplus}$ remain large---at roughly 30\%---encompassing the standard-mixing expectation and obscuring potential sub-leading high-energy perturbations like the ones we search for from HNLs (more on these later).  Figure~\ref{fig:hnl_flavor_contours} shows the approximation to the IceCube MESE flavor contour from \Refe~\cite{Abbasi:2025fjc} that we adopt in our work.

For our projections, we infer the flavor composition from a combination of High-Energy Starting Events (HESE)~\cite{Schonert:2008is, IceCube:2013low, Gaisser:2014bja, IceCube:2014stg, Arguelles:2018awr, IceCube:2020wum} and through-going muon tracks~\cite{IceCube:2019cia, IceCube:2021xar}, in the spirit of \Refe~\cite{IceCube:2015gsk}. We forecast multi-telescope detection in existing IceCube, Baikal-GVD, and KM3NeT, plus future telescopes~\cite{MammenAbraham:2022xoc, Ackermann:2022rqc} P-ONE~\cite{P-ONE:2020ljt}, IceCube-Gen2~\cite{IceCube-Gen2:2020qha}, NEON~\cite{Zhang:2024slv}, TRIDENT~\cite{TRIDENT:2022hql}, and HUNT~\cite{Huang:2023mzt}---up to 30 times the size of IceCube---by scaling IceCube event rates by the detector size, as in \Refe~\cite{Schumacher:2025qca} (also \Refes~\cite{Song:2020nfh, Fiorillo:2022rft, Telalovic:2023tcb, Liu:2023flr}).  For details, see \Refes~\cite{Liu:2023flr, Bustamante:2026aur, Bustamante:2026zst}, from which we adopt our benchmark  projections for 2040 and 2050 unchanged.

By 2040, combined multi-decade exposures from the baseline km$^3$-scale arrays (IceCube, Baikal-GVD, KM3NeT) will shrink current uncertainties. By 2050, the integration of additional km$^3$-scale P-ONE and multi-km$^3$ facilities---IceCube-Gen2 and HUNT---will yield precision contours capable of resolving deviations in the individual flavor fractions of order 5\% from standard-mixing expectations. Figure~\ref{fig:hnl_flavor_contours} shows our 2040 and 2050 flavor-measurement projections, for neutrino production via full and muon-damped pion decay (Sec.~\ref{sec:astro_production-prod}).

This measurement precision defines the testability threshold for the new physics. By establishing a tightly constrained null hypothesis around the standard-mixing expectation, our 2040 and 2050 multi-detector projections provide the resolving power necessary to detect the subtle, HNL-induced high-energy deviations in the flavor fractions at Earth, as we demonstrate later. 

(Future analyses could mitigate the $\nu_e$-$\nu_\tau$ degeneracy by distinguishing $\nu_e$-induced electromagnetic showers from the largely hadronic showers induced by $\nu_\tau$. This morphological separation leverages late-time Cherenkov-light ``echoes'' produced by low-energy muon decays and neutron captures~\cite{Li:2016kra}. While preliminary IceCube analyses have successfully demonstrated the neutron-capture technique~\cite{Steuer:2017tca, Farrag:2023jut, Dutta:2025qgk}, we conservatively exclude these echo-based enhancements from our projections.)


\section{Majorana corrections to the flavor composition}
\label{sec:astro_flavor_hnl_correction}

\textit{We derive the flavor shifts of the high-energy astrophysical neutrinos induced by the Majorana HNL mass-threshold correction, accounting for the momentum-dependent neutrino flavor-transition probabilities. We develop approximate expressions to help us understand the dependence of these flavor shifts on the active-sterile couplings and Majorana phases, followed by full numerical scans of the allowed parameter space.}


\subsection{Momentum dependence and averaging}
\label{sec:astro_flavor-exact_observables}

The standard flavor-mixing framework assumes that the leptonic mixing matrix, $U$, is constant. However, as established in Sec.~\ref{sec:threshold}, the presence of an HNL modifies the active-neutrino mass matrix at the momentum-transfer scale $Q \simeq M$. Consequently, the effective mixing matrix becomes dependent on $Q$, denoted by $U(Q)$.

Before proceeding to perturbative approximations in Sec.~\ref{sec:astro_flavor-first_order}, it is necessary to define the exact $Q$-dependent transition probabilities. In analogy to standard mixing [\equ{P_standard}], at any given $Q$, the transition probability is constructed directly from the modified mixing matrix,
\begin{equation}
 P_{\alpha\beta}(Q) = \sum_{i} |U_{\alpha i}(Q)|^2 |U_{\beta i}(Q)|^2 \,.
 \label{equ:P_Q_exact}
\end{equation}
Because active neutrinos are lost to the sterile sector during propagation, the raw active fractions at Earth, $\langle f_{\beta, \oplus} \rangle = \sum_{\alpha} f_{\alpha, {\rm S}} \langle P_{\alpha\beta} \rangle$, represent the absolute depleted flux. To map our theoretical predictions onto the detectable flavors, we must renormalize these fractions:
\begin{equation}
 f_{\beta, \oplus}^\prime = \frac{\langle f_{\beta, \oplus} \rangle}{\sum_{\gamma \in \{e, \mu, \tau\}} \langle f_{\gamma, \oplus} \rangle} \,.
 \label{equ:renormalized_fraction}
\end{equation}

In neutrino telescopes, detected events span a continuous distribution of momentum transfers, $\mathcal{P}(Q)$.  This distribution was derived in \Refe~\cite{Bustamante:2026zst}, and we adopt it here. It is determined for neutrino-nucleon DIS, computed by convolving the double-differential scattering cross section with the incoming astrophysical neutrino energy spectrum. Although TeV--PeV neutrinos can kinematically access vast momentum transfers, the steeply falling neutrino energy spectrum (taken to be $\propto E_\nu^{-2.5}$) and the parton distribution functions of the nucleon---which heavily favor lower values of the Bjorken-$x$ parameter---suppress these extremes. Consequently, the interactions are weighted toward moderate momentum transfers, causing $\mathcal{P}(Q)$ to peak broadly around $Q \simeq 20$~GeV. Across this distribution, the dynamic mixing matrix is governed by the step function at the kinematic threshold: $U(Q) \simeq U - \Theta(Q - M) \delta U$.

\begin{figure*}[t!]
 \centering
 \includegraphics[width=\textwidth]{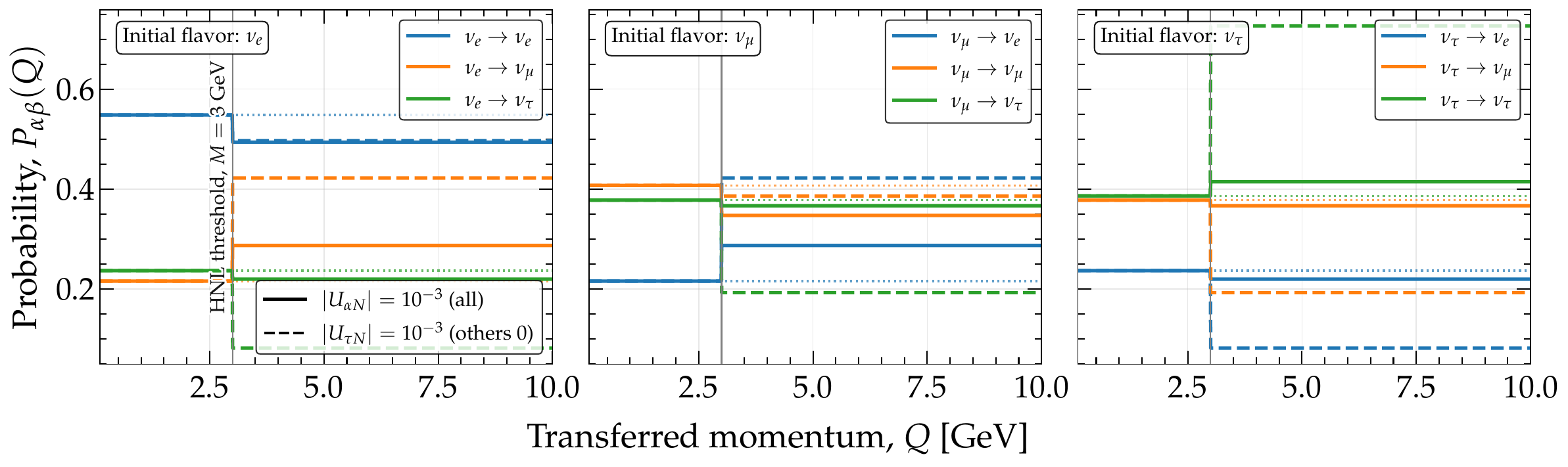}
 \caption{\textbf{Momentum-transfer dependence of the flavor-transition probabilities, $P_{\alpha\beta}(Q)$.} The exact $Q$-dependent probabilities evaluated across the HNL mass threshold ($M = 3$~GeV) with coupling magnitudes of $10^{-3}$. Below the kinematic threshold ($Q < M$), scattering fails to resolve the HNL, and the transitions match the standard-mixing expectation (dotted lines). At $Q = M$, the Majorana correction is triggered, instantly shifting the mixing matrix. A flavor-universal coupling induces modest, flavor-symmetric shifts, while a $\tau$-philic coupling reorganizes the matrix, driving $P_{\tau\tau}$ towards 1.}
 \label{fig:probabilities_vs_Q}
\end{figure*}

Figure~\ref{fig:probabilities_vs_Q} illustrates these $Q$-dependent probabilities across the mass threshold. Below the kinematic threshold ($Q < M$), the threshold correction vanishes, and the transition probabilities remain anchored to the standard-mixing expectation. Once the threshold is crossed ($Q > M$), the Majorana mass correction is triggered, abruptly rewiring the effective mixing matrix and forcing the probabilities onto a perturbed high-$Q$ plateau.

The integrated probability over the scattering distribution naturally separates into the standard low-energy expectation and a high-energy Majorana-induced shift:
\begin{equation}
 \langle P_{\alpha\beta} \rangle = P_{\alpha\beta}^{(0)} + \eta(M) \delta P_{\alpha\beta} \,,
 \label{equ:P_avg_expanded}
\end{equation}
where $\eta(M) \equiv \int_M^\infty dQ \, \mathcal{P}(Q)$ is the effective fraction of events with sufficient momentum transfer to resolve the HNL. Because $\mathcal{P}(Q)$ for TeV--PeV astrophysical neutrinos is broad and peaks around $Q \simeq 20$~GeV, the threshold suppression factor $\eta(M)$ is remarkably close to unity for the mass range targeted by SHiP [$\eta(5~\mathrm{GeV}) \simeq 0.985$]. Consequently, nearly the entirety of the  neutrino flux scatters with sufficient momentum to allow the full, unsuppressed Majorana threshold correction to apply.


\subsection{Perturbative expansion of observables}
\label{sec:astro_flavor-first_order}

Although our main results are fully numerical, we also develop approximate expressions of the relevant quantities from which we garner physical insight. To extract the explicit dependence of the observable flavor shifts on the HNL couplings and Majorana phases, we expand $\delta P_{\alpha\beta}$ to leading order in the threshold correction $\delta m_\nu$. Expanding \equ{P_standard} using the substitution $U \to U - \delta U$, the shift in the transition probability is
\begin{equation}
 \delta P_{\alpha\beta} \simeq \sum_{i=1}^3 \left( \Delta_{\alpha i} |U_{\beta i}|^2 + |U_{\alpha i}|^2 \Delta_{\beta i} \right) \,,
 \label{equ:delta_P_general}
\end{equation}
where $\Delta_{\alpha i} \equiv \delta(|U_{\alpha i}|^2)$ dictates the shift in the individual matrix elements. Using the general first-order expression for $\delta U_{\alpha i}$ from \equ{deltaU}, this is explicitly given by
\begin{equation}
 \Delta_{\alpha i} = -2 \varepsilon(M) \frac{M^3}{v^4} \sum_{j \neq i} \mathrm{Re}\left[ \frac{\xi_i \xi_j}{m_i - m_j} U_{\alpha i}^* U_{\alpha j} \right] \,,
 \label{equ:Delta_alpha_i}
\end{equation}
where $\xi_i \equiv (U^T)_{i,N}$ encapsulates the projection of the HNL couplings onto the active-neutrino basis [\equ{xi_explicit}]. The condition $\sum_\alpha \Delta_{\alpha i} = 0$ follows from the unitarity of the $4 \times 4$ mixing matrix, ensuring $\sum_{\beta \in \{e,\mu,\tau,s\}} \delta P_{\alpha\beta} = 0$, as required for probability conservation.

Propagating this into the flavor fractions at Earth, the predicted shift is
\begin{equation}
    \langle f_{\beta, \oplus} \rangle = f_{\beta, \oplus}^{(0)} + \eta(M) \delta f_{\beta, \oplus} \,,
\end{equation}
where the unsuppressed displacement is given by
\begin{equation}
 \delta f_{\beta, \oplus} = \sum_{\alpha} f_{\alpha, {\rm S}} \delta P_{\alpha\beta} = \sum_{i=1}^3 \left( W_i \Delta_{\beta i} + \delta W_i |U_{\beta i}|^2 \right) \,.
 \label{equ:delta_f_earth}
\end{equation}
Here, $W_i$ is the standard weight defined in \equ{f_earth_standard}, and $\delta W_i \equiv \sum_\alpha f_{\alpha, {\rm S}} \Delta_{\alpha i}$ is its perturbation.

Equation~(\ref{equ:delta_f_earth}) reveals the two ways in which the HNL modifies the flavor fractions at Earth. The first term, $W_i \Delta_{\beta i}$, represents the direct modification of the effective detection probability: the states arrive at Earth with their standard propagating weights, but their projection onto the active flavor $\beta$ is governed by the shifted mixing-matrix elements. The second term, $\delta W_i |U_{\beta i}|^2$, represents the modification of the effective flavor weights $W_i$ entering the propagation: the active flavors produced in the source are projected differently onto the propagating mass eigenstates due to the threshold correction.

\medskip

\textbf{\textit{Mass-based signal amplification:}} The presence of the neutrino mass differences $(m_i - m_j)$ in the denominator of \equ{Delta_alpha_i} dictates that the magnitude of the flavor shift is not exclusively limited by the feebleness of the HNL couplings. As the lightest neutrino mass increases and the mass spectrum approaches the quasi-degenerate regime, the absolute mass differences shrink, acting as a resonant amplifier. This mechanism means that even deeply suppressed couplings ($|U_{\alpha N}|^2 \ll 1$) can generate appreciable changes in the flavor composition.

\medskip

\textbf{\textit{Directional flavor alignment:}} Because the threshold correction to the mass matrix has a rank-one structure ($\delta m_{\alpha \beta} \propto U_{\alpha N} U_{\beta N}$), the perturbation to the light neutrino flavors is not random or isotropic. Instead, the analytical expansion demonstrates that the flavor shift acts along a single axis, aligned with the specific active flavors to which the HNL couples. The products of the projections $\xi_i$ in the numerator explicitly govern which rows of the PMNS matrix receive the largest corrections. Consequently, a muon-philic HNL ($U_{\mu N} \gg U_{eN}, U_{\tau N}$), for instance, selectively targets the $\mu$-row of the mixing matrix, predicting a depletion of the $\nu_\mu$ fraction at Earth and a consequent reduction in the track-to-cascade ratio.

\medskip

\textbf{\textit{Majorana phases:}} The intermediate variables $\xi_1$ and $\xi_2$ isolate the unconstrained Majorana phases $\rho$ and $\sigma$. Because the observable probability shifts $\Delta_{\alpha i}$ depend exclusively on the real parts of the cross-terms $\xi_i \xi_j$, the values of the high-energy flavor fractions are dictated by the constructive or destructive interference of these phases. This establishes the  astrophysical flavor composition as an interferometer for the Majorana phases---which remain  invisible to standard neutrino oscillations.


\subsection{Saturated regime limits}

While the first-order expansion describes the onset of the flavor shift, the saturated regime---where the threshold correction overwhelmingly dominates the intrinsic light-neutrino mass splittings ($\delta m_\nu \gg \Delta m_{ij}$)---admits an exact, analytical limit. In this regime, the mixing matrix completely reorganizes, forcing one physical mass eigenstate to align precisely with the HNL coupling vector. For a flavor-universal coupling assuming full pion decay production, the flavor composition at Earth is strictly bounded ($1/9 \le f_{\alpha, \oplus}^{\rm sat \prime} \le 5/9$), trapping the flavor fractions near the center of the flavor triangle and establishing a phenomenological blind spot. Conversely, single-flavor couplings drastically reorganize the flavors; for example, a muon-philic HNL enhances the muon fraction, while a tau-philic one dampens the tau fraction. 

We provide the analytical limits for both full and muon-damped pion decay production in Appendix~\ref{app:asymptotic_flavor_limits}.


\subsection{Impact of the neutrino mass ordering}
\label{sec:astro_flavor-mass_ordering}

The quantitative predictions for the astrophysical flavor shifts depend on the assumed neutrino mass ordering. Because the Majorana threshold correction is driven by the mass differences between the eigenstates [as isolated in \equ{Delta_alpha_i}], flipping the ordering from normal (NO, $m_1 < m_2 \ll m_3$) to inverted (IO, $m_3 \ll m_1 < m_2$) rewires the kinematic dynamics in three specific ways. 

First, IO acts as a kinematic amplifier: the intrinsic separation between the first and second mass eigenstates is roughly an order of magnitude smaller than in NO, which resonantly amplifies the mass-basis rotations and generates significantly larger observable flavor displacements for the exact same coupling strength. 

Second, switching the ordering swaps which flavors experience inertial damping. In NO, the heaviest state ($\nu_3$) contains the bulk of the muon and tau flavors, locking the $\mu$--$\tau$ sector in place and capping flavor shifts. In IO, this state becomes the lightest, possessing zero kinematic inertia, allowing HNLs coupled to the muon or tau sectors to induce far more extreme shifts. 

Finally, this amplification causes the flavor fractions in IO to be more sensitive to the Majorana phases. 

We provide the explicit algebraic expansions of the mass differences, rotational generators, and inertial suppression mechanisms in Appendix~\ref{app:mass_ordering}.


\subsection{Full numerical results}
\label{sec:astro_flavor-numerical_results}

To determine regions of parameter space that produce an observable signal in high-energy astrophysical neutrinos, we contextualize the absolute flavor shift against the scale of the standard-mixing expectation by defining the relative Euclidean flavor distance,
\begin{equation}
 \mathcal{D}_{\rm rel} = \frac{\sqrt{\sum_{\beta \in {e,\mu,\tau}} \left( f_{\beta, \oplus}' - f_{\beta, \oplus}^{(0)} \right)^2}}{\sqrt{\sum_\beta \left(f_{\beta, \oplus}^{(0)}\right)^2}} \,.
 \label{equ:euclidean_distance_rel}
\end{equation}

Although the analytical first-order expansions derived in the preceding section are helpful for isolating the underlying physical mechanisms, the results presented below are computed numerically using the exact, full-order expressions of the renormalized flavor ratios at Earth.  We show results for the nominal neutrino production mechanism---full pion decay, $\left( \frac{1}{3}, \frac{2}{3}, 0 \right)_{\rm S}$---and contrast them against muon-damped pion decay---$(0, 1, 0)_{\rm S}$, whose corresponding figures are provided in Appendix~\ref{app:muon_damped}.

To perform these numerical evaluations, we construct the absolute masses of the three active neutrino eigenstates ($m_1, m_2, m_3$). Standard neutrino oscillation experiments are sensitive only to the mass-squared splittings: the \textit{solar} splitting $\Delta m_{21}^2 \equiv m_2^2 - m_1^2 \approx 7.5 \times 10^{-5}$~eV$^2$ and the \textit{atmospheric} splitting $|\Delta m_{31}^2| \equiv | m_3^2- m_1^2 | \approx 2.5 \times 10^{-3}$~eV$^2$. Consequently, the entire light-neutrino spectrum is anchored by a single free parameter---the lightest neutrino mass, $m_{\rm lightest}$. In NO, $m_1$ is the lightest state, yielding heavier states $m_2 = \sqrt{m_1^2 + \Delta m_{21}^2}$ and $m_3 = \sqrt{m_1^2 + \Delta m_{31}^2}$. Conversely, in IO, $m_3$ is the lightest state, and the two remaining states are pushed to a higher absolute mass scale by the atmospheric splitting: $m_1 = \sqrt{m_3^2 + |\Delta m_{31}^2|}$ and $m_2 = \sqrt{m_1^2 + \Delta m_{21}^2}$.

\begin{figure}[t!]
 \centering
 \includegraphics[width=0.5\textwidth]{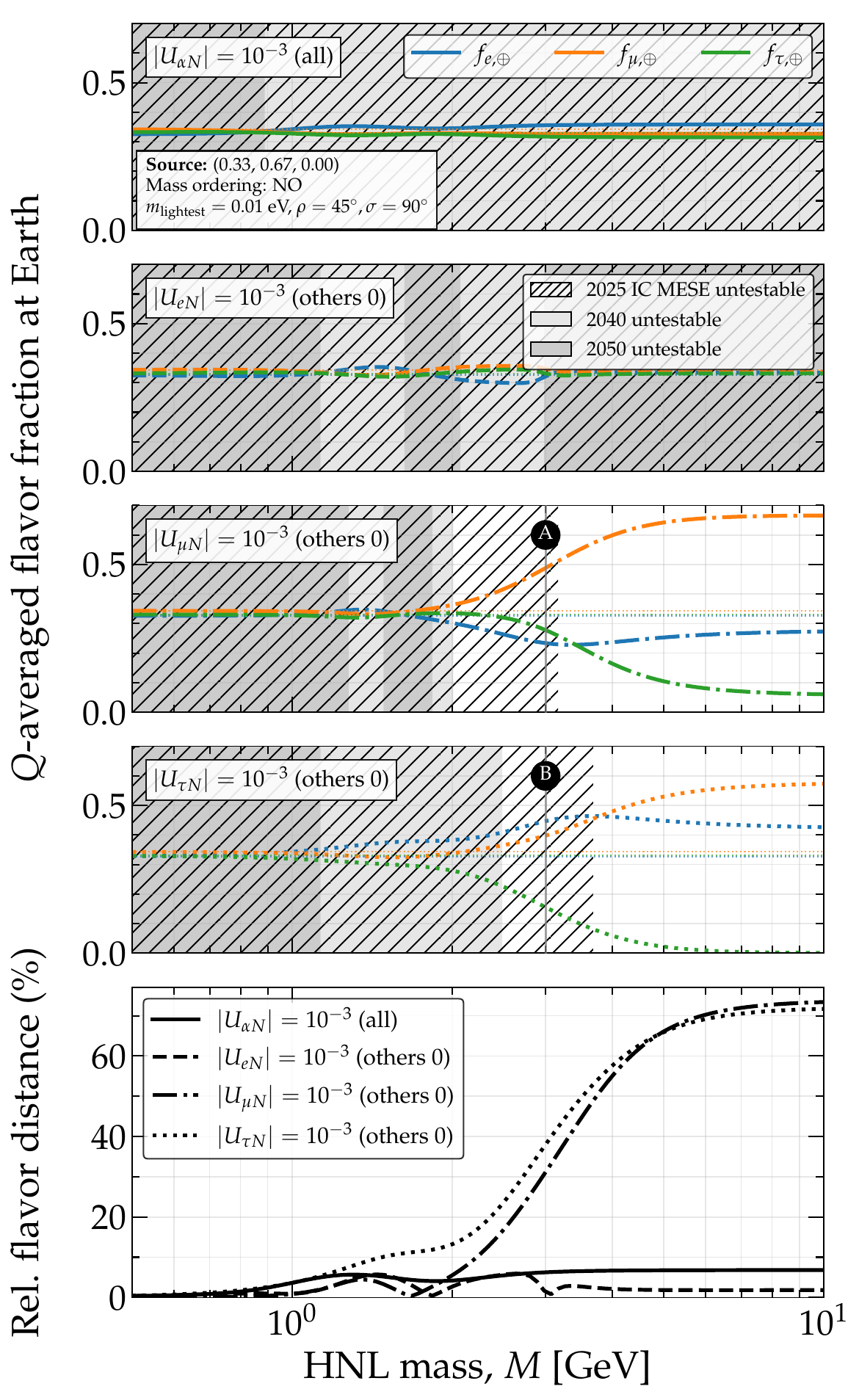}
 \caption{\textbf{Dependence of the flavor composition at Earth on the heavy neutral lepton (HNL) mass, $M$.}  In this plot, we assume neutrino production via full pion decay and set nonzero couplings to an illustrative value of $10^{-3}$.  \textit{Top panel:} Assuming flavor-universal couplings, \ie, $|U_{e N}| = |U_{\mu N}| = |U_{\tau N}|$.  \textit{Next three panels:} Assuming electron-philic (only $|U_{e N}|$ nonzero), muon-philic (only $|U_{\mu N}|$ nonzero), and tau-philic ($|U_{\tau N}|$ nonzero) couplings.  \textit{Bottom panel:} Relative flavor distance.  Shaded gray bands denote regions where the predicted flavor shift induced by the threshold correction falls below projected multi-detector flavor measurements in the years 2040 and 2050.  The mass ordering is normal (NO) and the standard mixing parameters are set to their NuFIT 6.1 best-fit values (with Super-Kamiokande atmospheric data).  \textbf{For production via full pion decay, the muon- and tau-philic scenarios yield large effects, while the flavor-universal and electron-philic scenarios remain untestable.}  Points A and B mark examples for which we produce results later.  See Sec.~\ref{sec:astro_flavor-numerical_results} for details.}
 \label{fig:flavor_vs_M}
\end{figure}

\begin{figure}[t!]
 \centering
 \includegraphics[width=0.5\textwidth]{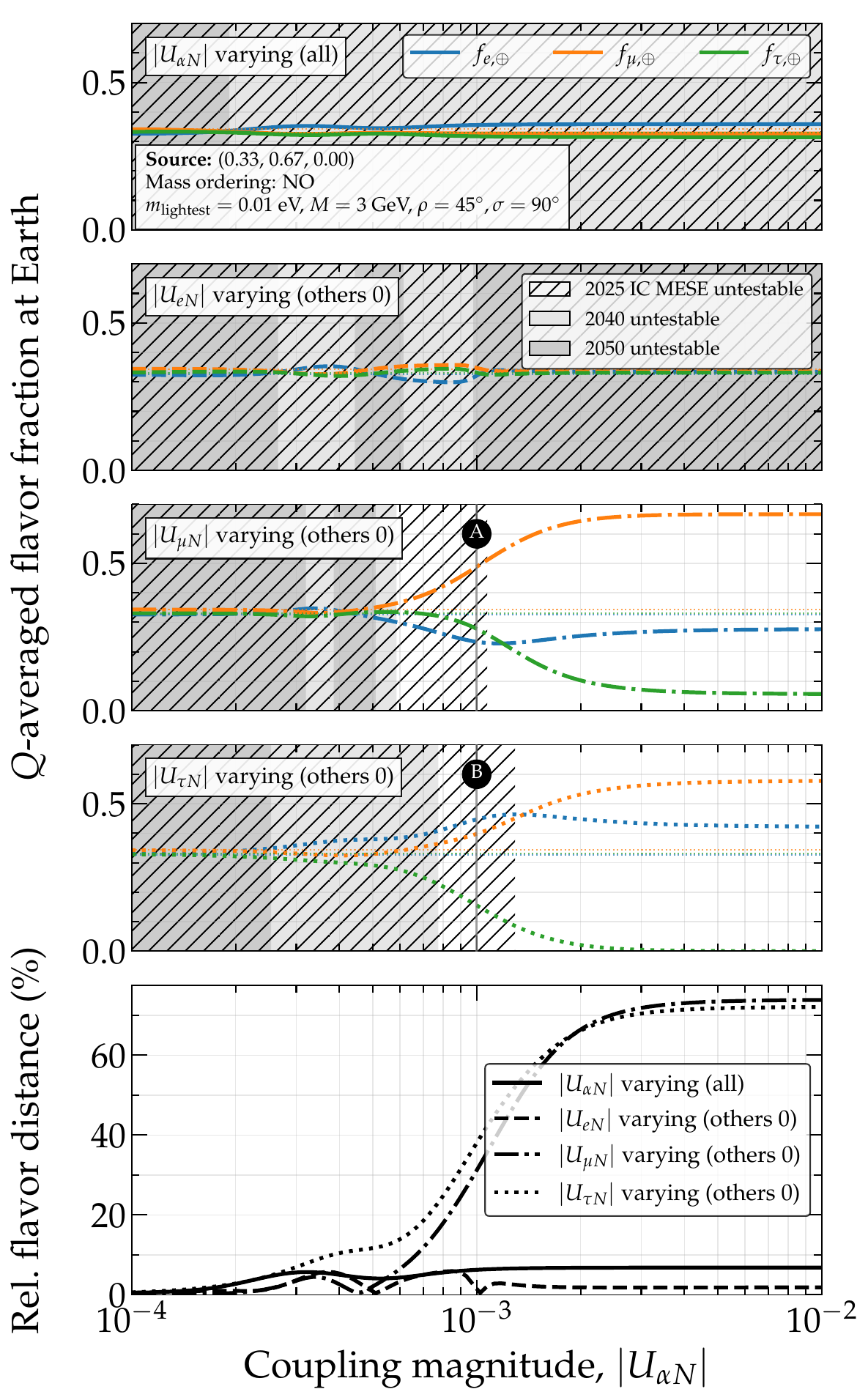}
 \caption{\textbf{Dependence of the flavor composition at Earth on the active-sterile coupling magnitude $|U_{\alpha N}|$.}  Similar to \figu{flavor_vs_M}, but varying $|U_{\alpha N}|$ while fixing the other model parameters to illustrative values.  See Sec.~\ref{sec:astro_flavor-numerical_results} for details.}
 \label{fig:flavor_vs_U}
\end{figure}

\begin{figure}[t!]
 \centering
 \includegraphics[width=0.5\textwidth]{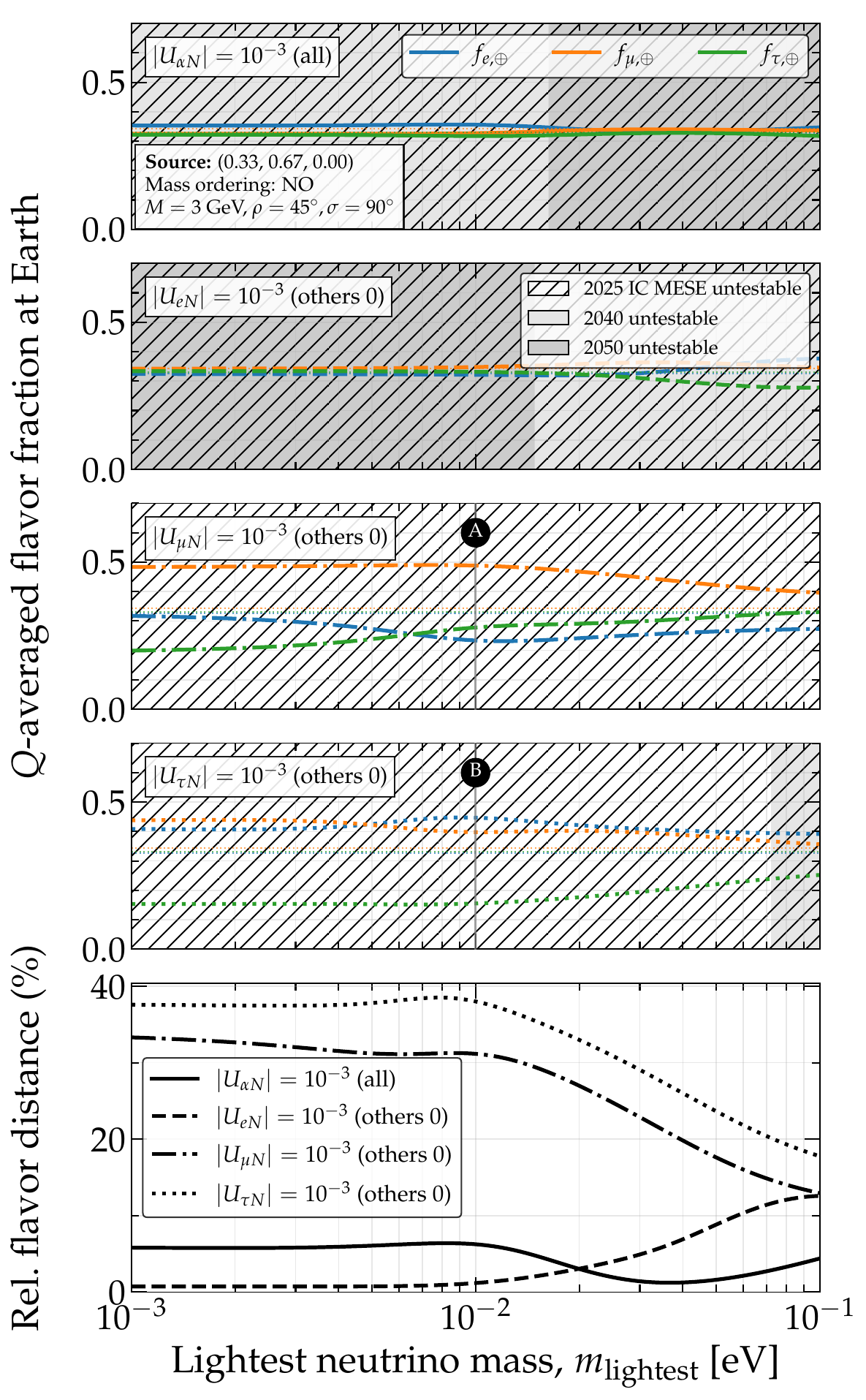}
 \caption{\textbf{Dependence of the flavor composition at Earth on the mass of the lightest neutrino mass eigenstate, $m_\text{lightest}$.}  Similar to \figu{flavor_vs_M}, but varying $m_\text{lightest}$, under NO, while fixing the other model parameters to illustrative values.  See Sec.~\ref{sec:astro_flavor-numerical_results} for details.}
 \label{fig:flavor_vs_m_lightest}
\end{figure}

\begin{figure}[t!]
 \centering
 \includegraphics[width=0.5\textwidth]{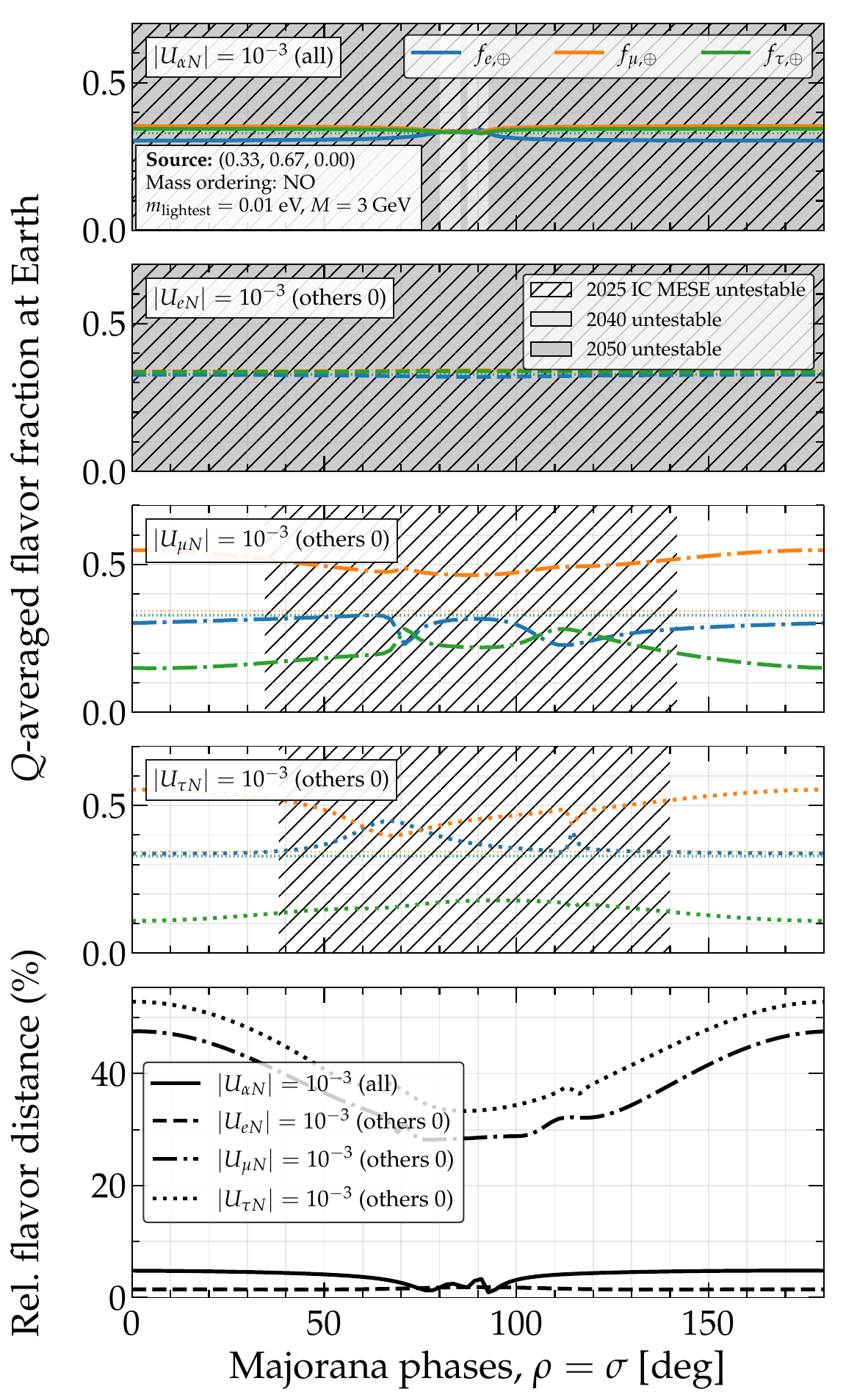}
 \caption{\textbf{Dependence of the flavor composition at Earth on the Majorana CP-violation phases, $\rho = \sigma$.}  Similar to \figu{flavor_vs_M}, but varying $\rho$ and $\sigma$ simultaneously and equal to each other while fixing the other model parameters to illustrative values.  See Sec.~\ref{sec:astro_flavor-numerical_results} for details.}
 \label{fig:flavor_vs_phases}
\end{figure}

We define four benchmark scenarios that isolate distinct flavor structures, and that we explore in Figs.~\ref{fig:flavor_vs_M}--\ref{fig:flavor_vs_phases}. The first scenario is flavor-universal, setting $|U_{eN}| = |U_{\mu N}| = |U_{\tau N}|$. The remaining three are maximally flavor-asymmetric, turning on only a single coupling at a time, $|U_{eN}|$, $|U_{\mu N}|$, or $|U_{\tau N}|$, while holding the others to zero. This  allows us to verify the directional alignment of the threshold correction that we pointed out earlier.

In Figs.~\ref{fig:flavor_vs_M}--\ref{fig:flavor_vs_phases} (also in Figs.~\ref{fig:flavor_vs_M_muon}--\ref{fig:flavor_vs_phases_muon}), we show the regions of parameter space that are testable and untestable by current and future flavor measurements. For all parameter combinations, the flavor-universal and electron-philic scenarios remain largely untestable throughout assuming neutrino production via full pion decay---due to their inducing only tiny flavor shifts---though they become resolvable assuming muon-damped pion decay.

\medskip

\textbf{\textit{Kinematic scale \& flavor saturation.---}}Figures~\ref{fig:flavor_vs_M} and \ref{fig:flavor_vs_U} demonstrate the onset and saturation of the flavor shift. The flavor shift is governed by the competition between the standard mass splittings ($\Delta m_{ij} \sim 10^{-3}$--$10^{-2}$~eV) and the HNL-induced correction ($\delta m_\nu \propto M^3 |U_{\alpha N}|^2 / v^4$). The competing mass scales become matched near $M \sim 1$--$3$~GeV for a fixed coupling of $10^{-3}$ (Figs.~\ref{fig:flavor_vs_M}, \ref{fig:flavor_vs_U}). Below this crossing, standard mixing dominates. Within the transition window, the competing scales induce interference, manifesting as fluctuations in the flavor-ratio curves. Once the mass threshold correction overpowers the standard mass splittings, the high-$Q$ mass eigenstates align with the HNL coupling vector, triggering a large flavor shift that becomes testable.

For example, in Figs.~\ref{fig:flavor_vs_M} and \ref{fig:flavor_vs_U}, a muon-philic HNL drives the expected muon fraction up from $\sim 33\%$ to nearly $70\%$, while depleting the tau fraction down to near zero. A tau-philic HNL induces a nearly identical shift, converging toward the same muon-heavy, tau-depleted composition at Earth, due to the angle $\theta_{23} \approx 45^\circ$ being nearly maximal. Consequently, both muon- and tau-philic HNLs yield large relative flavor distances at saturation ($\mathcal{D}_{\rm rel} \gtrsim 50\%$), showcasing their complementarity with neutrinoless double beta decay, while flavor shifts in the electron-philic scenario stay suppressed.

\medskip

\textbf{\textit{Inertial suppression.---}}Figure~\ref{fig:flavor_vs_m_lightest} maps the flavor fractions at Earth against the lightest neutrino mass, $m_{\rm lightest}$. As $m_{\rm lightest}$ grows and the light neutrino spectrum approaches the quasi-degenerate regime ($m_1 \approx m_2 \approx m_3$), the flavor shift is progressively suppressed. To see explicitly how this restricts the observables, we evaluate $\Delta_{\alpha i} = \delta(|U_{\alpha i}|^2)$ [Eqs.~(\ref{equ:delta_P_general}), (\ref{equ:Delta_alpha_i})] that governs the shift in the flavor-transition probability. Expanding it separates it into real and imaginary terms governed by opposite kinematic denominators:
\begin{align}
 \Delta_{\alpha i} 
 &= 
 2 \sum_{j \neq i} \left[ \text{Re}(U_{\alpha i}^* U_{\alpha j}) \frac{\text{Re}(A_{ij})}{m_i - m_j} \right.
 \nonumber \\
 & \qquad \left. +~ \text{Im}(U_{\alpha i}^* U_{\alpha j}) \frac{\text{Im}(A_{ij})}{m_i + m_j} \right] \,,
 \label{equ:Majorana_decoupling}
\end{align}
where $A_{ij} \propto M^3/v^4$ is defined in \equ{Aij}. While the real interference is amplified by the shrinking mass differences $(m_i - m_j)$, the complex interference is severely penalized by the absolute mass scale $(m_i + m_j)$. Therefore, as $m_{\rm lightest}$ increases, the active neutrinos effectively gain ``inertia'' against the mass perturbation $A_{ij}$. The large absolute masses damp the second term on the right-hand side of \equ{Majorana_decoupling}, forcing the flavor ratios at Earth toward their standard unperturbed expectations.

\medskip

\textbf{\textit{Resilience to Majorana phases.---}}Figure~\ref{fig:flavor_vs_phases} shows how the flavor composition varies with the Majorana phases. By sweeping the phases $\rho = \sigma$, the flavor fractions change as the projections $\xi_i \xi_j$ in \equ{Delta_alpha_i} undergo constructive and destructive interference. Allowing the phases to vary independently ($\rho \neq \sigma$, not shown in \figu{flavor_vs_phases}) would unlock further ways for them to interfere.

Crucially, while the unknown Majorana phases modulate the exact coordinates of the flavor shift, they cannot erase the signal of a muon- or tau-philic HNL. As demonstrated in  \figu{flavor_vs_phases}, the relative flavor distances for these specific scenarios oscillate but remain strictly bounded away from zero, never dropping below $\mathcal{D}_{\rm rel} \approx 30\%$. This guarantees that an unlucky configuration of the unknown Majorana CP-violation phases cannot accidentally rotate the observables back to the standard-mixing expectation. Therefore, if a heavy, muon- or tau-philic HNL exists, its flavor footprint remains robustly testable regardless of the underlying Majorana phase structure.

\medskip

\textbf{\textit{Contrast with muon-damped pion decay.---}}We conclude by contrasting the nominal full pion decay results above with the muon-damped scenario, $(0, 1, 0)_{\rm S}$, whose full numerical evaluations are in Appendix~\ref{app:muon_damped}. 

As shown analytically in Appendix~\ref{app:asymptotic_flavor_limits}, the footprint of the HNL-induced correction to the flavor composition at Earth depends critically on the flavor composition at the sources. Under the nominal neutrino production via full pion decay, the flavor fractions at Earth are weighted superpositions of transition channels ($f_{\beta, \oplus} = \frac{1}{3} P_{e\beta} + \frac{2}{3} P_{\mu\beta}$). Because unitarity imposes inverse correlations between these channels, the HNL-induced shifts partially cancel, diluting the flavor shift. This dilution anchors the flavor composition at Earth near the flavor-democratic expectation from standard mixing, creating blind spots for the flavor-universal and pure-electron coupling scenarios. 

In contrast, muon-damped production maps the flavor composition at Earth exclusively to a single probability row ($f_{\beta, \oplus} = P_{\mu\beta}$). Without an initial superposed electron flux to wash out the signal, the flavor fractions trace the exact, undiluted interference pattern of the mass eigenstates. Consequently, the HNL-induced shifts in the flavor-transition probabilities are transferred directly to the flavor composition at Earth, generating larger relative flavor distances and decisively breaking the inherent blind spots of the full pion decay scenario.


\subsection{Allowed flavor regions at Earth}
\label{sec:astro_flavor-flavor_regions}

\begin{figure*}[t]
 \centering
 \includegraphics[width=0.49\textwidth]{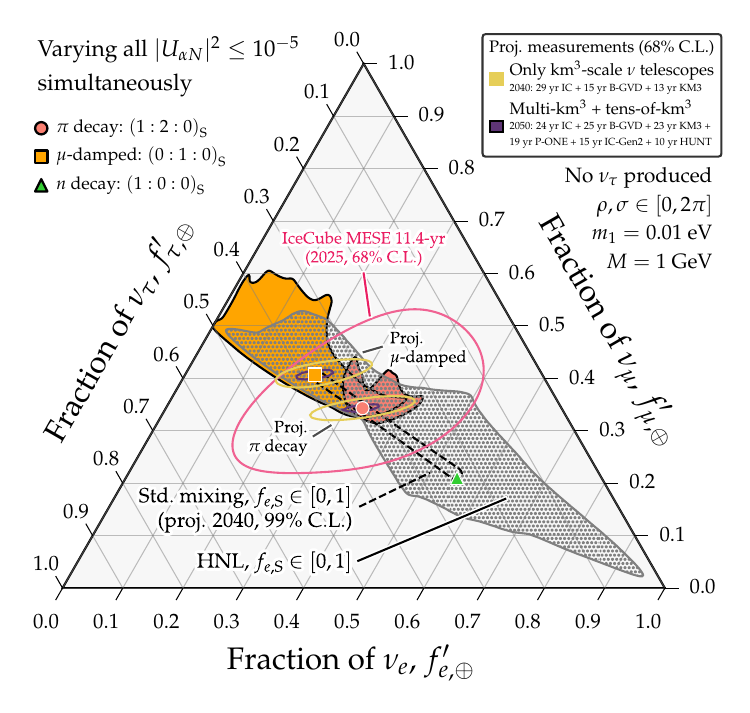}
 \hfill
 \includegraphics[width=0.49\textwidth]{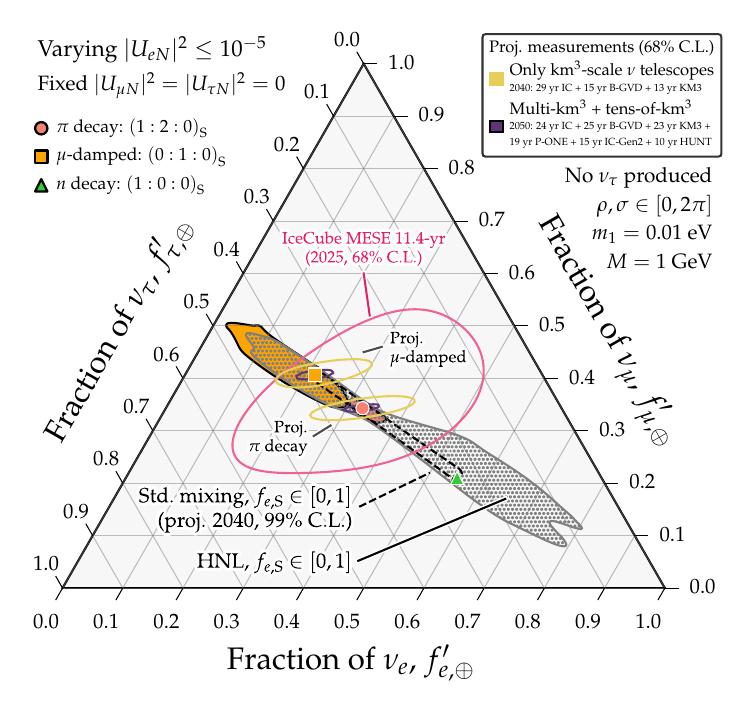} \\
 \vspace{0.02\textwidth}
 \includegraphics[width=0.49\textwidth]{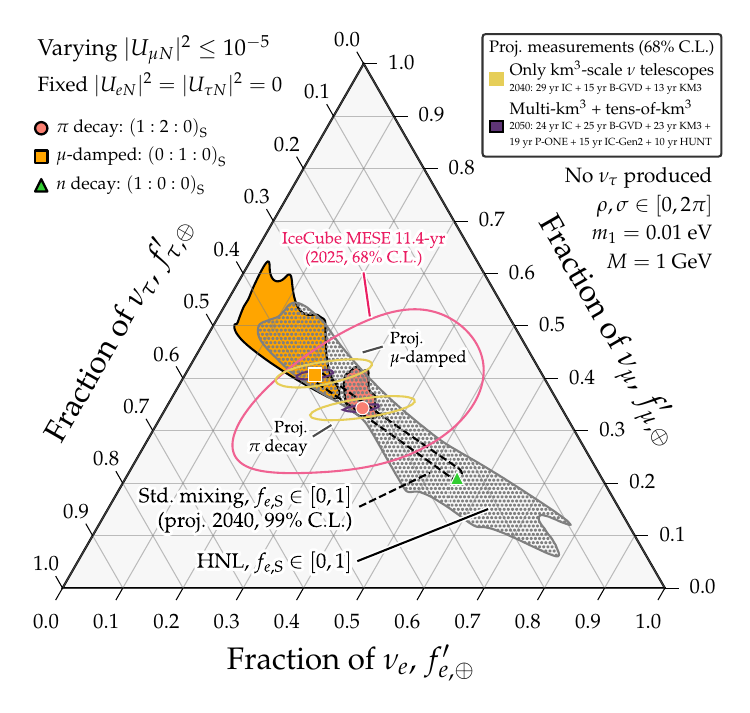}
 \hfill
 \includegraphics[width=0.49\textwidth]{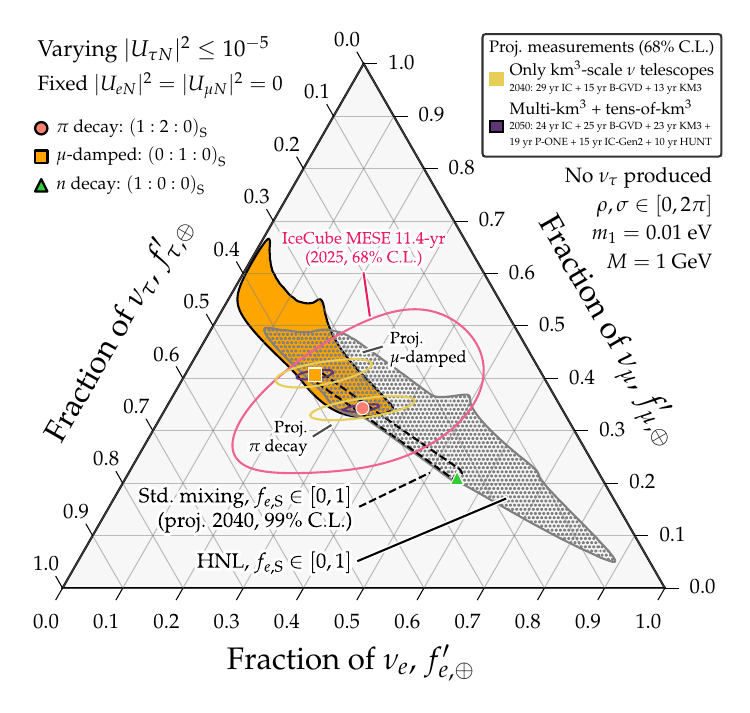}
 \caption{\textbf{Allowed regions of high-energy neutrino flavor composition at Earth induced by mixing with a heavy neutral lepton (HNL).} The electron flavor composition at the sources is varied over all possible values, $f_{e, {\rm S}} \in [0,1]$, assuming no initial $\nu_\tau$ production ($f_{\tau,S} = 0$). \textit{Top left:} All three active-sterile mixing parameters $|U_{\alpha N}|^2$ are varied simultaneously up to $10^{-5}$. \textit{Other panels:} Only a single parameter---$|U_{e N}|^2$ (\textit{top-right}), $|U_{\mu N}|^2$ (\textit{bottom-left}), or $|U_{\tau N}|^2$ (\textit{bottom-right})---is varied, while the other two are fixed to zero.  The flavor ratios, $f_{\alpha, \oplus}^\prime$, are averaged over $Q$ and renormalized following \equ{renormalized_fraction}. The allowed regions are generated for $M=1$~GeV and $m_1=0.01$~eV, varying the CP-violation Majorana phases, $\rho, \sigma \in [0, 2\pi]$.}
 \label{fig:hnl_flavor_contours}
\end{figure*}

To fully map the flavor-composition footprint of the HNL, we now transition from the isolated benchmark scenarios above to a global view of the accessible flavor space at Earth. By simultaneously floating all relevant model parameters---the HNL mass ($M$), flavor couplings ($|U_{\alpha N}|$), and Majorana phases ($\rho, \sigma$)---alongside the standard light-neutrino mixing parameters $\theta_{12}$, $\theta_{23}$, $\theta_{13}$, $\delta_\text{CP}$, $\Delta m_{21}^2$, and $\Delta m_{3l}^2$ (sampled from their present and projected experimentally allowed ranges), we predict the allowed flavor regions at Earth.

\medskip

\textbf{\textit{Standard-mixing flavor regions.---}}Under standard mixing, the allowed flavor regions---the \textit{theoretically palatable regions} from \Refes~\cite{Bustamante:2015waa, Song:2020nfh}---are notoriously restricted. For the nominal expectation of neutrino production via full pion decay, the values of the standard mixing parameter confine the expected composition to a highly localized region near the democratic center of the triangle, near $\left( \frac{1}{3}, \frac{1}{3}, \frac{1}{3} \right)_\oplus$. For muon-damped production, the standard-mixing allowed region forms a distinct, narrow band slightly off-center.

Figure~\ref{fig:hnl_flavor_contours} shows the nominal expectations for these two benchmarks, computed using the best-fit values of the standard mixing parameters from NuFIT 6.1~\cite{NuFIT6}.  In \Refe~\cite{Song:2020nfh}, Fig.~1 compares instead the allowed regions for these benchmarks, obtained using present and projected ranges of the standard mixing parameters. For the latter case, given the tighter ranges of the parameters (see Table~\ref{tab:free_parameters}), the resulting flavor regions are small enough to be considered point-like for all practical purposes.

As explained in Sec.~\ref{sec:astro_production}, abandoning these two benchmark production scenarios for a generic one without high-energy $\nu_\tau$ production  generalizes the flavor composition at the sources to $\left(f_{e, {\rm S}}, 1-f_{e, {\rm S}}, 0 \right)$.  In \figu{hnl_flavor_contours}, we show the theoretically palatable region in this case, obtained by varying the unknown value of $f_{e, {\rm S}} \in [0, 1]$ together with the standard mixing parameters within their projected ranges. The elongation of the region is due to the variation of $f_{e, {\rm S}}$---which remains the fundamental unknown---while its tiny transversal width is a reflection of the tight allowed ranges of the mixing parameters. 

\medskip

\textbf{\textit{HNL-modified flavor regions.---}}Figure~\ref{fig:hnl_flavor_contours} maps the footprint of the Majorana mass threshold correction on the predicted flavor space at Earth. By varying the flavor composition at the sources over all possible tau-depleted mixtures ($f_{e, {\rm S}} \in [0,1]$), we delineate the generic flavor boundaries reachable with the HNL. This envelope expands significantly beyond the narrow parameter space allowed by standard mixing---which represents the expectation for a Dirac neutrino. (All the modified flavor regions shown here fall strictly within the broader, model-independent $3+1$ unitarity bounds derived in \Refe~\cite{Ahlers:2020miq}.)

Individual couplings expand this space differently. An electron-philic HNL ($|U_{eN}|^2$, top-right panel) produces a generic boundary that largely mirrors the standard-mixing expectation. While the region does extend toward the electron-depleted left edge, the standard-mixing band already covers this space; the HNL merely introduces a marginal bulge. Under normal mass ordering, this perturbation attempts to pull the electron flavor into the lightest mass eigenstate ($\nu_1$). Because $\nu_1$ is natively electron-dominant ($\approx 68\%$) in the SM, the mixing matrix requires minimal structural rotation to accommodate this pull, keeping the resulting flavor shift confined.

In contrast, muon- and tau-philic HNLs ($|U_{\mu N}|^2, |U_{\tau N}|^2$, bottom panels) drive the allowed region into new territory. Unlike the electron flavor, the muon and tau flavors are natively scarce in the lightest state ($\nu_1$). Therefore, when the mass perturbation attempts to pull these specific flavors into $\nu_1$, it clashes with the established matrix structure. To satisfy this mass pull, the mixing matrix must forcefully extract the muon and tau content out of the heavier mass eigenstates ($\nu_2, \nu_3$) and reassign it to $\nu_1$. This massive structural rotation pushes the generic boundary sharply upward into the highly muon-rich regime ($f_\mu \gtrsim 0.6$), an expanse strictly forbidden by standard mixing. When all three couplings vary simultaneously (top-left panel), the parameter space synthesizes these individual expansions to fill the complete interior envelope.

How much of this generic expansion is actually realized depends entirely on the specific neutrino production mechanism. The generic boundary assumes total ignorance of the source; fixing a specific production mechanism restricts the accessible space within that envelope.

For the nominal full pion decay benchmark, the HNL-modified space remains tightly anchored near the center of the standard-mixing band across all coupling scenarios. Because this yields a mixed initial flavor composition, intrinsic probability cancellations dilute the flavor shifts. Even with muon- or tau-philic couplings active, this dilution prevents the observables from reaching the high-muon boundaries of the generic flavor regions. 

\begin{table*}[t!]
 \caption{\label{tab:benchmark_points} \textbf{Benchmark HNL points.} These points satisfy all three validity and testability criteria outlined in the text and are shown in \figu{testability_regions}. The Majorana CP-violation phases, $\rho$ and $\sigma$, have been optimized to maximize the Euclidean flavor distance. All points are generated assuming normal neutrino mass ordering and a lightest neutrino mass of $m_{\text{lightest}} = 0.01$ eV. The discovery significance shown is global, and is computed on the basis of projected flavor-composition measurements using combinations of HESE and through-going muons in multiple neutrino telescopes.}
 \begin{ruledtabular}
  \begin{tabular}{l c l l r r c c c}
   \multirow{2}{*}{Point} & \multirow{2}{*}{\makecell[l]{HNL mass,\\ $M$ [GeV]}} & \multirow{2}{*}{\makecell[l]{Single nonzero\\HNL coupling}} & \multirow{2}{*}{\makecell[l]{Astrophysical\\ production scenario}} & \multirow{2}{*}{$\rho$ [$^\circ$]} & \multirow{2}{*}{$\sigma$ [$^\circ$]} & \multirow{2}{*}{$(f_e, f_\mu, f_\tau)_\oplus$} & \multicolumn{2}{c}{Discovery significance ($\sigma$)} \\
   \cline{8-9} \noalign{\vspace{2pt}}
    & & & & & & & Year 2040\footnotemark[1] & Year 2050\footnotemark[2] \\
   \hline \noalign{\vspace{2pt}}
   A & 3 & $|U_{e N}|^2 = 5 \times 10^{-8}$ & Full pion & 0.00 & 16.24 & $(0.359, 0.327, 0.315)$ & 0.00 & 0.00 \\
   B & 3 & $|U_{\mu N}|^2 = 4 \times 10^{-8}$ & Muon-damped & 5.38 & 182.34 & $(0.152, 0.443, 0.405)$ & 2.61 & 7.45 \\
   C & 4 & $|U_{\mu N}|^2 = 1 \times 10^{-8}$ & Muon-damped & 5.14 & 181.26 & $(0.178, 0.429, 0.393)$ & 1.41 & 3.63 \\
   D & 1 & $|U_{\tau N}|^2 = 1.5 \times 10^{-6}$ & Full pion & 87.16 & 92.77 & $(0.358, 0.329, 0.313)$ & 0.00 & 0.00 \\
   E & 3 & $|U_{\tau N}|^2 = 6 \times 10^{-8}$ & Full pion & 90.61 & 91.23 & $(0.353, 0.331, 0.316)$ & 0.00 & 0.00 \\
   F & 1 & $|U_{\tau N}|^2 = 7 \times 10^{-7}$ & Muon-damped & 179.56 & 173.58 & $(0.157, 0.442, 0.401)$ & 2.42 & 8.75 \\
   G & 3 & $|U_{\tau N}|^2 = 3 \times 10^{-8}$ & Muon-damped & 179.74 & 174.05 & $(0.165, 0.437, 0.398)$ & 2.26 & 5.14 \\
  \end{tabular}
  \footnotetext[1]{Projections for 2040 using 29~yr of IceCube + 15~yr of Baikal-GVD + 13~yr of KM3NeT.}
  \footnotetext[2]{Projections for 2050 using 24~yr of IceCube + 25~yr of Baikal-GVD + 23~yr of KM3NeT + 19~yr of P-ONE + 15~yr of IceCube-Gen2 + 10~yr of HUNT.}
 \end{ruledtabular}
\end{table*}

Conversely, the muon-damped benchmark lacks these internal cancellations. While it remains relatively confined for an electron-philic HNL, it exhibits pronounced sensitivity to muon- and tau-philic HNLs. Undiluted by a superposed initial electron flux, the HNL-induced rotations transfer directly to the flavor fractions at Earth. The allowed flavor region balloons upward towards the muon-rich lobe of the generic flavor envelope and breaks away from the standard-mixing region.

Ultimately, the flavor space that is accessible under the actual neutrino production mechanisms dictates our statistical power in Section~\ref{sec:limits_and_discovery}: configurations that forcefully breach the standard-mixing envelope drive our high-significance discovery forecasts.

\medskip

\textbf{\textit{Impact of uncertainties and mass ordering.---}}The exact boundaries of these HNL-modified regions are further modulated by the precision of the standard oscillation parameters and the true neutrino mass ordering. While the anticipated improvements in the measurement of the standard mixing parameters adopted for \figu{hnl_flavor_contours} will marginally compress these global HNL envelopes compared to current measurements, flipping the mass hierarchy to IO expands them. Because IO forces the natively muon- and tau-heavy $\nu_3$ state to be the lightest eigenstate, it removes the kinematic inertia that suppresses muon- and tau-philic corrections, allowing the allowed region to extend even deeper into the flavor triangle. We detail this in Appendix~\ref{app:flavor_extra}.


\subsection{Example testable scenarios}
\label{sec:astro_flavor-testable_scenarios}

Table~\ref{tab:benchmark_points} defines seven benchmark HNL scenarios (A--G) that we use to evaluate our discovery forecasts in Sec.~\ref{sec:limits_and_discovery-discovery}. These points reside in a testable parameter space satisfying three criteria: compatibility with current IceCube MESE flavor measurements~\cite{Abbasi:2025fjc}, validity of our perturbative mass-threshold calculation, and resolvability by 2040 or 2050 multi-detector combinations. Current MESE measurements restrict the HNL-induced flavor shifts to 5--15\% relative to standard-mixing expectations, unresolvable by current measurements, but within reach of our projections.  We map the $M$--$|U_{\alpha N}|^2$ parameter space satisfying these conditions in Appendix~\ref{app:testable_scenarios}.


\section{Multi-experiment synergies}
\label{sec:synergies}

\textit{We place the astrophysical flavor shift within a broader experimental context. First, by probing all active-sterile mixings, it circumvents the $0\nu\beta\beta$ normal-ordering blind spot, enabling the discovery of muon- and tau-philic HNLs. Second, alongside a \ship\ discovery and an LHC null result eliminating alternative mediators, it completes a three-experiment framework yielding model-independent evidence for the Majorana neutrino.}


\subsection{Neutrinoless double beta decay}
\label{subsec:0vbb}

\textbf{\textit{The $0\nu\beta\beta$ effective mass.---}}Neutrinoless double beta decay ($0\nu\beta\beta$) probes the Majorana nature of neutrinos through the effective mass
\begin{equation}
 m_{\beta\beta} = \left| m_{\beta\beta}^{(\nu)} + m_{\beta\beta}^{(N)} \right| \;,
 \label{equ:mbb_full}
\end{equation}
where the first and second terms on the right-hand side are the contributions from light neutrinos and the HNL, respectively.  The light-neutrino contribution is
\begin{equation}
 m_{\beta\beta}^{(\nu)} \equiv \sum_{i=1}^{3} U_{ei}^2\, m_i p(m_i) \;,
\end{equation}
where $p(m)$ is a dimensionless (in natural units) nuclear-scale propagator factor that encodes the dependence of the decay amplitude on the neutrino mass.  For light neutrinos, with $m_i \ll p_F$, where $p_F \sim 100~\mathrm{MeV}$ is the typical Fermi momentum, $p(m_i) \simeq 1$. The HNL contribution is 
\begin{equation}
 m_{\beta\beta}^{(N)} \equiv U_{eN}^2\, M\, p(M) \;,
\label{equ:mbb_heavy}
\end{equation}
where now $M \gg p_F$ and $p(M) \simeq p_F^2 / M^2$.

\medskip

\textbf{\textit{The $0\nu\beta\beta$ blind spot.---}}A central observation is that $m_{\beta\beta}$ depends \emph{only} on $U_{eN}$ among the three HNL flavor couplings.  The muon and tau couplings, $U_{\mu N}$ and $U_{\tau N}$, enter $0\nu\beta\beta$ only through their effect on the light-neutrino mixing matrix via the seesaw relation, encoded in $m_{\beta\beta}^{(\nu)}$, independently of any new mass-threshold correction.

In contrast, as shown in Eqs.~(\ref{equ:delta_s13})--(\ref{equ:delta_s23}), the HNL-induced mass-threshold corrections to the active mixing angles---and, therefore, the HNL-induced flavor ratios at Earth, $f_{\alpha, \oplus}^\prime$---depend on the three mixing elements, $U_{e N}$, $U_{\mu N}$, and $U_{\tau N}$. This complementarity has an important consequence: there are regions of HNL parameter space where the HNL-induced shift to $m_{\beta\beta}$ is suppressed while the shift to $f_{\alpha, \oplus}^\prime$ remains large, and vice versa.

\medskip

\textbf{\textit{Complementarity regions.---}}The complementarity between $0\nu\beta\beta$ and astrophysical neutrinos depends primarily on two factors: the flavor structure of the HNL, \ie, the relative sizes of $|U_{eN}|^2$, $|U_{\mu N}|^2$, and $|U_{\tau N}|^2$, and the HNL mass.  Together, they govern the $0\nu\beta\beta$ effective mass and the direction of the flavor shift of high-energy astrophysical neutrinos at Earth.  In addition, the flavor composition of the high-energy astrophysical neutrinos at their sources also modulates how large a flavor shift is for a given set of active-sterile mixings.

To map this, we define a ratio that compares the two probes in terms of their HNL-modified observables,
\begin{equation}
 \mathcal{R} \equiv \frac{\mathcal{D} / \sigma_{\mathcal{D}}}{|m_{\beta\beta}^{(N)}| / \sigma_{0\nu\beta\beta}} \;,
 \label{equ:ratio_R}
\end{equation}
where $\mathcal{D}$ is the numerator of the relative Euclidean flavor distance [\equ{euclidean_distance_rel}] between the standard and HNL-modified flavor composition at Earth, $\sigma_{\mathcal{D}}$ is the projected flavor measurement sensitivity of the 2050 multi-detector network, $m_{\beta\beta}^{(N)}$ is the HNL contribution to the effective Majorana mass [\equ{mbb_heavy}], and $\sigma_{0\nu\beta\beta} = 5$~meV is the projected nEXO sensitivity~\cite{nEXO}. Both numerator and denominator are dimensionless signal-to-noise ratios, so $\mathcal{R} = 1$ defines genuine equal discovery power rather than equality at an arbitrary reference scale. 

Two features of this definition are worth noting. First, the denominator depends only on $|U_{eN}|^2$ through $m_{\beta\beta}^{(N)} \propto |U_{eN}|^2/M$, while the numerator depends on all three flavor couplings through the full flavor composition. Second, because $\mathcal{D}$ depends on the flavor composition at the sources, so does $\mathcal{R}$: we evaluate $\mathcal{R}$ below for full pion decay, but comment on muon-damped production. 

\begin{figure}[t!]
 \centering
 \includegraphics[width=\columnwidth]{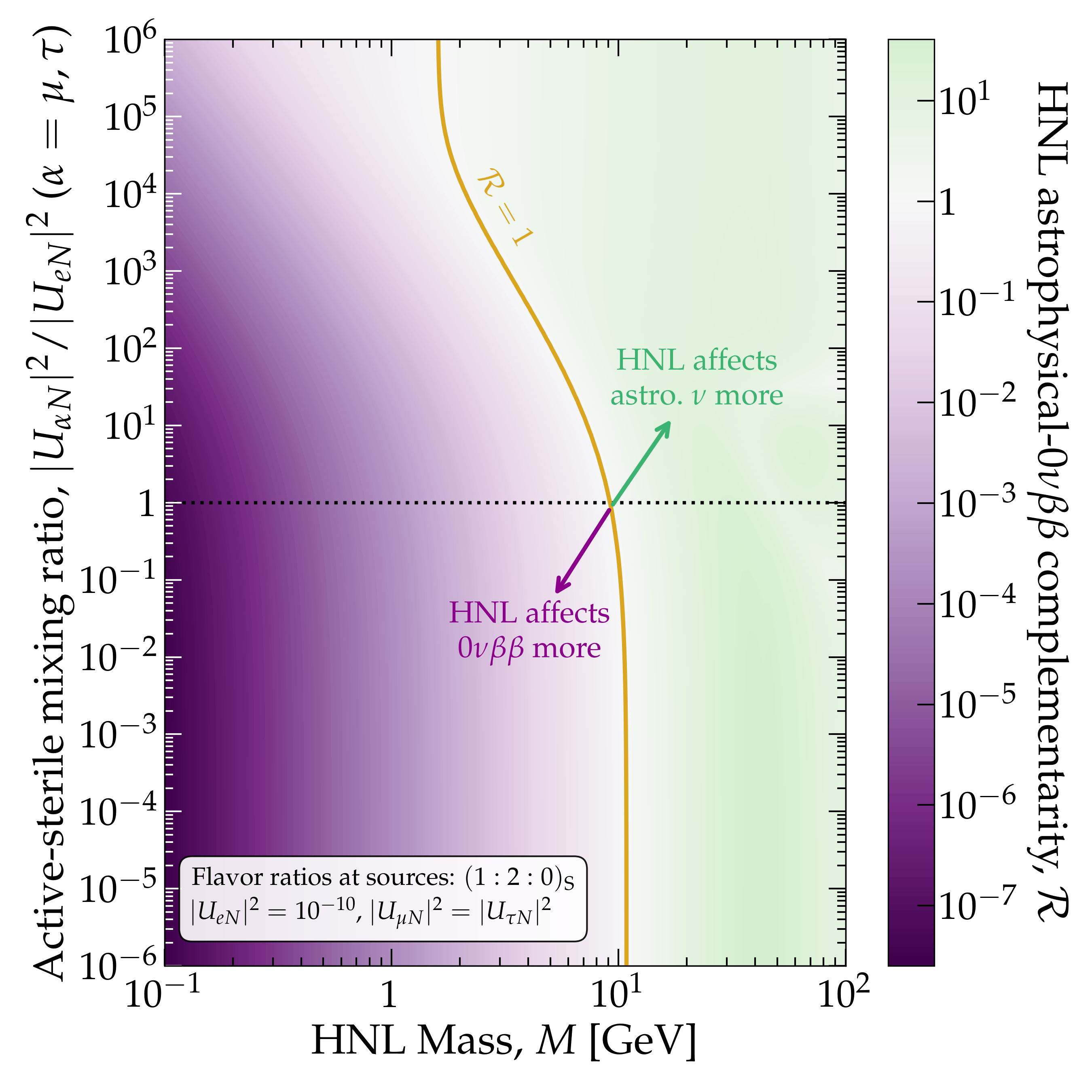}
 \caption{\textbf{Complementarity of astrophysical flavor and $0\nu\beta\beta$ in HNL searches.} The contour of complementarity [$\mathcal{R} = 1$, \equ{ratio_R}] delineates the boundary where the HNL-sensitivity of high-energy astrophysical neutrino flavor measurements by 2050 overtakes that of neutrinoless double-beta decay ($0\nu\beta\beta$) searches (\ie, nEXO). This plot assumes nominal neutrino production via full pion decay [$(1:2:0)_{\rm S}$] and fixed $|U_{eN}|^2 = 10^{-10}$. \textbf{\textit{While $0\nu\beta\beta$ bounds the low-mass and electron-philic HNLs, the flavor composition of high-energy astrophysical neutrinos provides a discovery window for heavier HNLs and muon- and tau-philic HNLs. }} See Sec.~\ref{subsec:0vbb} for details.}
 \label{fig:ratio_R}
\end{figure}

Figure~\ref{fig:ratio_R} shows $\mathcal{R}$ as a function of $|U_{\alpha N}|^2 / |U_{e N}|^2$ and $M$ plane ($\alpha = \mu, \tau$). We identify three distinct regions:
\begin{description}
 \item[$\mathbf{\mathcal{R} \ll 1}$] \textbf{The $0\nu\beta\beta$ mass is sensitive, astrophysical neutrinos are insensitive.} The $0\nu\beta\beta$ mass is dominantly sensitive to an HNL when $|U_{e N}|^2 \gg |U_{\mu N}|^2, |U_{\tau N}|^2$, \ie, in \figu{ratio_R}, when $|U_{\alpha N}|^2 / |U_{e N}|^2$ is small.  This happens at relative low masses $M \lesssim 10$~GeV that enhance $m_{\beta\beta}^{(N)} \propto 1/M$ [\equ{mbb_heavy}]. Simultaneously, $|U_{\mu N}|^2$ and $|U_{\tau N}|^2$ are small, which suppresses the more striking HNL-induced modifications of the flavor composition at Earth (Sec.~\ref{sec:astro_flavor_hnl_correction}). This is the regime where $0\nu\beta\beta$ experiments like KamLAND-Zen~\cite{KamLAND-Zen} and nEXO~\cite{nEXO} have optimal and essentially exclusive sensitivity. 
 \item[$\mathbf{\mathcal{R} \gg 1}$] \textbf{Astrophysical neutrinos are sensitive, the $0\nu\beta\beta$ mass is insensitive.} As the HNL mass increases past about $10$~GeV, the HNL modification to the $0\nu\beta\beta$ mass is suppressed, but the flavor shift is enhanced $\propto M^3$ [\equ{deltamnu_approx}]. The flavor shift is more sensitive where the active-sterile mixing is weighted toward the muon or tau sectors, since these sectors introduce the most dramatic deviations from standard-mixing expectations. Figure~\ref{fig:ratio_R} assumes neutrino production via full pion decay, $\left( \frac{1}{3}, \frac{2}{3}, 0 \right)_\text{S}$, where cancellations in the flavor-transition probabilities dampen the HNL-induced flavor shifts.  Assuming instead production via muon-damped pion decay, $\left( 0, 1, 0 \right)_\text{S}$, bypasses these cancellations and widens the HNL astrophysical sensitivity region, but preserves our conclusions. This is the regime where neutrino telescopes have optimal sensitivity. 
 \item[\textbf{$\mathcal{R} \approx 1$}] \textbf{Both $0\nu\beta\beta$ and astrophysical neutrinos are sensitive.} When the three flavor couplings are comparable, both $m_{\beta\beta}$ and $\mathcal{D}$ are large. The two measurements that enter $\mathcal{R}$ carry independent information: $m_{\beta\beta}^{(N)}$ is governed by $U_{eN}$, while $\mathcal{D}$ depends on all three couplings (with a weighting that varies with source composition). The flavor couplings can be extracted from SHiP, the predicted $m_{\beta\beta}$ computed and compared with experiment, and the predicted flavor shift compared with astrophysical measurements. A consistent picture across these stages constitutes robust evidence for the Majorana nature of $N$. In \figu{ratio_R}, this regime is traced by the contour of complementarity [\(\mathcal{R} = 1\), \equ{ratio_R}], which delineates the boundary where the discovery potential of high-energy astrophysical neutrino flavor measurements by 2050 overtakes that of neutrinoless $0\nu\beta\beta$ searches.
\end{description}

The complementarity landscape depends on the neutrino mass ordering. For the \emph{inverted ordering} (IO), the light-neutrino contribution, $m_{\beta\beta}^{(\nu)}$, has a well-known lower bound, $m_{\beta\beta}^{(\nu)} \gtrsim 10$~meV, making $0\nu\beta\beta$ experiments generally more powerful. For the \emph{normal ordering} (NO), $m_{\beta\beta}^{(\nu)}$ can vanish when the Majorana phases are tuned such that $\left|\sum_{i} U_{ei}^2 m_i \right| \simeq 0$, a well-known blind spot of $0\nu\beta\beta$~\cite{Pascoli:2002xq}. This leaves the astrophysical-neutrino measurement as the \emph{only} available probe of the Majorana nature of the HNL, and galvanizes the complementarity between the two measurements.


\subsection{The role of LHC null results}
\label{subsec:LHC}

\textbf{\textit{The general problem of interpretation.---}}An observed deviation of $U(Q)$ from the low-energy PMNS matrix $U$ at some transferred momentum $Q \sim M \lesssim 20~\mathrm{GeV}$ is a necessary but not sufficient signature of a Majorana HNL. In principle, any beyond-the-Standard-Model (BSM) scenario that modifies the neutrino mixing matrix at high energies could produce a similar effect, including, for instance, non-standard neutrino interactions~\cite{NSI}, Lorentz-invariance violation~\cite{LIV}, or other new physics that acts during neutrino propagation. The central question is therefore: under what conditions can a deviation in $U(Q)$ be unambiguously attributed to a threshold correction from a heavy Majorana state, rather than to some other BSM effect? We argue that a null result at the LHC, combined with a SHiP discovery, provides the conditions needed to resolve this ambiguity. 

\medskip

\textbf{\textit{What a null LHC result excludes.---}}LHC searches constrain BSM physics through three channels:
\begin{description}
 \item[Direct searches] Production of new colored or electroweakly charged states are constrained up to masses of a few TeV, depending on the decay topology. For states that couple to the neutrino sector, the most relevant are heavy $W'$ and $Z'$ bosons, leptoquarks, and charged Higgs bosons, all of which can generate effective neutrino interactions at lower energies after being integrated out. A null result in these searches excludes such states as the origin of a deviation in $U(Q)$ at $Q \lesssim 20~\mathrm{GeV}$.
 \item[Precision electroweak observables] Loop-level contributions of new charged particles to oblique corrections ($S$, $T$, $U$ parameters) are constrained at the per-mille level by LEP and LHC data~\cite{PDG}. Any new state with electroweak quantum numbers and mass below a few TeV that contributes to $U(Q)$ at one loop would generically shift the oblique parameters beyond current bounds, unless its contributions cancel precisely. Therefore, a null result in precision electroweak tests excludes a broad class of BSM scenarios as the source of the deviation.
 \item[Lepton flavor violation (LFV)] Processes such as $\mu \to e + \gamma$~\cite{MEG:2016leq, MEGII:2023ltw}, $\mu \to 3e$, and $\mu$-to-$e$ conversion in nuclei are sensitive to off-diagonal couplings in the lepton sector at loop level. Any new particle with flavor-violating couplings large enough to shift the PMNS mixing angles at the level detectable by astrophysical neutrino telescopes would, in most models, also generate LFV rates in tension with current bounds. The exception is precisely the case of a feebly coupled HNL, for which LFV rates scale as $|U_{\alpha N}|^4$ and can be naturally suppressed below current experimental sensitivity while still producing observable mass threshold corrections.
\end{description}

\textbf{\textit{The surviving scenario.---}}After simultaneously imposing the three sets of constraints above, the BSM scenarios capable of generating a physical threshold correction in $U(Q)$ at scales $Q \lesssim 20~\mathrm{GeV}$ while remaining consistent with a comprehensive LHC null result reduce to a single class: feebly coupled neutral leptons with masses in the GeV range and mixing angles $|U_{\alpha N}|^2 \lesssim 10^{-3}$. This is precisely the parameter space targeted by SHiP~\cite{SHiP:2018xqz}.

More formally, let $\mathcal{S}_{\rm LHC}$ denote the set of BSM scenarios consistent with all LHC and precision electroweak null results, and let $\mathcal{S}_{{\rm HE}\nu}$ denote the set of scenarios capable of generating a deviation in $U(Q)$ at $Q \sim M$ that induces a flavor shift of the size detectable by next-generation neutrino telescopes. We then have $\mathcal{S}_{\rm LHC} \cap \mathcal{S}_{{\rm HE}\nu} \simeq \mathcal{S}_{\rm HNL}$, where $\mathcal{S}_{\rm HNL}$ denotes the set of models containing a \emph{Majorana} HNL in the GeV mass range. The intersection is not exact---exotic scenarios involving, for instance, secret neutrino interactions mediated by a new light gauge boson could, in principle, contribute---but such scenarios do not generate a threshold correction to the Majorana mass matrix and would therefore not produce the specific flavor fingerprint computed in Sec.~\ref{sec:threshold}.

\medskip

\textbf{\textit{The three-experiment logic.---}}Combining the above, the central interpretive result of this paper can be stated as follows. Suppose the following three conditions are simultaneously satisfied:
\begin{enumerate}
 \item \textbf{Null LHC results:} The LHC, together with precision electroweak and LFV searches, returns a null result for BSM physics up to scales of a few TeV.
 \item \textbf{SHiP discovery:} SHiP discovers an HNL with mass $M$ and couplings $|U_{eN}|^2$, $|U_{\mu N}|^2$, $|U_{\tau N}|^2$.
 \item \textbf{Flavor at neutrino telescopes:} Neutrino telescopes measure a high-energy flavor deviation matching the HNL-induced mixing modifications at $Q \sim M$ predicted by SHiP via Eqs.~\eqref{equ:delta_s13}--\eqref{equ:delta_s23}.
\end{enumerate}
Condition (1) implies that the only viable source of the deviation in $U^\prime$ is an HNL. Under condition (2), the HNL exists and its properties are measured. Under condition (3), the observed flavor shift matches the prediction under the Majorana hypothesis and is inconsistent with the Dirac hypothesis, which predicts no shift.

The joint satisfaction of all three conditions therefore constitutes \textbf{\textit{model-independent evidence for the Majorana nature of the HNL}}, and, by extension, for the Majorana nature of the light neutrinos generated by the seesaw mechanism. This argument does not rely on the observation of lepton number violation in any single experiment, but on the consistency of three measurements with entirely uncorrelated systematic uncertainties across three complementary experimental programs.

\medskip

\textbf{\textit{Robustness against LHC discoveries.---}}Our argument for combining SHiP and neutrino telescopes to assert the Majorana nature of the HNL does not \emph{require} a null result at the LHC---it is merely sharpened by one. If the LHC discovers new BSM states, those states can be characterized and their potential contribution to $U(Q)$ computed and subtracted. The residual deviation, if any, would be attributable to the threshold correction from the HNL discovered by SHiP. While an LHC null result maximizes clarity, a positive result does not invalidate the program---it merely complicates the interpretation.

Furthermore, the program is robust against the specific form of the LHC null result. Even if the LHC excludes supersymmetry, leptoquarks, and all other standard BSM scenarios up to multi-TeV scales, the feebly coupled HNL sector remains viable because its signatures at the LHC are structurally suppressed: production cross sections scale as $|U_{\alpha N}|^2 \ll 1$, placing these states below LHC sensitivity by construction. The SHiP plus astrophysical neutrino program therefore occupies a region of theory space genuinely inaccessible to hadron colliders, regardless of their energy or luminosity reach.


\section{HNL limits and discovery}
\label{sec:limits_and_discovery}

\begin{table*}[t!]
  \caption{\label{tab:free_parameters} \textbf{Free model parameters and applied priors.} Summary of the continuous parameters varied during the profile likelihood minimization. For our present results, the priors on the standard three-flavor oscillation parameters ($\theta_{12}$, $\theta_{23}$, $\theta_{13}$, $\delta_{\rm CP}$, $\Delta m_{21}^2$, $\Delta m_{3l}^2$, with $l=1$ in NO and $l=2$ in IO) are given by pairwise joint $\chi^2$ functions from the NuFIT 6.1 global fit. For our projections, the priors are Gaussian penalty terms centered on the NuFIT 6.1 best-values with tightened uncertainties reflecting the anticipated sensitivities of next-generation facilities (DUNE, Hyper-Kamiokande, and JUNO) taken from \Refe~\cite{Song:2020nfh} for the angles and Dirac phase, and assuming 5\% relative uncertainty for the mass splittings. The Majorana CP-violation phases ($\rho$, $\delta$) and the flavor composition at sources ($f_{e, {\rm S}}$) are floated freely without external penalties. The lightest neutrino mass ($m_\text{lightest}$) is also floated freely, but is indirectly penalized by imposing a pull term on the sum of neutrino masses ($\sum_\nu \equiv m_1 + m_2 + m_3$) via a half-Gaussian bounded at $\sum_\nu = 0$, representing present~\cite{ParticleDataGroup:2024cfk} and projected cosmological upper limits (which we take to be represented by tight recent limits; \eg, \Refes~\cite{Palanque-Delabrouille:2019iyz, DiValentino:2021hoh}). The HNL mass ($M$) and couplings ($|U_{\alpha N}|^2$) are floated freely.}
  \begin{ruledtabular}
    \begin{tabular}{l l c c c c c}     
      \multirow{3}{*}{Parameter} & \multirow{3}{*}{Description} & \multirow{3}{*}{Units} & \multicolumn{4}{c}{Prior} \\
      \cline{4-7} \noalign{\vspace{2pt}}   
      & & & \multicolumn{2}{c}{Present} & \multicolumn{2}{c}{Projections} \\
      \cline{4-5} \cline{6-7} \noalign{\vspace{2pt}}  
      & & & \makebox[2.5cm]{NO} & \makebox[2.5cm]{IO} & NO & IO \\
      \hline \noalign{\vspace{2pt}}
      $M$                & HNL mass              & GeV & \multicolumn{4}{c}{None} \\
      $|U_{\alpha N}|^2$ & Active-sterile mixing & --  & \multicolumn{4}{c}{None} \\
      \hline \noalign{\vspace{2pt}}
      \multicolumn{7}{c}{Nuisance parameters} \\
      \hline \noalign{\vspace{2pt}}
      $\sin^2\theta_{12}$ & Solar mixing angle & -- & 
      \multirow{2}{*}{\begin{tabular}{@{}c@{}}Joint $\chi^2$ \\ NuFIT 6.1 NO\end{tabular}} & 
      \multirow{2}{*}{\begin{tabular}{@{}c@{}}Joint $\chi^2$ \\ NuFIT 6.1 IO\end{tabular}} & 
      \begin{tabular}{@{}c@{}}Normal \\ $\mu = 0.3088$ \\ $\sigma = 0.0016416$\end{tabular} & 
      \begin{tabular}{@{}c@{}}Normal \\ $\mu = 0.3088$ \\ $\sigma = 0.0016416$\end{tabular} \\
      $\sin^2\theta_{13}$ & Reactor mixing angle & -- & 
      & 
      & 
      \begin{tabular}{@{}c@{}}Normal \\ $\mu = 0.02248$ \\ $\sigma = 0.00059$\end{tabular} & 
      \begin{tabular}{@{}c@{}}Normal \\ $\mu = 0.02262$ \\ $\sigma = 0.00059$\end{tabular} \\
      $\sin^2\theta_{23}$ & Atmospheric mixing angle & -- & 
      \multirow{2}{*}{\begin{tabular}{@{}c@{}}Joint $\chi^2$ \\ NuFIT 6.1 NO\end{tabular}} & 
      \multirow{2}{*}{\begin{tabular}{@{}c@{}}Joint $\chi^2$ \\ NuFIT 6.1 IO\end{tabular}} & 
      \begin{tabular}{@{}c@{}}Normal \\ $\mu = 0.470$ \\ $\sigma = 0.00418$\end{tabular} & 
      \begin{tabular}{@{}c@{}}Normal \\ $\mu = 0.550$ \\ $\sigma = 0.00418$\end{tabular} \\
      $\delta_{\rm CP}$ & Dirac CP-violation phase & $^\circ$ & 
      & 
      & 
      \begin{tabular}{@{}c@{}}Normal \\ $\mu = 212$ \\ $\sigma = 6.687$\end{tabular} & 
      \begin{tabular}{@{}c@{}}Normal \\ $\mu = 274$ \\ $\sigma = 6.687$\end{tabular} \\
      $\Delta m^2_{21}$ & Solar mass splitting & eV$^2$ & 
      \multirow{2}{*}{\begin{tabular}{@{}c@{}}Joint $\chi^2$ \\ NuFIT 6.1 NO\end{tabular}} & 
      \multirow{2}{*}{\begin{tabular}{@{}c@{}}Joint $\chi^2$ \\ NuFIT 6.1 IO\end{tabular}} & 
      \begin{tabular}{@{}c@{}}Normal \\ $\mu = 7.537 \times 10^{-5}$ \\ $\sigma = 3.7685 \times 10^{-7}$\end{tabular} & 
      \begin{tabular}{@{}c@{}}Normal \\ $\mu = 7.537 \times 10^{-5}$ \\ $\sigma = 3.7685 \times 10^{-7}$\end{tabular} \\
      $\Delta m^2_{3l}$ & Atmospheric mass splitting & eV$^2$ & 
      & 
      & 
      \begin{tabular}{@{}c@{}}Normal \\ $\mu = 2.511 \times 10^{-3}$ \\ $\sigma = 1.2555 \times 10^{-5}$\end{tabular} & 
      \begin{tabular}{@{}c@{}}Normal \\ $\mu = -2.483 \times 10^{-3}$ \\ $\sigma = 1.2415 \times 10^{-5}$\end{tabular} \\
      $\rho$   & Majorana CP-violation phase & rad & \multicolumn{4}{c}{Uniform (allowed range $[0,2\pi]$)} \\
      $\sigma$ & Majorana CP-violation phase & rad & \multicolumn{4}{c}{Uniform (allowed range $[0,2\pi]$)} \\
      $m_{\rm lightest}$ & Lightest neutrino mass          & eV & \multicolumn{2}{c}{$\sum_\nu < 0.12$ (95\% C.L.)} & \multicolumn{2}{c}{$\sum_\nu < 0.09$ (95\% C.L.)} \\
      $f_{e,{\rm S}}$    & Source electron flavor fraction & -- & \multicolumn{4}{c}{Uniform (allowed range $[0,1]$)} \\
    \end{tabular}
  \end{ruledtabular}
\end{table*}

\textit{We establish the frequentist profile-likelihood framework used to extract HNL limits and discovery prospects from simulated multi-detector projections. To ensure these constraints do not artificially benefit from specific source models, we float the initial electron fraction ($f_{e, {\rm S}}$) as a nuisance parameter, robustly accounting for astrophysical ignorance. Present IceCube flavor measurements are too broad to yield meaningful constraints, but projected multi-detector combinations will be able to.}


\subsection{General setup}

We quantify the capability of neutrino telescopes to resolve HNL-induced shifts in the high-energy astrophysical neutrino flavor composition, using a frequentist statistical approach, accounting for parameter uncertainties.

The free parameters in our model comprise the standard mixing parameters, the Majorana CP-violation phases, and the lightest neutrino mass, $\boldsymbol{\theta} = \{ \theta_{12}, \theta_{23}, \theta_{13}, \delta_{\rm CP}, \Delta m^2_{21}, |\Delta m^2_{3l}|, \rho, \sigma, m_{\rm lightest} \}$ (evaluated for a chosen mass ordering, with $l = 1$ for NO and $l = 2$ for IO), the HNL parameters, $\boldsymbol{\eta} = \{ M, |U_{e N}|^2, |U_{\mu N}|^2, |U_{\tau N}|^2 \}$, and the fraction of high-energy $\nu_e$ emitted by astrophysical sources, $f_{e, {\rm S}}$.

To extract sensitivities, we employ the $\chi^2$ function
\begin{equation}
 \chi_{\rm total}^2 (\boldsymbol{\theta}, \boldsymbol{\eta}, f_{e, {\rm S}}) = \chi_{\rm data}^2 (\boldsymbol{\theta}, \boldsymbol{\eta}, f_{e, {\rm S}}) + \chi_{\rm prior}^2 (\boldsymbol{\theta}) \;,
 \label{equ:chi2_total_generic}
\end{equation}
where the comparison of our theoretical flavor predictions against the flavor measurements is performed via 
\begin{equation}
\chi_{\rm data}^2 (\boldsymbol{\theta}, \boldsymbol{\eta}, f_{e, {\rm S}}) = -2 \ln \mathcal{L} (f_{\alpha, \oplus}(\boldsymbol{\theta}, \boldsymbol{\eta}, f_{e, {\rm S}})) \;.
\end{equation}
Here, $\mathcal{L}$ is the experimental likelihood of the flavor-composition measurements in IceCube and future multi-detector combinations, which tacitly reflects the challenges inherent to flavor measurements presented in Sec.~\ref{sec:astro_production-flavor_measurement}.  Following Sec.~\ref{sec:astro_production-flavor_present_future}, we use the 11.4-year IceCube MESE flavor likelihood for present results and our 2040 and 2050 multi-detector combinations for  projections. The projected flavor measurements are different depending on whether we use them to place limits on HNL parameters (Sec.~\ref{sec:limits_and_discovery-limits}) or to discover them (Sec.~\ref{sec:limits_and_discovery-discovery}); we elaborate on them in each case below. The term $\chi_{\rm prior}^2$ in \equ{chi2_total_generic} applies penalty terms (pulls) to the standard mixing parameters, as we explain next.

Table~\ref{tab:free_parameters} lists the free parameters of our analysis and the priors we use on them.  For present-day results, we use the NuFIT 6.1 joint $\chi^2$ profiles on $\sin^2 \theta_{23}$--$\delta_{\rm CP}$, $\sin^2 \theta_{13}$--$\sin^2 \theta_{12}$, and $\Delta m_{21}^2$--$\Delta m_{3l}^2$. For our projections, we use individual Gaussian priors on each of these parameters, with tighter uncertainties, inspired by the improvements expected from measurements at DUNE, Hyper-Kamiokande, and JUNO, as presented in \Refe~\cite{Song:2020nfh}. We let $f_{e, {\rm S}}$ float unconstrained in $[0, 1]$ to account for the unknown astrophysical neutrino production mechanism.  We let $\rho$ and $\sigma$ float unconstrained in $[0, 2\pi]$, their full physically allowed range.  Finally, we also let $m_\text{lightest}$ float (with $m_\text{lightest} = m_1$ in NO and $m_\text{lightest} = m_3$ in IO), but apply a penalty to the sum of the neutrino masses, $\sum_\nu \equiv m_1 + m_2 + m_3$, representing present and future cosmological upper limits (the heavier masses are computed dynamically from $m_\text{lightest}$, $\Delta m_{21}^2$, and $\Delta m_{3l}^2$). 

Depending on the assumed true flavor composition at Earth, we perform two distinct statistical analyses: extracting limits or assessing discovery potential. While projecting stringent upper limits on the HNL yields valuable supplementary constraints, limit-setting is not our ultimate goal. The primary objective of our proposed program is the definitive discovery of the HNL and the establishment of its Majorana nature.


\subsection{Results using present flavor measurements}

To probe for HNL-modified flavor shifts using present-day IceCube MESE flavor measurements, we adopt the corresponding flavor likelihood from Fig.~A1 in \Refe~\cite{Bustamante:2026aur}, which approximates the result reported by the IceCube Collaboration in \Refe~\cite{Abbasi:2025fjc} (the approximate 68\% C.L.~contours are shown in \figu{hnl_flavor_contours}).

We have found that the present-day flavor measurement provides no constraint on the HNL parameter space: its uncertainty  exceeds appreciably the predicted size of the HNL-induced flavor shifts. It yields neither meaningful HNL exclusion limits nor discovery signatures. The remainder of this section focuses exclusively on projected limits and discovery prospects.


\subsection{Projected limits on HNL parameters}
\label{sec:limits_and_discovery-limits}

In this analysis, we assume there is no experimental evidence for the mass-threshold correction. The simulated flavor measurement at Earth is centered strictly on the standard-mixing expectation from either full or muon-damped pion decay. Our constraints represent limits on the values of the HNL mass and mixing elements.


\subsubsection{Statistical procedure}
\label{sec:limits_and_discovery-statistics}

To derive meaningful bounds, we constrain a single mixing element, $|U_{\alpha N}|^2$, as a function of the HNL mass, $M$. (We have tested that attempting to constrain multiple elements simultaneously yields no meaningful constraints.)  We achieve this by setting the remaining mixing elements to zero and profiling the likelihood over all the standard mixing parameters, Majorana CP phases, and $f_{e, {\rm S}}$. At each fixed mass $M$, we compute the one-dimensional profiled likelihood relative to the local minimum at that mass:
\begin{align}
 &\Delta \chi^2(M, |U_{\alpha N}|^2) = \min_{\boldsymbol{\theta}, f_{e, {\rm S}}} [\chi_{\rm total}^2(M, |U_{\alpha N}|^2, \boldsymbol{\theta}, f_{e, {\rm S}})] \nonumber \\
 &\qquad - \min_{|U_{\alpha N}|^2, \boldsymbol{\theta}, f_{e, {\rm S}}} [\chi_{\rm total}^2(M, |U_{\alpha N}|^2, \boldsymbol{\theta}, f_{e, {\rm S}})] \;.
 \label{equ:Delta_chi2_limits}
\end{align}
We determine the $1\sigma$, $2\sigma$, and $3\sigma$ confidence intervals by applying Wilks' theorem~\cite{Wilks:1938dza} and requiring $\Delta\chi^2 \leq 1$, 4, and 9, respectively.  This yields limits on $|U_{\alpha N}|^2$ as a function of $M$. This asymptotic treatment is valid given the large event statistics of our multi-detector projections and yields conservative constraints compared to the computationally intensive Feldman-Cousins construction or full Monte-Carlo coverage evaluations, which are beyond the scope of our projections.

For the flavor likelihood, $\mathcal{L}$, we adopt the 2040 and 2050 projections shown in Figs.~A1 and A2 in \Refe~\cite{Bustamante:2026aur} (from Fig.~A2, we adopt the most ambitious projection, using HUNT). We explore separately flavor measurements centered on the standard-mixing expectation from full and muon-damped pion decay.


\subsubsection{Results}
\label{sec:limits_and_discovery-results}

\begin{figure*}[t!]
 \centering
 \includegraphics[width=\textwidth]{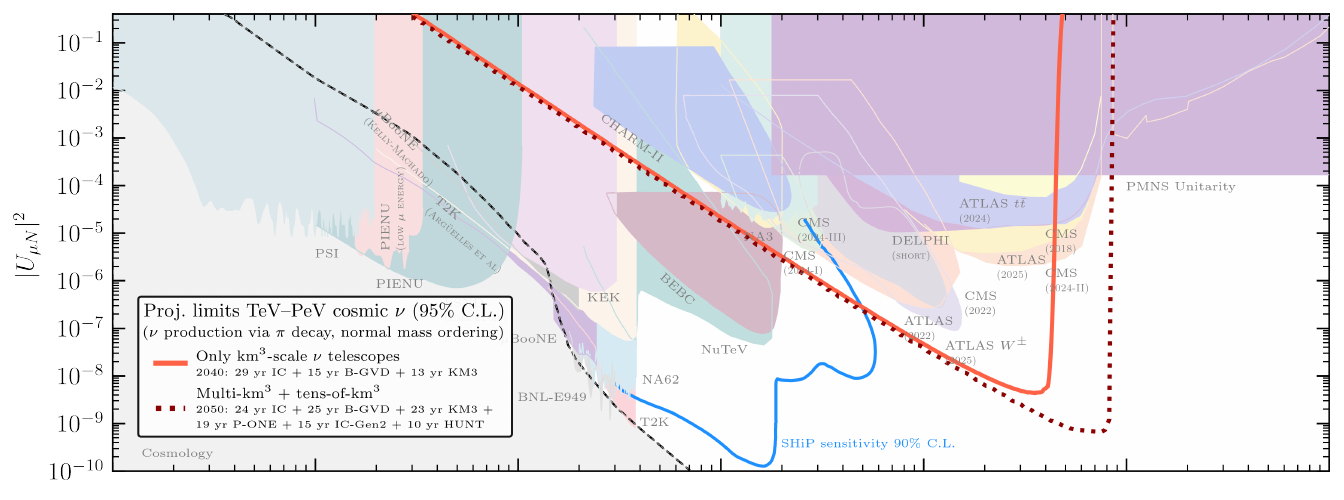}
 \vspace{5pt}
 \includegraphics[width=\textwidth]{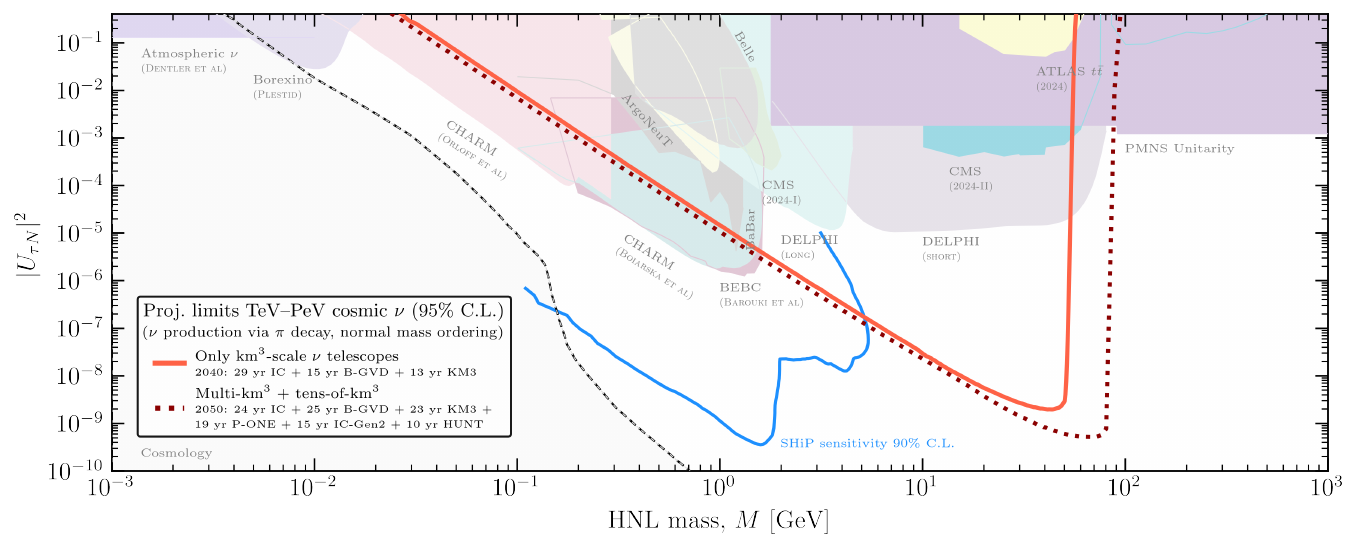}
 \vspace{-20pt}
 \caption{\textbf{Existing constraints and future sensitivities for heavy neutral leptons.} At the GeV scale and above, existing constraints on the HNL mixing are predominantly from colliders.  Previous limits are from colliders and cosmology, as summarized in \Refe~\cite{Fernandez-Martinez:2023phj}; we use \Refe~\cite{hnl_github} to import them. See Sec.~\ref{sec:limits_and_discovery-results} for details. \textbf{\textit{Above a few GeV, our projected constraints from TeV--PeV astrophysical neutrinos become the world's best, improving over even upcoming fixed-target experiments like SHiP and DUNE.}}}
 \label{fig:stacked_hnl_limits}
\end{figure*}

Figure~\ref{fig:stacked_hnl_limits} compares our limits derived from the TeV--PeV neutrino flavor composition to existing limits, across a wide HNL mass range. We show the resulting projected upper limits on the mixing elements $|U_{\mu N}|^2$ and $|U_{\tau N}|^2$. No limits can be placed on $|U_{e N}|^2$. Above masses of a few GeV, our projected limits outclass existing ones from colliders~\cite{DELPHI:1996qcc, L3:1992xaz, Abdullahi:2022jlv} by up to several orders of magnitude, becoming the most stringent limits achievable on heavy HNLs.  The SHiP sensitivity remains the strongest projected one for HNL masses under 6~GeV~\cite{SHiP:2018xqz}.

A key aspect of our results is that \textbf{\textit{world's best limits on GeV HNLs at and above the GeV scale could be achievable by 2040 using exclusively existing neutrino telescopes IceCube, KM3NeT, and Baikal-GVD}}, with improvements possible by 2050.  Further, as we show later, the limits will be achievable regardless of whether neutrinos are produced via the nominal mechanism of full pion decay or the alternative mechanism of muon-damped pion decay.

Crucially, these limits are achieved despite embracing our full ignorance on the flavor composition at the sources, by treating $f_{e, {\rm S}} \in [0,1]$ as a completely unconstrained nuisance parameter that we profile over (together with the other parameters). This is because---as prefigured in Figs.~\ref{fig:flavor_vs_U}, \ref{fig:hnl_flavor_contours}, \ref{fig:hnl_flavor_uncertainties}, and \ref{fig:hnl_flavor_orderings_compare}---for sufficiently large $|U_{\alpha N}|^2$, the flavor composition at Earth is pushed so far from the standard-mixing expectation that it cannot be mimicked by it, regardless of the value of $f_{e, {\rm S}}$.   

\medskip

\textbf{\textit{Strength of the limits.---}}This exceptional sensitivity at high masses arises from a shift in the probed physics. Traditional direct searches rely on the physical production and subsequent decay of the HNL within a detector volume, incurring a severe $|U_{\alpha N}|^4$ event-rate penalty~\cite{Gorbunov:2007ak} and suffering rapid phase-space suppression as the HNL mass approaches the kinematic limit of the accelerator, because the finite center-of-mass energy is overwhelmingly consumed by the HNL rest mass, shrinking the available phase space for the final-state particles, making them harder to detect.

In contrast, because our approach uses TeV--PeV neutrinos, the $Q > M$ threshold is comfortably exceeded for masses above a few tens of GeV, where traditional searches struggle.  The mass-threshold correction scales only $\propto |U_{\alpha N}|^2$, and grows $\propto M^3$---up to a maximum value, as we explain below.  This combination of favorable kinematics and scaling allows the astrophysical flavor measurement to bypass the bottlenecks of direct searches and probe deeply into the feeble-coupling regime.

\medskip

\textbf{\textit{Loss of sensitivity.---}}The limits in \figu{stacked_hnl_limits} terminate sharply at $M \approx 60$~GeV (2040) and $M \approx 90$~GeV (2050) due to flavor saturation and kinematic flux suppression. As the HNL mass grows, the mass correction $\delta m_{\alpha\alpha}$ [\equ{deltamnu_LEFT}] overwhelms standard mass splittings. In the effective mixing matrix $\hat{U}$, this dominant diagonal term causes the perturbed flavor state to decouple, freezing the mixing structure into a fixed asymptotic state. Consequently, the flavor fractions at Earth hit a physical cap and cannot shift any further. Because this flavor saturation stops the signal from changing, the test statistic $\Delta\chi^2$ [\equ{Delta_chi2_limits}] hits a ceiling where further coupling increases yield no additional statistical significance.

The specific truncation boundaries at $M \approx 60$~GeV and $90$~GeV emerge from a competition between detector exposure and the steeply falling astrophysical flux. The high-$Q$ tail of the neutrino momentum distribution falls as $\mathcal{P}(Q) \propto Q^{-2.86}$ (see Figs.~1 and B3 in \Refe~\cite{Bustamante:2026zst}), which restricts the surviving flux fraction above the kinematic threshold to $\eta(M) \propto M^{-1.86}$. Because the test statistic scales quadratically with this available flux ($\propto \eta^2$), the statistical significance of the saturated flavor shift plummets as $M^{-3.72}$. Consequently, the nearly nine-fold exposure upgrade of the 2050 network~\cite{Liu:2023flr} extends the mass reach by only a factor of $8.84^{1/3.72} \approx 1.80$, pushing the kinematic cutoff from 60~GeV to 90~GeV before the signal drops below the detection threshold.

\medskip

\begin{figure}[t!]
 \centering
 \includegraphics[width=\columnwidth]{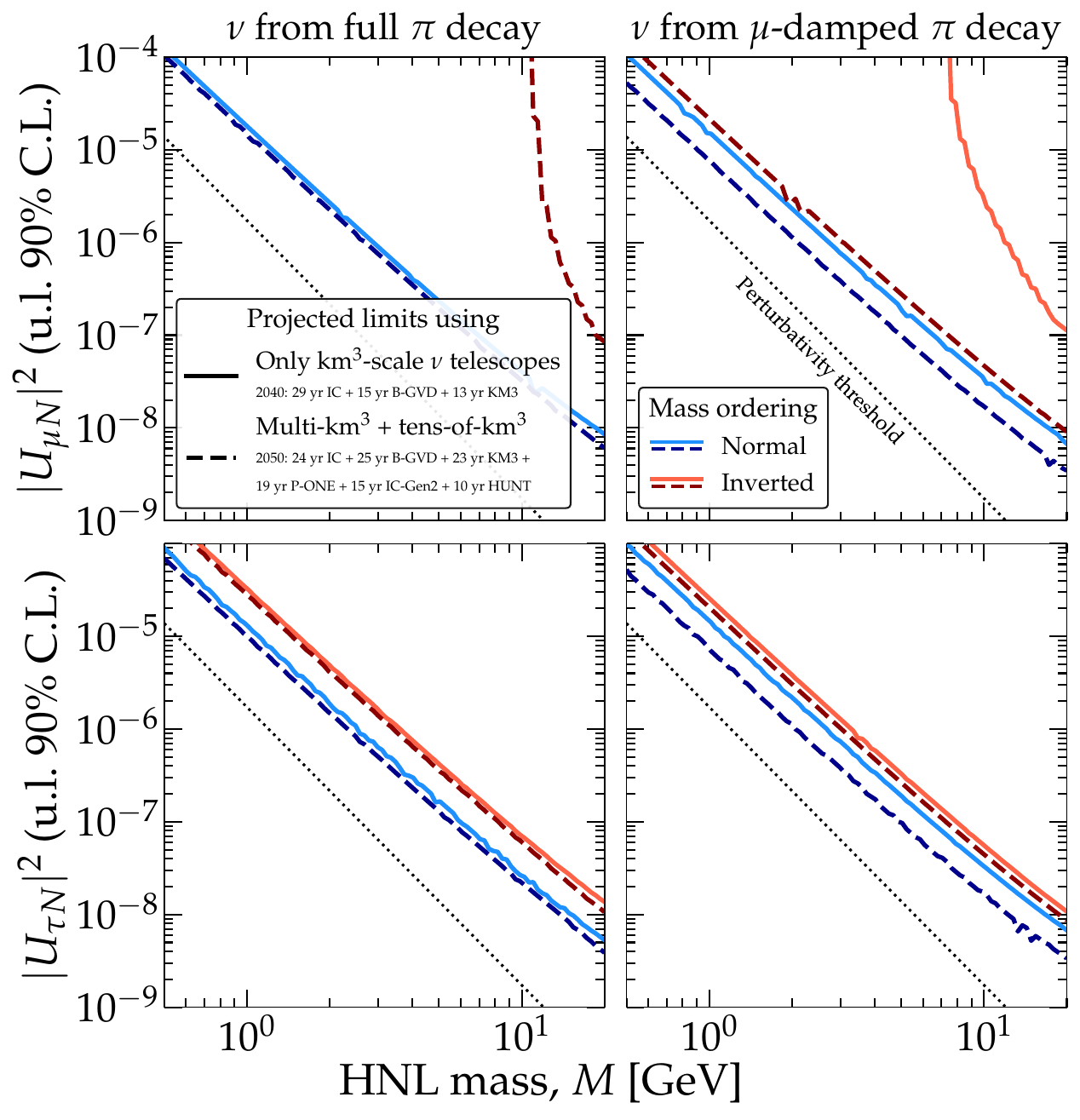}
 \caption{\textbf{Projected 90\% C.L. upper limits on the heavy active-sterile mixing elements.} Unlike standard bump-hunt constraints, these limits are heavily profiled over the completely free initial electron-flavor fraction at the sources, $f_{e, {\rm S}}$, ensuring robustness against astrophysical unknowns. The flat sensitivity plateau for $M \gtrsim 1$ GeV reflects the saturation of the Majorana-induced kinematic shift; here, the center-of-mass energy of the TeV--PeV astrophysical flux universally satisfies $s \gg M^2$, maximizing the distortion of the Earth-measured flavor composition.}
 \label{fig:limits_Usq_vs_M}
\end{figure}

\textbf{\textit{The effect of mass ordering.---}}Figure~\ref{fig:limits_Usq_vs_M} shows that the projected upper limits are stronger under NO than under IO. While HNL-induced flavor shifts are larger under IO (Sec.~\ref{sec:astro_flavor-flavor_regions}), they align more closely with the standard-mixing allowed region, allowing them to be partially absorbed when profiling over $f_{e, {\rm S}}$. 

This alignment stems from the native flavor content of the lightest mass eigenstate, which the HNL coupling $|U_{\alpha N}|^2$ attempts to populate with flavor $\alpha$. In IO, the lightest eigenstate ($\nu_3$) is natively muon-rich ($|U_{\mu 3}|^2 \approx 0.46$). Adding an HNL muon-flavor coupling ($|U_{\mu N}|^2$) induces no major structural reorganization of the mixing matrix, leaving the direction of the flavor shift collinear with standard-mixing expectations and highly absorbable. However, because standard atmospheric mixing ($\theta_{23}$) prefers the second octant, $\nu_3$ natively contains less tau flavor ($|U_{\tau 3}|^2 \approx 0.42$). Due to this asymmetry, accommodating a tau-flavor coupling ($|U_{\tau N}|^2$) forces a larger structural rotation of the mixing matrix than a muon perturbation does. This pushes the tau-induced shift further out of alignment, making it significantly harder to absorb via profiling. This explains why the IO limits are stronger for tau-philic HNLs than for muon-philic ones.

In contrast, in NO, the lightest eigenstate ($\nu_1$) is heavily electron-rich ($|U_{e 1}|^2 \approx 0.68$). Introducing a muon or tau perturbation forces a drastic structural rotation of the mixing matrix. Consequently, the resulting flavor shift points  away from standard-mixing expectations, rendering the HNL signal unabsorbable by the profiling over the unknown $f_{e, {\rm S}}$, which explains the tighter NO contours. We detail the exact numerical mechanics of this eigenstate reorganization in Appendix~\ref{app:limits_details}. The remainder of our analysis focuses entirely on the NO case.

\medskip

\textbf{\textit{The effect of the source composition.---}}Figure~\ref{fig:limits_Usq_vs_M} demonstrates that the limits remain robust under alternative astrophysical source mechanisms, and in fact become visibly stronger when assuming neutrino production via muon-damped pion decay rather than full pion decay. The muon-damped source composition, $(0, 1, 0)_{\rm S}$, lacks the initial electron-flavor component present in full pion decay, $(1/3, 2/3, 0)_{\rm S}$. This absence bypasses cancellations in the flavor-transition probabilities that otherwise dampen the HNL-induced flavor shifts. By maximizing the observable deviation from standard-mixing expectations at Earth, the muon-damped scenario widens the HNL sensitivity region, though the fundamental mass reach of the multi-detector network remains independent of the specific source mixture.


\subsection{Projected HNL discovery}
\label{sec:limits_and_discovery-discovery}

In this analysis, we base our projections on the Asimov data set~\cite{Cowan:2010js} corresponding to one of our non-standard benchmark points A--G from Table~\ref{tab:benchmark_points}. These data sets represent the exact theoretical expectation without statistical fluctuations, such that the best-fit parameters perfectly match the injected true HNL parameters ($\boldsymbol{\eta}_{\rm best} \equiv \boldsymbol{\eta}_{\rm true}$) and the unconstrained global minimum evaluates identically to zero ($\chi^2_\text{global~min} = 0$). To evaluate the test statistic, we construct experimental likelihood functions using the same projected multi-detector exposures as in our limit-setting procedure, but centered on the HNL-modified flavor composition of the injected benchmark A--G rather than the standard-mixing expectation. Under this assumption, the flavor composition exhibits evidence of the Majorana HNL-induced mass-threshold correction. We evaluate the experimental capacity to discover this new physics in two sequential steps: hypothesis testing and parameter estimation.

\subsubsection{Statistical procedure}

\textbf{\textit{Hypothesis testing.---}}First, we quantify the global significance of rejecting the standard-mixing null hypothesis ($|U_{\alpha N}|^2 = 0$). We compute the local test statistic,
\begin{equation}
q_0(\boldsymbol{\eta}_{\rm true}) \!=\! \min_{\boldsymbol{\theta}, f_{e, {\rm S}}} [\chi_{\rm total}^2(|U_{\alpha N}|^2 = 0, \boldsymbol{\theta}, f_{e, {\rm S}}; \boldsymbol{\eta}_{\rm true})] \;.
\end{equation}
While a real experiment must scan the unknown mass $M$ to find the maximum test statistic, our use of an Asimov data set guarantees this maximum occurs exactly at the injected true value, $M_{\rm true}$, making this direct evaluation a simplification for sensitivity projections.

(Because the HNL mass $M$ is unknown \textit{a priori}, a statistical fluctuation could mimic a signal anywhere within the probed mass range. Accounting for such fluctuation would be pertinent in a statistical analysis performed at the event-rate level, where we would need to compute the look-elsewhere effect to report the global $p$-value, $p_{\rm global}$, say, via the Gross-Vitells method~\cite{Gross:2010qma}.  However, our analysis is performed at the level of flavor composition, where there are no such random fluctuations and, therefore, no need to account for them.)

We assume only one mixing element is non-zero but float which flavor it is. The null and alternative hypotheses differ by exactly one degree of freedom. We enforce the physical constraint $|U_{\alpha N}|^2 \geq 0$ during minimization. The null hypothesis ($|U_{\alpha N}|^2 = 0$) thus lies exactly on the boundary of the allowed parameter space. Under the null hypothesis, random fluctuations in the reconstructed flavor coordinates would cause the unconstrained best-fit estimator of the squared mixing element to mathematically prefer a negative value. However, because the boundary condition intercepts these unphysical fits and pegs the estimator to zero, the resulting test statistic piles up exactly at $q_0 = 0$. 

According to Chernoff's theorem~\cite{Chernoff:1954eli}, this modifies the asymptotic distribution of the test statistic from a pure $\chi^2_1$ distribution (a $\chi^2$ distribution with one degree of freedom) to a mixture distribution, $\frac{1}{2}\delta(0) + \frac{1}{2}\chi^2_1$, where $\delta$ is the Dirac delta. The local $p$-value is therefore half the standard Wilks expectation. By applying this factor to the standard $\chi^2_1$ right-tail probability, $1 - \mathrm{erf}(\sqrt{q_0/2})$ (which enters because a $\chi^2_1$ variable is the square of a standard normal variable), we obtain
\begin{equation}
 p_{\rm local} = \frac{1}{2} \left[ 1 - \mathrm{erf}\left(\sqrt{\frac{q_0}{2}}\right) \right] \;.
\end{equation}
Because the active flavor channel is not specified \textit{a priori}, we apply a discrete factor-of-3 trials factor in the global $p$-value, \ie, 
\begin{equation}
 p_\text{global} = 3 \, p_\text{local} \;.
\end{equation}
A discovery is claimed if the resulting $p_{\rm global}$ falls below the $5\sigma$ significance threshold ($2.87 \times 10^{-7}$), decisively rejecting the standard-mixing hypothesis and establishing the Majorana nature of the HNL.

\medskip

\textbf{\textit{Parameter estimation.---}}Second, having established HNL discovery, we attempt to measure the HNL parameters by computing how well the values of $M$ and $|U_{\alpha N}|^2$ can be simultaneously constrained. We evaluate the two-dimensional profiled likelihood test statistic against the assumed true benchmark values:
\begin{align}
&\Delta \chi^2(M, |U_{\alpha N}|^2; \boldsymbol{\eta}_{\rm true}) \nonumber \\
&\qquad = \chi_{\rm total}^2(M, |U_{\alpha N}|^2, \hat{\hat{\boldsymbol{\theta}}}, \hat{\hat{f}}_{e, {\rm S}}; \boldsymbol{\eta}_{\rm true}) \;,
\end{align}
where $\hat{\hat{\boldsymbol{\theta}}}$ and $\hat{\hat{f}}_{e, {\rm S}}$ denote the parameter values that minimize the test statistic for fixed $M$ and $|U_{\alpha N}|^2$, and the subtraction of $\chi^2_\text{global~min}$ is omitted as it vanishes on the Asimov data set. We map the allowed regions in the parameter space of $M$ and $|U_{\alpha N}|^2$ by computing this test statistic across the grid. We delineate the $1\sigma$, $2\sigma$, and $3\sigma$ confidence regions by drawing closed contours at $\Delta \chi^2 = 2.30$, 6.18, and 11.83, respectively. The resulting contours demonstrate the experimental capability to pinpoint the mass and couplings of the HNL.


\subsubsection{Results}
\label{sec:limits_and_discovery-discovery-results}

\textbf{\textit{Hypothesis testing.---}}Table~\ref{tab:benchmark_points} shows the resulting p-values for our benchmark points A--G. Points A, D, and E---generated assuming neutrino production via full pion decay---yield null discovery significance.  Points B, C, F, and G---generated assuming neutrino production via muon-damped pion decay---yield high discovery significance, exceeding $5\sigma$ in our 2050 projections.  

The null significance of Point A ($|U_{eN}|^2$) is due to the same reason as the lack of constraints on $|U_{eN}|^2$ we found earlier: in this case, under NO, the effect of the mass threshold correction is to marginally increase the already large $\nu_e$ content of the lightest mass eigenstate, $\nu_1$, rendering the change small. 

For Points D and E, despite the coupling ($|U_{\tau N}|^2$) successfully forcing a larger correction to the mass matrix, the resulting flavor shift occurs in the same direction as the standard-mixing flavor region. This occurs because Points D and E assume neutrino production via full pion decay ($f_{e, {\rm S}} \approx 1/3$), for which the standard-mixing flavor region  resides in the interior of the HNL-allowed flavor region (as a result of full pion decay being relatively impervious to the HNL-induced modifications; see Sec.~\ref{sec:astro_flavor-numerical_results} and Appendix~\ref{app:asymptotic_flavor_limits}), allowing the likelihood profiling over $f_{e, {\rm S}}$ to absorb the HNL-induced flavor shifts. These arguments are further elaborated in Appendix~\ref{app:limits_details}.

In stark contrast, Points B, C, F, and G are generated assuming neutrino production via muon-damped pion decay ($f_{e, {\rm S}} \approx 0$), which positions the standard-mixing flavor expectation on the boundary of the generic standard-mixing region spanned by varying $f_{e, {\rm S}} \in [0,1]$; see \figu{hnl_flavor_contours}. The HNL-induced mass correction takes the flavor composition at Earth into a region that cannot be mimicked by any standard-mixing variation of $f_{e, {\rm S}}$. Unable to absorb the displacement, the standard-mixing hypothesis incurs a severe $\chi^2$ penalty, yielding global significances that approach or cross the $5\sigma$ discovery threshold in our 2050 multi-detector projections.  \textbf{\textit{Reaching this threshold constitutes confirmation of the Majorana nature of the HNL}}.

\medskip

\begin{figure*}[t!]
 \centering
 \includegraphics[width=\textwidth]{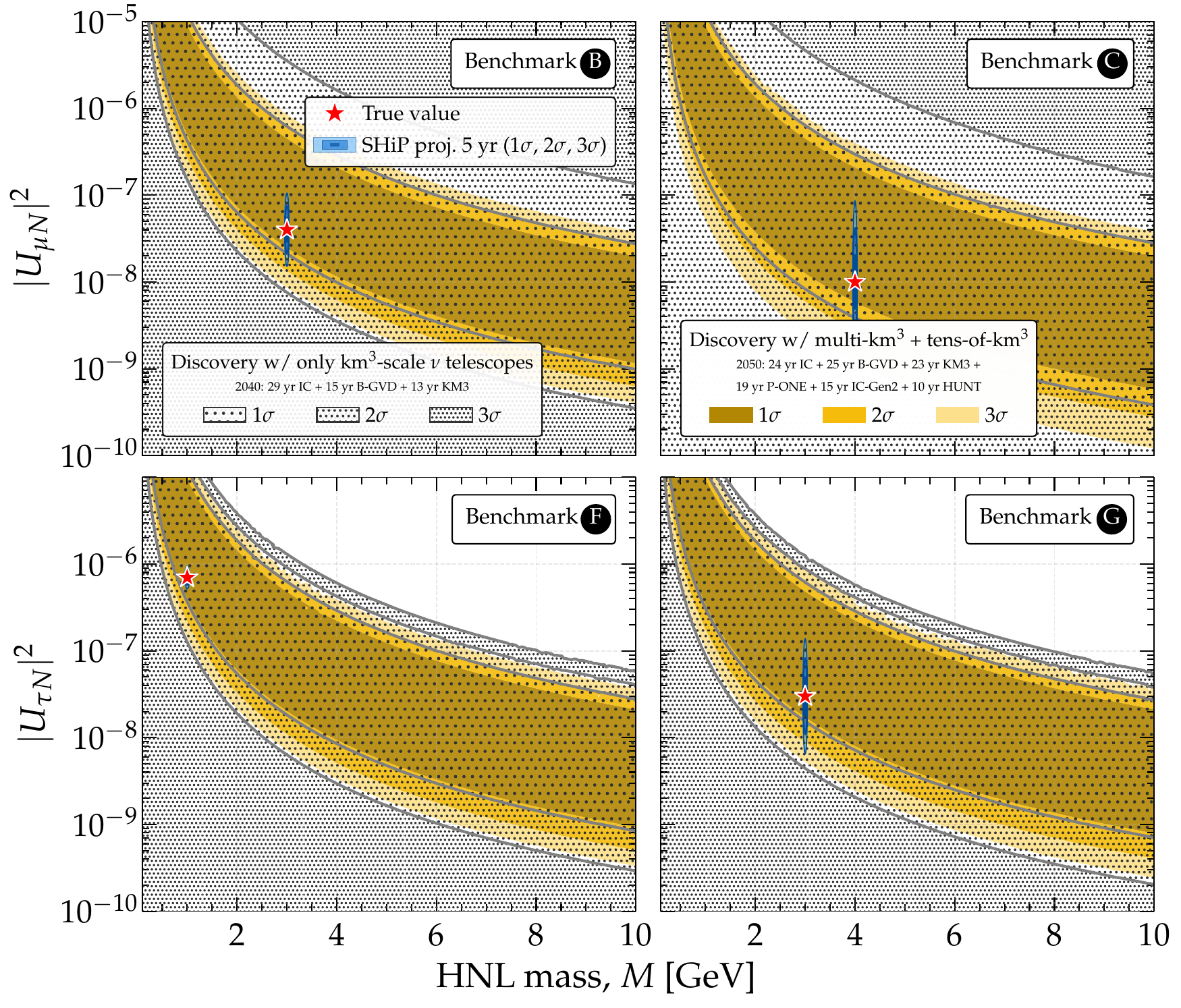}
 \caption{\textbf{Projected sensitivity for post-discovery HNL parameter estimation.} The $1\sigma$, $2\sigma$, and $3\sigma$ allowed regions in the $M$--$|U_{\alpha N}|^2$ parameter space for the benchmark points yielding high discovery significance (B, C, F, and G). The contours reflect profiling over all standard oscillation parameters and the initial source composition, $f_{e, {\rm S}}$; see Table~\ref{tab:free_parameters}. Results are compared between the 2040 and 2050 multi-detector projections. The shape of the allowed regions reflects a parametric degeneracy between the HNL mass and coupling.  Overlaid are estimated SHiP HNL measurement regions.  See Sec.~\ref{sec:limits_and_discovery-discovery-results} for details. \textbf{\textit{Guided by fixed-target experiments, present and future TeV--PeV neutrino telescopes will be able to establish the Majorana nature of neutrinos.}}}
 \label{fig:param_estimation}
\end{figure*}

\textbf{\textit{Parameter estimation.---}}Figure~\ref{fig:param_estimation} shows the resulting allowed regions in the $M$--$|U_{\alpha N}|^2$ parameter space for the observable benchmark points B, C, F, and G. Consistent with our use of Asimov data sets, the test statistic evaluates to exactly zero at the injected true parameters. However, the profile likelihood does not converge on a single, unique best-fit point. Instead, it yields a continuous trough of equally good best-fit values that stretches across the parameter space. This is the manifestation of an unbroken parametric degeneracy: because the flavor shift at Earth is governed by the mass-threshold correction, $\delta m_{\alpha\alpha} \propto |U_{\alpha N}|^2 M^3$ at low masses [\equ{deltamnu_approx}], an infinite set of anti-correlated $M$ and $|U_{\alpha N}|^2$ combinations induce the same structural reorganization of the mixing matrix, yielding identical experimental signatures.

This parametric degeneracy is resolved by incorporating complementary terrestrial constraints. Figure~\ref{fig:param_estimation} superimposes projected confidence intervals for SHiP, calculated via a simple semi-analytic projection that extrapolates official $90\%$ C.L. limits using $|U_{\alpha N}|^4$ event scaling; see Appendix~\ref{app:ship_forecasts} for full details. SHiP extracts $M$ through the kinematic reconstruction of decay products and $|U_{\alpha N}|^2$ via the independent total event yield. These measurements produce localized contours that intersect the astrophysical troughs at the true injected parameters. This explicitly exposes the central tenet of our proposal: \textbf{\textit{SHiP pinpoints the HNL parameters, while high-energy neutrino telescopes confirm the Majorana nature of the underlying flavor shift.}}

Despite the benchmark points targeting different coupling channels ($|U_{\mu N}|^2$ in B and C versus $|U_{\tau N}|^2$ in F and G) and different mass scales, the physical size of the uncertainty bands on the inferred parameters is remarkably comparable across all points. This uniformity occurs because the resolving power of the measurement is not intrinsically limited by the specific choice of flavor coupling. Rather, it is limited by the unknown astrophysical source composition, $f_{e, {\rm S}}$. Floating freely without external penalty, $f_{e, {\rm S}}$ overwhelmingly dominates the systematic uncertainty budget of the parameter estimation, equally smearing the predictive power across the parameter space regardless of the active HNL coupling.

The extraction of parameters is dependent on both the true HNL mass and the accumulated detector exposure. As shown in \figu{param_estimation}, the constraints degrade as $M$ increases. At higher masses, the constant-correction degeneracy dictates that the required coupling $|U_{\alpha N}|^2$ becomes correspondingly smaller. As the magnitude of the coupling drops, the active-sterile mixing signature becomes increasingly difficult to disentangle from the dominant $f_{e, {\rm S}}$ uncertainty and the standard-mixing nuisance parameters, widening the allowed regions. Overcoming this requires vast statistical samples. While the 2040 multi-detector projections correctly identify the presence of HNL mixing and bound the parameter space at $1\sigma$ and $2\sigma$, they lack the resolving power to close the $3\sigma$ intervals. In contrast, the 2050 multi-detector projections successfully close the $3\sigma$ bands, \textbf{\textit{demonstrating that future neutrino telescopes may meaningfully identify the Majorana nature of neutrinos.}}


\section{Conclusions}
\label{sec:conclusions}

We proposed a new probe of the Majorana nature of neutrinos, complementary to traditional neutrinoless double beta decay ($0\nu\beta\beta$) searches, by integrating beam-dump experiment discoveries (\ship) with TeV--PeV astrophysical neutrino flavor measurements at IceCube, KM3NeT, Baikal-GVD, and future telescopes. A GeV-scale Majorana heavy neutral lepton (HNL) induces a correction to the neutrino mass and mixing parameters, shifting the flavor composition of high-energy astrophysical neutrinos at Earth. Because this correction vanishes identically for Dirac fermions, correlating an HNL discovery at \ship\ with a high-energy flavor shift provides an unambiguous test of its Majorana nature. Our main results are as follows.

\medskip

\textbf{\textit{A clean Majorana discriminator.---}}A Majorana HNL modifies the masses of the light, active neutrinos [\equ{deltamnu_LEFT}]. This perturbation is determined entirely by Standard-Model masses, the HNL mass $M$, and the active-sterile mixing elements ($U_{e N}$, $U_{\mu N}$, $U_{\tau N}$), yielding a precisely calculable effect once these physical properties are measured. Because this correction evaluates to identically zero for Dirac fermions, it serves as a definitive physical discriminator of the Majorana nature.

\medskip

\textbf{\textit{An astrophysical observable.---}}This correction alters the effective mixing of neutrinos scattering at momentum transfers $Q > M$. Because the $Q$-distribution of TeV--PeV astrophysical neutrinos peaks around $Q \simeq 20$~GeV, the astrophysical neutrinos comfortably scatter above this threshold. The HNL correction applies unsuppressed, producing a characteristic shift in the observable proportions of $\nu_e$, $\nu_\mu$, and $\nu_\tau$ at Earth. The direction of this shift is dictated by the active-sterile couplings, making the astrophysical signal predictive once couplings are measured at \ship.

\medskip

\textbf{\textit{Complementarity with $0\nu\beta\beta$.---}}The astrophysical flavor shift is governed by all three flavor couplings, whereas $0\nu\beta\beta$ depends exclusively on $|U_{eN}|$. In the normal neutrino mass ordering, the light-neutrino contribution to the $0\nu\beta\beta$ effective mass can vanish for specific values of the Majorana CP-violation phases, creating a blind spot. Here, the astrophysical flavor measurement is the \emph{only} available probe of the Majorana nature. Furthermore, where muon or tau couplings dominate ($|U_{\mu N}|^2, |U_{\tau N}|^2 \gg |U_{e N}|^2$), the discovery potential of the astrophysical measurement eclipses that of $0\nu\beta\beta$.

\medskip

\textbf{\textit{World-leading high-mass sensitivity.---}}By bypassing the severe phase-space suppression that bottlenecks terrestrial experiments, our projected limits on the HNL couplings are the most stringent constraints globally. Even when profiling over the unknown source composition ($f_{e, {\rm S}}$), projected multi-detector networks eclipse both existing collider bounds and upcoming fixed-target sensitivities on $|U_{\mu N}|^2$ and $|U_{\tau N}|^2$ by orders of magnitude for HNL masses above a few GeV.

\medskip

\textbf{\textit{The joint discovery window.---}}The joint parameter space where \ship\ discovers the HNL and next-generation neutrino telescopes detect the induced flavor shift simultaneously is strictly viable and compatible with current IceCube data. Within this joint window, the HNL induces a flavor shift of 5--15\% relative to standard-mixing expectations. While currently obscured by experimental uncertainties, this deviation falls within the resolving power of projected multi-detector networks by 2040--2050 for both muon- and tau-philic HNLs.

\medskip

\textbf{\textit{The three-experiment logic.---}}An LHC null result at the TeV scale eliminates other heavy BSM mediators; a direct HNL discovery at \ship\ fixes the mass and active-sterile couplings; and a correlated high-energy flavor shift at neutrino telescopes confirms the consequence unique to Majorana fermions. Because these three independent measurements possess uncorrelated systematic uncertainties, their joint observation constitutes model-independent evidence for the Majorana nature of the heavy neutral lepton, bypassing the need to observe direct lepton number violation.\\

\medskip

To our knowledge, this is the first proposal to use the high-energy flavor composition of astrophysical neutrinos as a physical discriminator between the Majorana and Dirac nature of neutrinos.  By exploiting the unique kinematic reach of TeV--PeV astrophysical neutrino telescopes, this mechanism opens a new window onto one of the deepest open questions in particle physics: the origin of neutrino mass.


\acknowledgements

MB is supported by {\sc Villum Fonden} under project no.~29388. This work used the Tycho supercomputer hosted at the SCIENCE High Performance Computing Center at the University of Copenhagen.  QL is supported by Canada First Research Excellence Fund and Natural Sciences and Engineering Research Council of Canada through the Arthur B.~McDonald Canadian Astroparticle Physics Research Institute.  GB is supported by the Spanish grants CIPROM/2021/054 (Generalitat Valenciana), PID2023-151418NB-I00 funded by MCIU/AEI/10.13039/501100011033/, and by the European ITN project HIDDeN (H2020-MSCA-ITN-2019/860881-HIDDeN).


\bibliography{refs}


\appendix


\section{SHiP sensitivity to mixing-angle corrections}
\label{app:var_angles_UeN_UtauN}

\renewcommand{\theequation}{A\arabic{equation}}
\renewcommand{\thefigure}{A\arabic{figure}}
\renewcommand{\thetable}{A\arabic{table}}
\setcounter{figure}{0} 
\setcounter{table}{0}

Figure~\ref{fig:discovery_window_var_angles_UmuN} showed the variation of the mixing-angle corrections, $\delta (\sin^2 \theta_{ij})$, with $M$ and $|U_{\mu N}|^2$ within the SHiP sensitivity region, for a muon-philic HNL under NO.

\begin{figure*}[t]
 \centering
 \includegraphics[width=\textwidth]{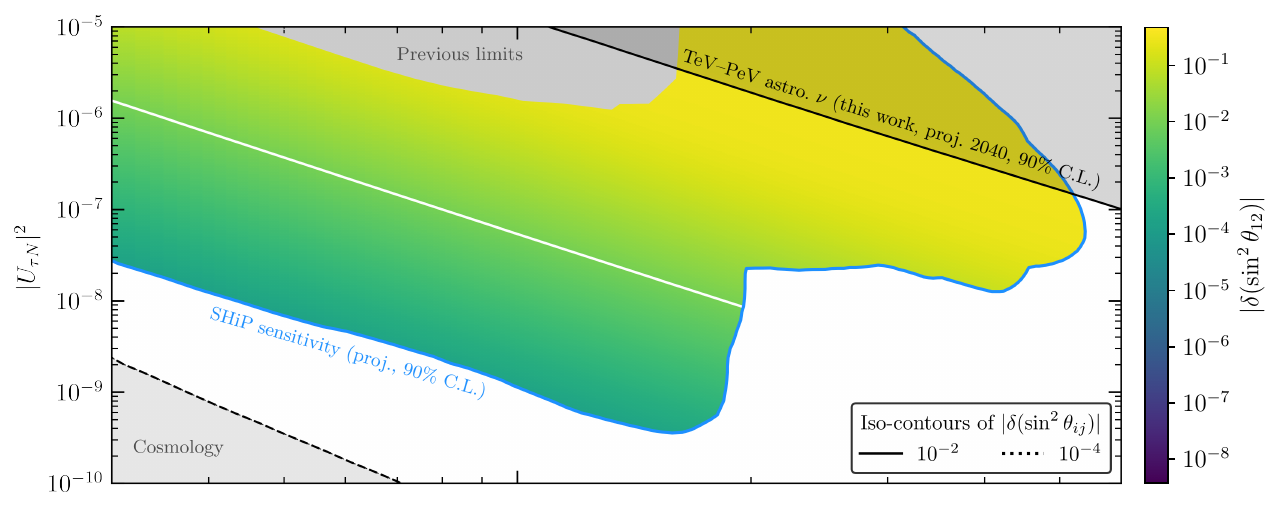}
 \\[-7pt]
 \includegraphics[width=\textwidth]{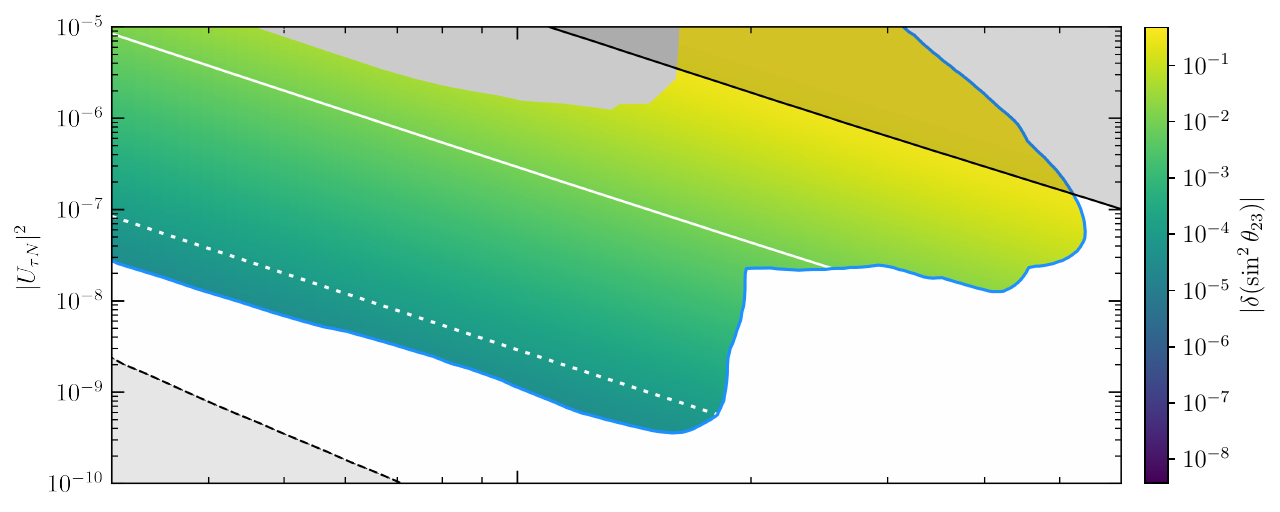}
 \\[-7pt]
 \includegraphics[width=\textwidth]{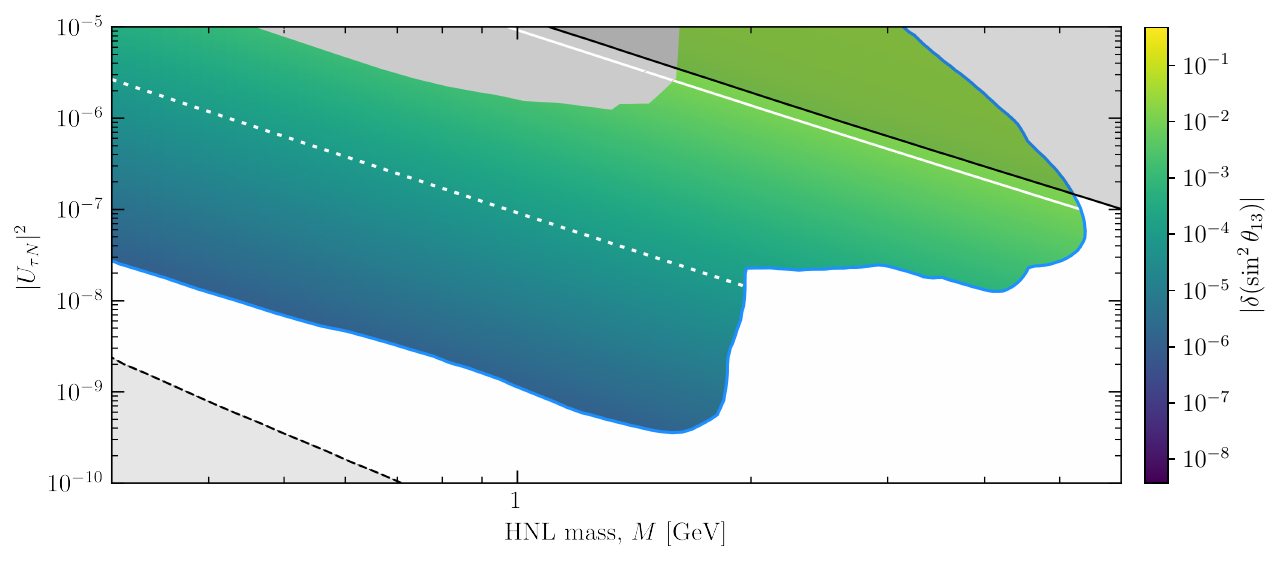}
 \caption{\textbf{Mass-threshold correction of the active neutrino mixing angles for a tau-philic HNL.} Same as \figu{discovery_window_var_angles_UmuN} in the main text, but for a tau-philic HNL where $|U_{\tau N}|$ is the only non-zero active-sterile mixing element. See \figu{discovery_window_var_angles_UeN} for the electron-philic case.}
 \label{fig:discovery_window_var_angles_UtauN}
\end{figure*}

\begin{figure*}[t]
 \centering
 \includegraphics[width=\textwidth]{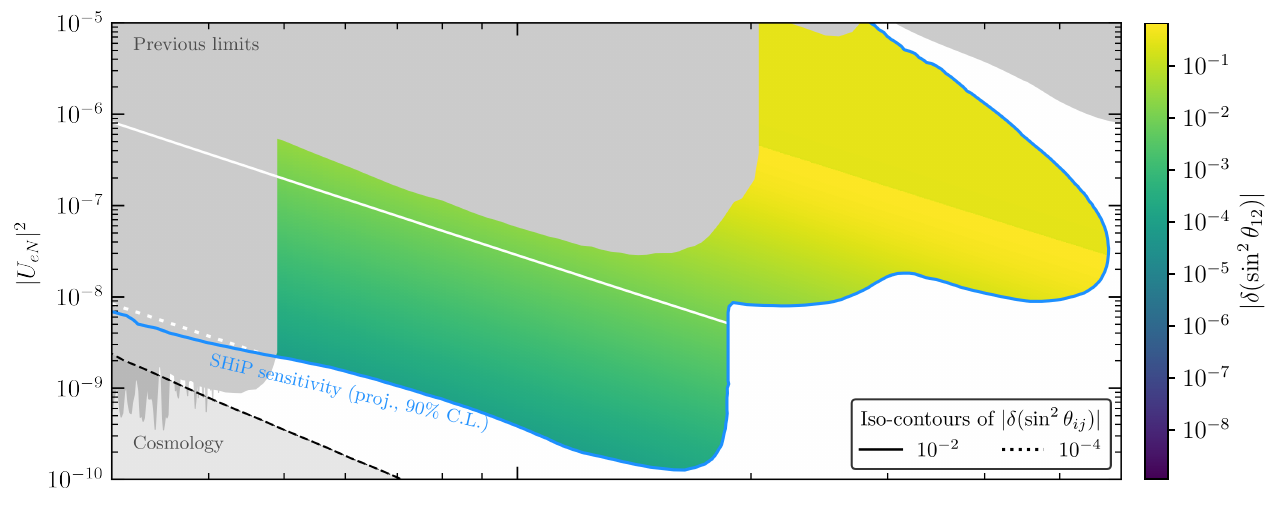}
 \\[-7pt]
 \includegraphics[width=\textwidth]{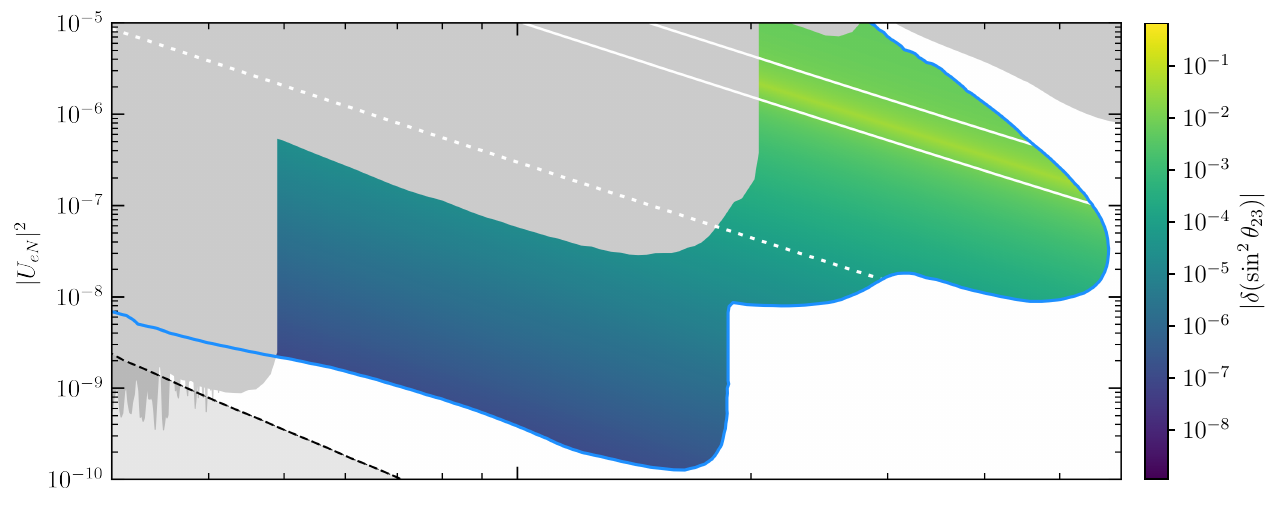}
 \\[-7pt]
 \includegraphics[width=\textwidth]{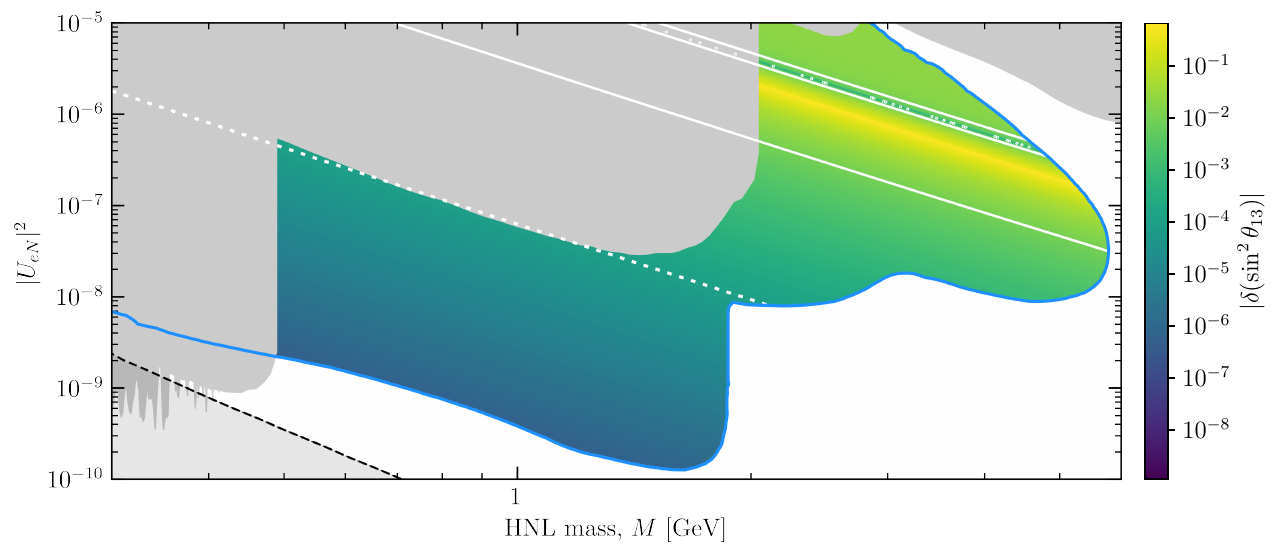}
 \caption{\textbf{Mass-threshold correction of the active neutrino mixing angles for an electron-philic HNL.} Same as \figu{discovery_window_var_angles_UmuN} in the main text, but for an electron-philic HNL where $|U_{e N}|$ is the only non-zero active-sterile mixing element. See \figu{discovery_window_var_angles_UtauN} for the tau-philic case.}
 \label{fig:discovery_window_var_angles_UeN}
\end{figure*}

Figures \ref{fig:discovery_window_var_angles_UeN} and \ref{fig:discovery_window_var_angles_UtauN} show analogous results for electron- and tau-philic HNLs, respectively, still under NO.  
While the results for a tau-philic HNL are similar to the results for the muon-philic HNL---because of $\theta_{23} \approx 45^\circ$---the results for an electron-philic HNL display a novel feature: a diagonal band of high $\delta(\sin^2 \theta_{ij})$ visible in the top-right quadrant of the $(M, |U_{e N}|^2)$ space in \figu{discovery_window_var_angles_UeN}.

This band is a signature of an \textit{avoided} level crossing. Under NO, the lighter mass eigenstates ($\nu_1$ and $\nu_2$) carry the bulk of the electron flavor content, whereas the heaviest eigenstate ($\nu_3$) is predominantly composed of muon and tau flavors. Consequently, the mass-threshold correction from an electron-philic HNL, $\delta m_{ee} \propto M^3 |U_{e N}|^2$, preferentially raises the masses of the lighter eigenstates while leaving the unperturbed $\nu_3$ largely stationary. As the HNL mass or mixing increases, the masses of the perturbed light eigenstates approach $m_3$. In the unperturbed mass basis, $\delta m_{ee}$ generates an off-diagonal mixing term $\Delta_{i3} \simeq U_{ei}^* U_{e3} \delta m_{ee}$ between a light state $\nu_i$ ($i=1,2$) and the heavy state $\nu_3$.  This coupling defines an effective $2 \times 2$ subspace whose new mass eigenvalues are
\begin{equation}
 m_{\pm} \simeq \frac{\tilde{m}_i + \tilde{m}_3}{2} \pm \frac{1}{2}\sqrt{(\tilde{m}_i - \tilde{m}_3)^2 + 4|\Delta_{i3}|^2} \;,
 \label{equ:mass_eigenvalues_new}
\end{equation}
where $\tilde{m}_k \approx m_k + |U_{ek}|^2 \delta m_{ee}$ (for $k = i, 3$) are the perturbed diagonal masses. 

If $U_{e3} = 0$, the coupling $\Delta_{i3}$ vanishes, allowing a true degeneracy ($m_+ = m_-$) where the states trivially pass through one another. However, because in reality $U_{e3} \neq 0$, the strictly positive $4|\Delta_{i3}|^2$ term prevents the eigenvalues from ever becoming equal, enforcing a minimum mass gap of $2|\Delta_{i3}|$. (Because the light states are inherently electron-rich, $U_{ei}$ is $\mathcal{O}(1)$, making $U_{e3}$ the critical bottleneck parameter here.) At the point of closest approach ($\tilde{m}_i = \tilde{m}_3$), the off-diagonal term dominates the subspace, rendering the effective mixing angle maximal ($\pi/4$) and driving the drastic exchange of flavor content that manifests as resonance bands in \figu{discovery_window_var_angles_UeN}.

This resonance is exclusive to electron-philic HNLs because the correction $\delta m_{ee}$ is the only correction that acts in opposition to the normal mass hierarchy. It preferentially targets the electron-rich lighter states $\nu_1$ and $\nu_2$, forcing them to be heavier to converge with $\nu_3$. Conversely, a muon- or tau-philic correction, $\delta m_{\mu\mu}$ or $\delta m_{\tau\tau}$, couples predominantly to the already-heavy $\nu_3$, pushing it further away from $\nu_1$ and $\nu_2$ and precluding any level crossing. (Under IO, not shown, this paradigm reverses: the resonance exclusively manifests for muon- and tau-philic HNLs, as their perturbations drive the now-lightest $\nu_3$ state upward to cross the heavier $\nu_1$ and $\nu_2$.) 

We can derive the location and shape of this resonance band analytically by determining where the rapidly shifting mass of a light state, $\tilde{m}_1$ or $\tilde{m}_2$, converges with the slightly shifted heavy state, $\tilde{m}_3$ (it is only slightly  shifted because $|U_{e3}|^2 \ll |U_{e1}|^2, |U_{e2}|^2$). Because both $\nu_1$ and $\nu_2$ possess substantial electron flavor, the crossing occurs in two stages. For our choice of $m_1 = 0.01$~eV and the NuFIT~6.1 best-fit values of the mixing parameters, there is one degeneracy ($\tilde{m}_1 = \tilde{m}_3$) at $\delta m_{ee} \approx 0.063$~eV, and another one ($\tilde{m}_2 = \tilde{m}_3$) at $\delta m_{ee} \approx 0.136$~eV, together establishing the continuous width of the resonance region. 

To map this threshold onto the parameter space $(M, |U_{e N}|^2)$ of \figu{discovery_window_var_angles_UeN}, we recall that $\delta m_{ee} \propto M^3 |U_{eN}|^2$ for light HNLs. Because of this scaling, maintaining $\delta m_{ee}$ fixed to the value required for resonance defines a linear trajectory in logarithmic space with a slope of $-3$, matching the diagonal path of the resonance bands seen in \figu{discovery_window_var_angles_UeN}. Further, analytically evaluating the mixing-angle correction at an illustrative value of $M = 3$~GeV places one resonance peak at $|U_{eN}|^2 \approx 1.2 \times 10^{-6}$ and the other near $2.5 \times 10^{-6}$, which roughly encompasses the profile of the resonance bands in \figu{discovery_window_var_angles_UeN}. (While this location is universally fixed by the crossing of the mass eigenvalues, the intensity of the resonance spike varies across $\delta(\sin^2 \theta_{12})$, $\delta(\sin^2 \theta_{23})$, and $\delta(\sin^2 \theta_{13})$ because each angle is defined by a different mathematical projection of the PMNS matrix elements, such as $|U_{e3}|^2$ for $\theta_{13}$ versus $|U_{\mu 3}|^2/|U_{\tau 3}|^2$ for $\theta_{23}$.)

\section{Asymptotic flavor limits of the Majorana mass threshold correction}
\label{app:asymptotic_flavor_limits}

\renewcommand{\theequation}{D\arabic{equation}}
\renewcommand{\thefigure}{D\arabic{figure}}
\renewcommand{\thetable}{D\arabic{table}}
\setcounter{figure}{0} 
\setcounter{table}{0}

While the first-order expansion of the HNL-induced corrections presented in Sec.~\ref{subsec:corrections_mixing_angles} in the main text describes the onset of the flavor shift, the saturated regime---where the threshold correction overwhelmingly dominates the intrinsic light-neutrino mass splittings ($\delta m_\nu \gg \Delta m_{ij}$)---admits an exact, analytical limit. Because the dominant mass term takes a rank-one form, $\delta m_\nu \propto \vec{U}_N \vec{U}_N^T$, the high-energy mixing matrix completely reorganizes. Specifically, one physical mass eigenstate is forced to align precisely with the HNL coupling vector, $\vec{U}_N \equiv (U_{eN}, U_{\mu N}, U_{\tau N})^T$. Geometrically, this means the mass eigenstate rotates in flavor space until its $(e, \mu, \tau)$ composition exactly mirrors the active-flavor interactions of the HNL. (Note that all asymptotic limits derived below reflect the renormalized flavor fractions, $f_{\alpha, \oplus}^{\rm sat \prime}$, to properly map onto the ternary flavor space.)


\subsection{Flavor-universal couplings}

\textbf{\textit{The geometric limit.---}}In the most symmetric scenario---where the HNL couples universally to all three flavors ($|U_{eN}| = |U_{\mu N}| = |U_{\tau N}|$)---the coupling vector points evenly along all three flavor axes. Consequently, the mass eigenstate that aligns with this vector is forced into a perfectly tri-maximal composition ($|U_{ei}|^2 = |U_{\mu i}|^2 = |U_{\tau i}|^2 = 1/3$). 

By isolating the specifically aligned tri-maximal state (arbitrarily labeled $i=3$) from the two remaining orthogonal states ($i=1, 2$), the sum decomposes into $P_{\alpha\beta} = \left(|U_{\alpha 3}|^2 |U_{\beta 3}|^2\right) + \sum_{i=1}^{2} |U_{\alpha i}|^2 |U_{\beta i}|^2$. Because the tri-maximal state's projection onto every single active flavor is exactly one-third, the first term evaluates identically to $1/3 \times 1/3 = 1/9$. The probability therefore simplifies to an exact analytical form: $P_{\alpha\beta} = 1/9 + \tilde{P}_{\alpha\beta}$, where the rigid $1/9$ baseline is the unavoidable symmetric contribution from the tri-maximal state, and $\tilde{P}_{\alpha\beta}$ encapsulates the residual mixing of the orthogonal subspace. 

Crucially, because unitarity within the active $3 \times 3$ subspace demands that the remaining orthogonal projections for any given flavor sum to exactly $1 - 1/3 = 2/3$, the residual contribution to the final flavor fraction is bounded. When calculating the asymptotic flavor fractions at Earth, $f_{\alpha, \oplus}^{\rm sat \prime} = \sum_\gamma P_{\gamma\alpha} f_{\gamma, {\rm S}}$, the $1/9$ probability baseline evaluates to a fixed $1/9$ contribution to every flavor (since the initial source fractions sum to unity). The maximum possible residual contribution depends on the orthogonal subspace. Because the tri-maximal state projects equally onto all flavors, the fraction of the initial flux it carries is invariant: $w_3 = \sum_\gamma (1/3) f_{\gamma, {\rm S}} = (1/3)(1) = 1/3$.  This leaves exactly $2/3$ of the initial flux to be distributed among the two orthogonal mass eigenstates ($w_1 + w_2 = 2/3$). Similarly, the available orthogonal projections onto the final flavor $\alpha$ must also sum to $2/3$ ($|U_{\alpha 1}|^2 + |U_{\alpha 2}|^2 = 2/3$). The residual fraction is determined by how these fluxes align with these projections: $\tilde{f}_\alpha = w_1 |U_{\alpha 1}|^2 + w_2 |U_{\alpha 2}|^2$. 

To maximize this sum, the system must perfectly align the largest possible flux with the largest possible projection. This occurs if all the available flux is concentrated entirely in one mass eigenstate (\eg, $w_1 = 2/3, w_2 = 0$), and that exact same state simultaneously possesses all the available projection onto flavor $\alpha$ (\eg, $|U_{\alpha 1}|^2 = 2/3, |U_{\alpha 2}|^2 = 0$). Substituting these extreme limits back into the equation for the residual fraction yields exactly $\tilde{f}_\alpha = (2/3)(2/3) + (0)(0) = 4/9$. This geometry reveals that an HNL with flavor-universal couplings injects a rigid lower bound of $1/9 \approx 11\%$ into every possible flavor fraction, while imposing an absolute upper bound of $1/9 + 4/9 = 5/9 \approx 55\%$.

This confinement ($1/9 \le f_{\alpha, \oplus}^{\rm sat  \prime} \le 5/9$)  forbids any flavor from reaching the extreme 0\% or 100\% limits, guaranteeing that the flavor composition remains perpetually trapped near the center of the flavor triangle.

\medskip

\textbf{\textit{Application to full pion decay.---}}Applying this geometric limit to the nominal case of neutrino production via full pion decay, $\boldsymbol{f}_S = (1/3, 2/3, 0)$, analytically yields the exact asymptotic coordinates of the flavor plateaus. As established above, this symmetric geometry structurally confines every flavor fraction to the strict analytical window $1/9 \le f_{\alpha, \oplus}^{\rm sat \prime} \le 5/9$, with the exact coordinate determined by the residual term $\tilde{f}_\alpha = \sum_{i=1}^{2} |U_{\alpha i}|^2 w_i$. When evaluated for a standard full pion decay flux, the flux carried by the remaining orthogonal states expands explicitly to $w_i = \frac{1}{3}|U_{ei}|^2 + \frac{2}{3}|U_{\mu i}|^2$. Substituting this into the residual fraction yields the exact analytical form: $\tilde{f}_\alpha = \sum_{i=1}^{2} |U_{\alpha i}|^2 \left( \frac{1}{3}|U_{ei}|^2 + \frac{2}{3}|U_{\mu i}|^2 \right)$. 
  
We can evaluate this explicitly by projecting the standard light-neutrino mass matrix---using NuFIT 6.1 best-fit parameters (Table~\ref{tab:free_parameters})---into the orthogonal subspace to determine the exact remaining elements $|U_{\alpha 1}|^2$, $|U_{\alpha 2}|^2$. This numerical evaluation reveals that the available flux does not concentrate into a single state, but rather distributes roughly evenly ($w_1 \approx 0.34$, $w_2 \approx 0.33$). When multiplied by their corresponding fractional projections onto the final flavors, the residual terms symmetrically evaluate to $\tilde{f}_e \approx 0.23$, $\tilde{f}_\mu \approx 0.22$, and $\tilde{f}_\tau \approx 0.22$. Adding the rigid $1/9 \approx 0.11$ baseline prevents the system from ever approaching the $5/9$ upper bound or the $1/9$ floor, forcing the asymptotic composition at Earth to settle firmly in the center of the bounded region, $\boldsymbol{f}_{\oplus}^{\rm sat \prime} \approx (0.34, 0.33, 0.33)$. Because this saturated state is virtually identical to the standard unperturbed expectation, it analytically proves why a flavor-universal HNL generates a blind spot under full pion decay.

\medskip

\textbf{\textit{Application to muon-damped pion decay.---}}Applying these same analytical limits to neutrino production via muon-damped pion decay, $f_S = (0, 1, 0)$, reveals a more extreme landscape. Because this production scenario lacks the intrinsic symmetric cancellations of the full pion decay case, the geometric asymptotes become cleanly resolvable. With the source restricted entirely to muon neutrinos, the flux carried by the remaining orthogonal states reduces simply to their muon projections: $w_i = |U_{\mu i}|^2$. Substituting this into the residual fraction yields the exact analytical form: $\tilde{f}_\alpha = \sum_{i=1}^{2} |U_{\alpha i}|^2 |U_{\mu i}|^2$. Evaluating this numerically using the NuFIT 6.1 projections ($w_1 \approx 0.45$, $w_2 \approx 0.22$) yields residual terms $\tilde{f}_e \approx 0.15$, $\tilde{f}_\mu \approx 0.25$, and $\tilde{f}_\tau \approx 0.27$. Adding the rigid $1/9 \approx 0.11$ baseline establishes the asymptotic plateau at exactly $\boldsymbol{f}_{\oplus}^{\rm sat \prime} \approx (0.26, 0.36, 0.38)$. Unlike the mixed full pion case, this structurally bounded plateau measurably drags the observable composition away from the standard unperturbed expectation ($\sim 0.20, 0.39, 0.41$) toward the democratic center, analytically proving why a muon-damped source breaks the  blind spot.


\subsection{Single-flavor couplings}

\textbf{\textit{The geometric limit.---}}For the extreme cases of single-flavor coupling, this geometry yields trivial asymptotes in the flavor-transition probabilities and flavor fractions. If the HNL couples exclusively to flavor $\alpha$, that flavor decouples and becomes an exact mass eigenstate. The transition probabilities collapse to $P_{\alpha\alpha} \to 1$ and $P_{\alpha \gamma} \to 0$ ($\gamma \neq \alpha$), while the remaining two flavors mix in an orthogonal $2 \times 2$ subspace with an effective angle $\tilde{\theta}$. Consequently, the general expression for the asymptotic flavor fractions at Earth reduces exactly to $f_{\alpha, \oplus}^{\rm sat \prime} = f_{\alpha, {\rm S}}$ for the decoupled flavor, and $f_{\beta, \oplus}^{\rm sat \prime} = \sum_{\gamma \neq \alpha} f_{\gamma, {\rm S}} P_{\gamma \beta}$ for the remaining orthogonal flavors ($\beta \neq \alpha$).

\medskip

\textbf{\textit{Pure electron coupling ($|U_{eN}|$ only).---}}For an electron-philic HNL:
\begin{itemize}
 \item
  \textbf{Under full pion decay:} The electron flavor decouples, yielding $f_{e, \oplus}^{\rm sat \prime} = 1/3(1) + 2/3(0) = 1/3$. Because the rank-one perturbation is confined entirely to the electron sector, the orthogonal $\mu$--$\tau$ subspace is unperturbed and retains the nearly maximal atmospheric mixing of standard oscillations ($\tilde{\theta} \approx \theta_{23} \approx 45^\circ$), forcing $P_{\mu\mu} \approx P_{\mu\tau} \approx 1/2$. Consequently, $f_{\mu, \oplus}^{\rm sat \prime} \approx f_{\tau, \oplus}^{\rm sat \prime} \approx 1/3$. The saturated state is $\boldsymbol{f}_{\oplus}^{\rm sat \prime} \approx (1/3, 1/3, 1/3)$, mimicking the standard unperturbed expectation and analytically explaining the blind spot for this scenario.
 \item
  \textbf{Under muon-damped pion decay:} The electron flavor dynamically decouples ($P_{\mu e} \to 0$). Because there is no initial electron flux at the source to populate this decoupled state, the electron fraction at Earth is annihilated: $f_{e, \oplus}^{\rm sat \prime} = 0$. The initial muon flux entirely populates the unperturbed $\mu$--$\tau$ orthogonal subspace. Driven by the maximal atmospheric mixing ($\tilde{\theta} \approx \theta_{23} \approx 45^\circ$), the remaining flux is divided virtually evenly, yielding a strict, saturated coordinate of $\boldsymbol{f}_{\oplus}^{\rm sat \prime} \approx (0, 1/2, 1/2)$.
\end{itemize}

\medskip

\textbf{\textit{Pure muon coupling ($|U_{\mu N}|$ only).---}}For a muon-philic HNL:
\begin{itemize}
 \item
  \textbf{Under full pion decay:} The muon flavor decouples ($P_{\mu\mu} \to 1, P_{e\mu} \to 0$). The muon fraction at Earth becomes an invariant of the source flux: $f_{\mu, \oplus}^{\rm sat \prime} = 1/3(0) + 2/3(1) = 2/3$. The threshold correction acts as a massive enhancement, doubling the muon fraction from $\sim 33\%$ to $67\%$.
 \item
  \textbf{Under muon-damped pion decay:} The muon flavor decouples ($P_{\mu\mu} \to 1, P_{\mu e} \to 0, P_{\mu \tau} \to 0$). Because the entirety of the initial flux is comprised of $\nu_\mu$, the specifically aligned high-energy mass eigenstate simply locks the flux in place. Active-flavor mixing is fundamentally shut off, and the initial source flux is perfectly preserved at Earth: $\boldsymbol{f}_{\oplus}^{\rm sat \prime} = (0, 1, 0)$. This extreme geometric limit cleanly forces an event sample comprised exclusively of track-like events.
\end{itemize}

\medskip

\textbf{\textit{Pure tau coupling ($|U_{\tau N}|$ only).---}}For a tau-philic HNL:
\begin{itemize}
 \item \textbf{Under full pion decay:} The tau flavor decouples ($P_{\tau\tau} \to 1, P_{e\tau} \to 0, P_{\mu\tau} \to 0$). The high-energy tau fraction at Earth is annihilated: $f_{\tau, \oplus}^{\rm sat \prime} = 1/3(0) + 2/3(0) = 0$. The observable flux is partitioned entirely between the electron and muon flavors, yielding a striking $0\%$ tau neutrino fraction.
 \item \textbf{Under muon-damped pion decay:} The tau flavor decouples ($P_{\mu\tau} \to 0$). Because there is no initial tau flux, the resulting high-energy tau fraction at Earth is annihilated: $f_{\tau, \oplus}^{\rm sat \prime} = 0$. The initial muon flux is forced to mix entirely within the orthogonal $e$--$\mu$ subspace, which remains governed by the unperturbed standard solar mixing angle ($\tilde{\theta} \approx \theta_{12} \approx 33^\circ$). The flux partitions cleanly into electron and muon flavors according to this remaining geometry, yielding an asymptotic state confined to the bottom edge of the flavor triangle.
\end{itemize}


\section{Results for neutrino production via muon-damped pion decay}
\label{app:muon_damped}

\renewcommand{\theequation}{B\arabic{equation}}
\renewcommand{\thefigure}{B\arabic{figure}}
\renewcommand{\thetable}{B\arabic{table}}
\setcounter{figure}{0} 
\setcounter{table}{0}

In the main text, Figs.~\ref{fig:flavor_vs_M}--\ref{fig:flavor_vs_phases} show the behavior of the $Q$-averaged flavor composition at Earth as functions of $M$, $|U_{\alpha N}|^2$, $m_\text{lightest}$, and $\rho = \sigma$, assuming nominal neutrino production via full pion decay. In this appendix, we provide results for muon-damped pion decay, $(0, 1, 0)_{\rm S}$. 

As analytically derived in Appendix~\ref{app:asymptotic_flavor_limits} and discussed in Sec.~\ref{sec:astro_flavor-numerical_results} in the main text, the flavor footprint of the threshold correction is highly sensitive to the initial source composition. Because muon-damped neutrino production lacks a superposed initial electron neutrino flux, the intrinsic probability cancellations that dilute the signal in the full pion decay scenario are completely absent. Consequently, the HNL-induced shifts to the transition probabilities are transferred directly and undiluted to the observable active flavor fractions at Earth, leading to larger flavor shifts and relative flavor distances. 

\begin{figure}[t!]
 \centering
 \includegraphics[width=0.5\textwidth]{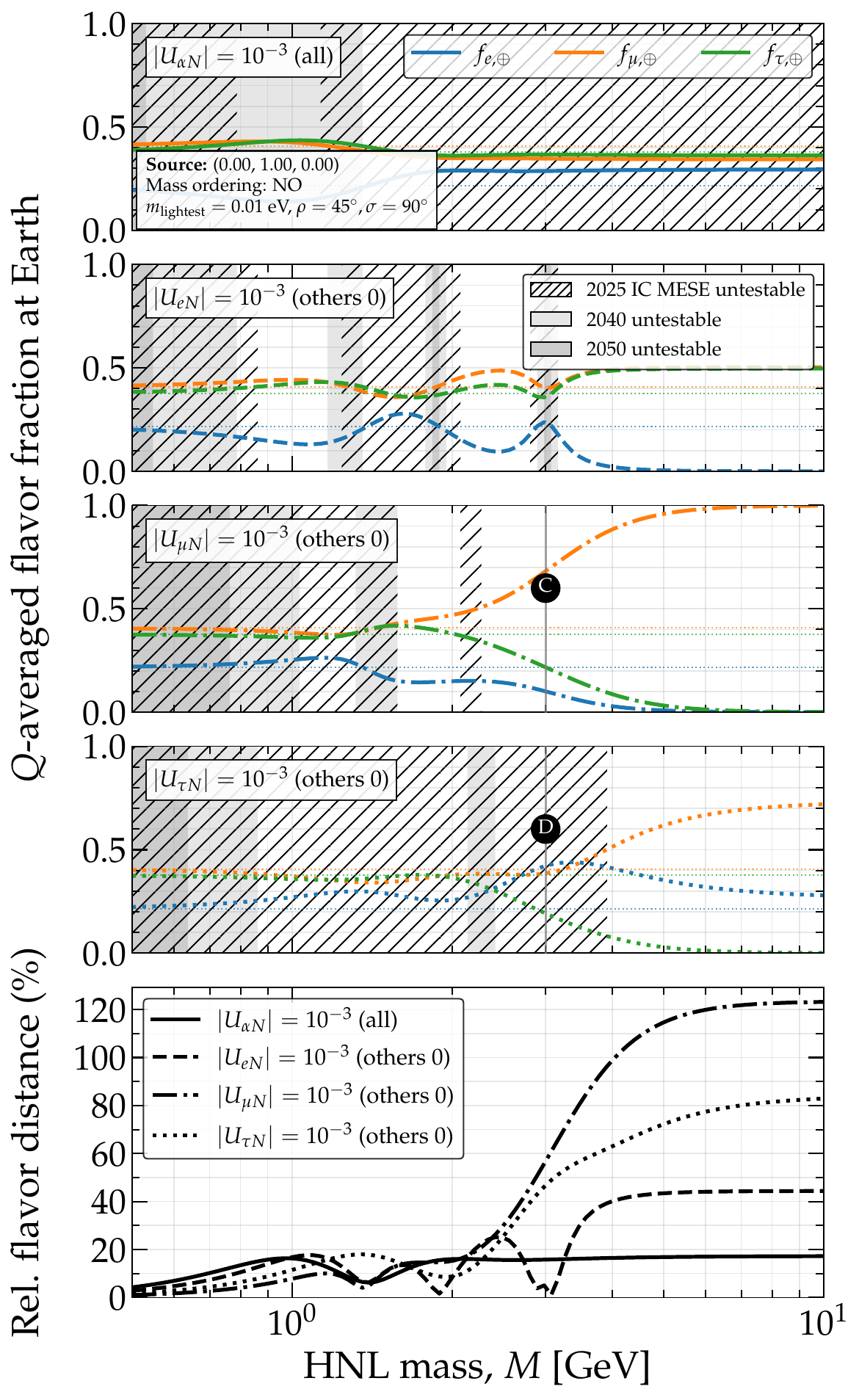}
 \caption{\textbf{Dependence of the flavor composition on the HNL mass, $M$, for a muon-damped source.} Similar to Fig.~\ref{fig:flavor_vs_M}, but assuming an initial source composition of $f_{\rm S} = (0, 1, 0)$. Because the signal is not diluted by overlapping transition channels, the relative flavor distances (bottom panel) are drastically larger than in the full pion decay scenario, reaching up to $\sim 120\%$ and pushing the electron-philic scenario ($|U_{eN}|$ only) into the testable regime.}
 \label{fig:flavor_vs_M_muon}
 \vspace*{-0.5cm}
\end{figure}

\begin{figure}[t!]
 \centering
 \includegraphics[width=0.5\textwidth]{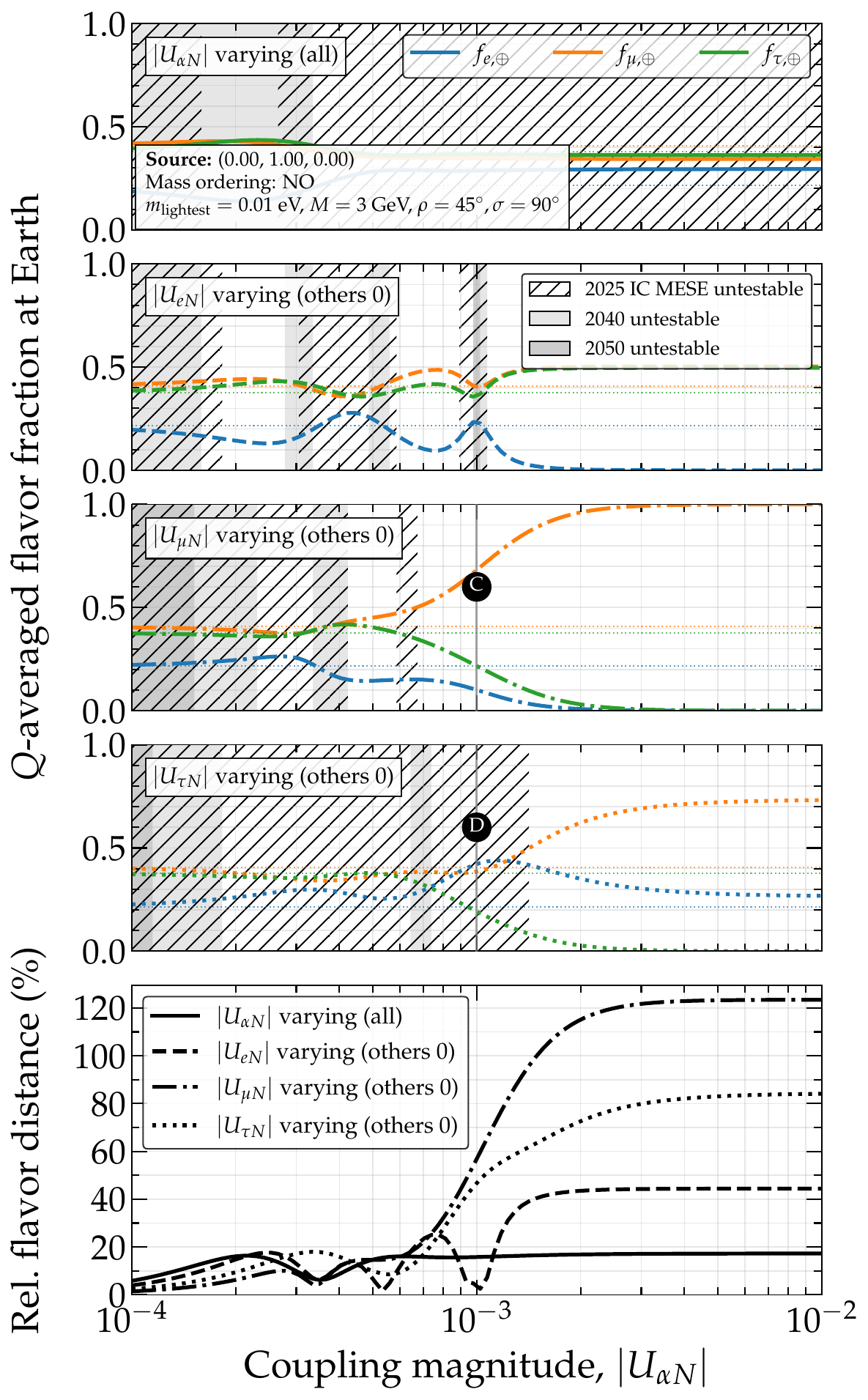}
 \caption{\textbf{Dependence of the flavor composition on the coupling magnitude, $|U_{\alpha N}|$, for a muon-damped source.} Similar to Fig.~\ref{fig:flavor_vs_U}, but assuming an initial source composition of $f_{\rm S} = (0, 1, 0)$. Once the threshold correction overcomes the standard mass splittings near $|U_{\alpha N}| \sim 10^{-3}$, the flavor fractions are violently displaced toward their geometric asymptotes.}
 \label{fig:flavor_vs_U_muon}
\end{figure}

\begin{figure}[t!]
 \centering
 \includegraphics[width=0.5\textwidth]{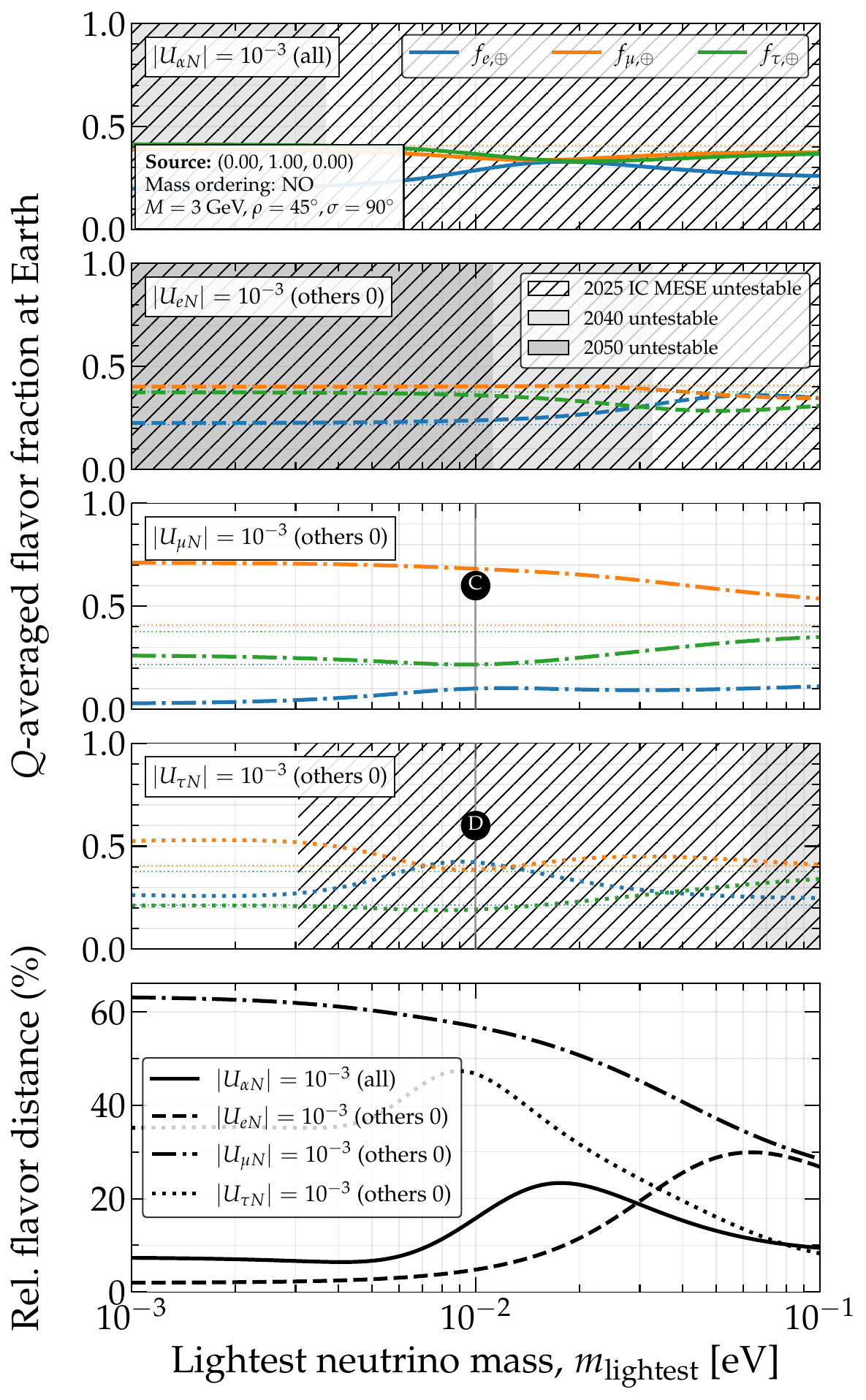}
 \caption{\textbf{Dependence of the flavor composition on the lightest neutrino mass, $m_{\rm lightest}$, for a muon-damped source.} Similar to Fig.~\ref{fig:flavor_vs_m_lightest}, but assuming an initial source composition of $f_{\rm S} = (0, 1, 0)$. The inertial suppression of mixing corrections in the quasi-degenerate mass regime ($m_{\rm lightest} \gtrsim 10^{-2}$~eV) remains a universal feature, driving the observables back toward standard expectations.}
 \label{fig:flavor_vs_m_lightest_muon}
 \vspace*{-0.2cm}
\end{figure}

\begin{figure}[t!]
 \centering
 \includegraphics[width=0.5\textwidth]{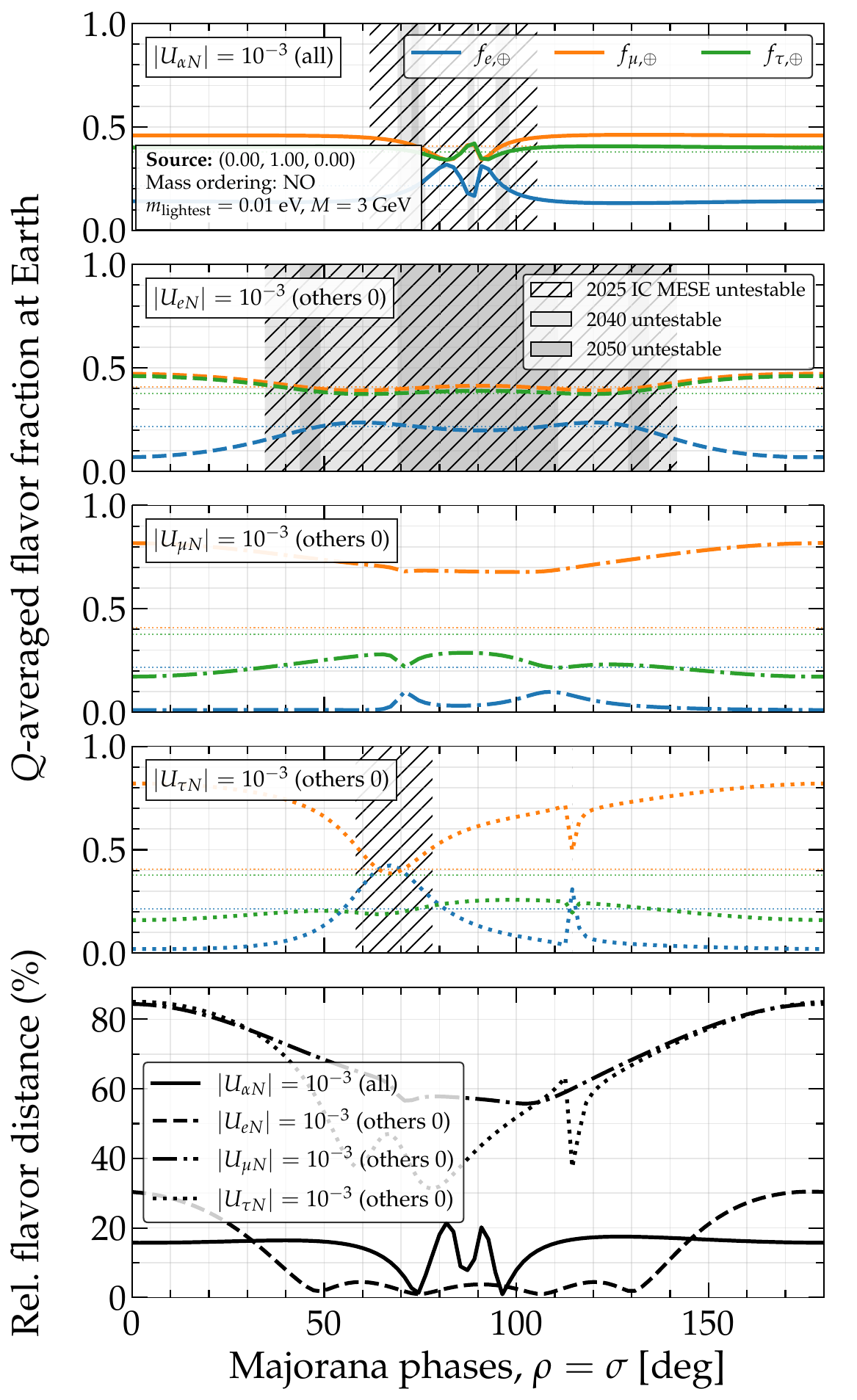}
 \caption{\textbf{Dependence of the flavor composition on the Majorana CP-violation phases, $\rho = \sigma$, for a muon-damped source.} Similar to Fig.~\ref{fig:flavor_vs_phases}, but assuming an initial source composition of $f_{\rm S} = (0, 1, 0)$. Because muon-damped pion decay maps the observables exclusively to the $P_{\mu\beta}$ probability row, the phase-dependent interference pattern of the mass eigenstates is preserved without cancellation, resulting in larger flavor shifts compared to full pion decay.}
 \label{fig:flavor_vs_phases_muon}
\end{figure}

Figures~\ref{fig:flavor_vs_M_muon}--\ref{fig:flavor_vs_phases_muon} show the consequences of this amplified signal. By breaking the inherent phenomenological blind spots of the mixed full pion decay, the undiluted transitions render the flavor-universal and electron-philic scenarios potentially testable at next-generation neutrino telescopes, and allow the Majorana CP-violation phases to induce larger modulation.  

\section{Details on limits on $|U_{\alpha N}|^2$}
\label{app:limits_details}

\renewcommand{\theequation}{C\arabic{equation}}
\renewcommand{\thefigure}{C\arabic{figure}}
\renewcommand{\thetable}{C\arabic{table}}
\setcounter{figure}{0} 
\setcounter{table}{0}

This appendix elaborates on the differences in the limits on $|U_{\alpha N}|^2$ between normal (NO) and inverted neutrino mass ordering (IO), and between different flavors shown in Sec.~\ref{sec:limits_and_discovery-results} in the main text.


\subsection{Normal vs.~inverted neutrino mass ordering}

The computation of the limits $|U_{\alpha N}|^2$ relies on comparing the HNL-induced flavor shift against the standard-mixing expectation. Because $f_{e, {\rm S}}$ is treated as a completely unconstrained nuisance parameter, the test statistic, $\Delta \chi^2$ in \equ{Delta_chi2_limits} in the main text, does not simply measure the absolute magnitude of the flavor shift; it measures the orthogonal distance of that shift from the band allowed by standard mixing, as seen in \figu{hnl_flavor_contours} in the main text.  (Other model parameters are also varied when computing $\Delta \chi^2$, but $f_{e, {\rm S}}$ remains the dominant nuisance parameter even in our projections.)

Assuming no $\nu_\tau$ production, the source flavor vector is $\boldsymbol{f}_{\rm S} = (f_{e, {\rm S}}, 1 - f_{e, {\rm S}}, 0)^{\rm T}$. The expected standard-mixing flavor composition at Earth can be written as a linear combination of the standard flavor-transition probability columns, $\boldsymbol{P}_\alpha \equiv (P_{\alpha e}, P_{\alpha \mu}, P_{\alpha \tau})$:
\begin{align}
 \boldsymbol{f}_{\oplus}^{\rm std}(f_{e, {\rm S}}) 
 &= f_{e, {\rm S}} \boldsymbol{P}_e + (1 - f_{e, {\rm S}}) \boldsymbol{P}_\mu \nonumber \\
 &= \boldsymbol{P}_\mu + f_{e, {\rm S}} \boldsymbol{v}_{\rm std} \;,
\end{align}
where $\boldsymbol{v}_{\rm std} \equiv \boldsymbol{P}_e - \boldsymbol{P}_\mu$ is the direction vector defining the ``standard allowed band'' in the flavor triangle traced by varying $f_{e, {\rm S}} \in [0, 1]$, visible in \figu{hnl_flavor_contours}.

The flavor composition at Earth in the presence of an HNL contains a flavor shift, $\delta \boldsymbol{f}_{\rm HNL}$, with respect to the standard-mixing expectation from the true value of the flavor composition at the sources, $f_{e, {\rm S}}^{\rm true}$, yielding $\boldsymbol{f}_{\oplus}^{\rm true} = \boldsymbol{f}_{\oplus}^{\rm std}(f_{e, {\rm S}}^{\rm true}) + \delta \boldsymbol{f}_{\rm HNL}$. To extract a limit, the profile likelihood procedure (Sec.~\ref{sec:limits_and_discovery-statistics} in the main text) minimizes the penalty over the unknown values of $f_{e, {\rm S}}$. Assuming a simplified identity covariance matrix, the test statistic is proportional to the minimum squared Euclidean distance to the standard-mixing band, \ie,
\begin{align}
 \Delta \chi^2 
 &\propto
 \min_{f_{e, {\rm S}}} \left| \boldsymbol{f}_{\oplus}^{\rm true} - \boldsymbol{f}_{\oplus}^{\rm std}(f_{e, {\rm S}}) \right|^2 
 \nonumber \\
 & = \min_{\Delta f} \left| \delta \boldsymbol{f}_{\rm HNL} - \Delta f \boldsymbol{v}_{\rm std} \right|^2 \;,
\end{align}
where $\Delta f \equiv f_{e, {\rm S}} - f_{e, {\rm S}}^{\rm true}$. Minimizing this yields
\begin{equation}
\Delta \chi^2 \propto |\delta \boldsymbol{f}_{\rm HNL}|^2 \sin^2 \gamma \;.
\end{equation}
Here, $\gamma$ is the angle between the HNL shift vector ($\delta \boldsymbol{f}_{\rm HNL}$) and the standard allowed band ($\boldsymbol{v}_{\rm std}$). The strength of the limit on $|U_{\alpha N}|^2$ depends on this orthogonal projection: if $\sin^2 \gamma \to 0$, the shift can be completely absorbed by the $f_{e, {\rm S}}$ uncertainty.

To evaluate $\gamma$, we map the flavor composition of the $i$-th mass eigenstate at Earth, $\boldsymbol{w}_i = (|U_{ei}|^2, |U_{\mu i}|^2, |U_{\tau i}|^2)^{\rm T}$. For our estimates here, we adopt the tribimaximal (TBM) mixing approximation ($\theta_{23} = 45^\circ$, $\sin^2\theta_{12} = 1/3$, $\theta_{13} \approx 0$) for analytical clarity, which captures the dominant geometry of the mixing:
\begin{align}
 & \boldsymbol{w}_1 = \left(\frac{2}{3}, \frac{1}{6}, \frac{1}{6}\right)^{\rm T}, \quad 
 \boldsymbol{w}_2 = \left(\frac{1}{3}, \frac{1}{3}, \frac{1}{3}\right)^{\rm T}, 
 \nonumber \\
 &\boldsymbol{w}_3 = \left(0, \frac{1}{2}, \frac{1}{2}\right)^{\rm T} \;.
\end{align}

The standard-mixing band direction is $\boldsymbol{v}_{\rm std} = \sum_{i=1}^3 ( |U_{ei}|^2 - |U_{\mu i}|^2 ) \boldsymbol{w}_i$. Under TBM, $\nu_2$ is maximally mixed ($|U_{e2}|^2 = |U_{\mu 2}|^2$), so its contribution vanishes, \ie,
\begin{equation}
 \boldsymbol{v}_{\rm std}
 = \frac{1}{2} \boldsymbol{w}_1 - \frac{1}{2} \boldsymbol{w}_3
 = \left( \frac{1}{3}, -\frac{1}{6}, -\frac{1}{6} \right)^{\rm T} \;.
 \label{equ:flavor_direction_vector_std}
\end{equation}

In IO, the mass spectrum is $m_2 \gtrsim m_1 \gg m_3$. Because the mass-threshold correction scales inversely with the mass separation, the HNL drives a complex rotation in the quasi-degenerate 1-2 subspace; see Sec.~\ref{sec:astro_flavor-numerical_results} in the main text. This transfers probability between state 1 and state 2, making the HNL flavor shift  proportional to their difference:
\begin{equation}
\delta \boldsymbol{f}_{\rm HNL}^{\rm (IO)} \propto (\boldsymbol{w}_1 - \boldsymbol{w}_2) = \left( \frac{1}{3}, -\frac{1}{6}, -\frac{1}{6} \right)^{\rm T} \;.
\end{equation}
By inspection, $\delta \boldsymbol{f}_{\rm HNL}^{\rm (IO)}$ is perfectly collinear with $\boldsymbol{v}_{\rm std}$. Therefore, $\gamma = 0$ and $\sin^2 \gamma = 0$. Even though the absolute magnitude of the IO flavor shift is large, its trajectory points exactly along the standard-mixing allowed band. The profile likelihood easily absorbs this HNL signal by sliding the nuisance parameter $f_{e, {\rm S}}$, resulting in significantly weaker exclusion limits on $|U_{\alpha N}|^2$.

In NO, the mass spectrum is hierarchical ($m_3 \gg m_2 \gg m_1 \approx 0$). The mass correction is dominated by $\nu_1$, leading to a shift proportional to its flavor vector,
\begin{equation}
\delta \boldsymbol{f}_{\rm HNL}^{\rm (NO)} \propto \boldsymbol{w}_1 = \left(\frac{2}{3}, \frac{1}{6}, \frac{1}{6}\right)^{\rm T} \;.
\end{equation}
Comparing this vector to $\boldsymbol{v}_{\rm std}$, the alignment is
\begin{equation}
\cos^2 \gamma = \frac{(\delta \boldsymbol{f}_{\rm HNL}^{\rm (NO)} \cdot \boldsymbol{v}_{\rm std})^2}{|\delta \boldsymbol{f}_{\rm HNL}^{\rm (NO)}|^2 |\boldsymbol{v}_{\rm std}|^2} = \frac{(1/6)^2}{(1/2)(1/6)} = \frac{1}{3} \;.
\end{equation}
Consequently, $\sin^2 \gamma = 2/3$. Unlike IO, the NO shift is largely orthogonal to the standard-mixing band. Because no physical variation of the astrophysical source composition ($f_{e, {\rm S}}$) can replicate this perpendicular displacement, the profiling of the test statistic fails to absorb the shift. The $\Delta \chi^2$ penalty accumulates, leading to more stringent exclusion limits at smaller values of $|U_{\alpha N}|^2$.


\subsection{Muon-philic vs.~tau-philic couplings}

While the TBM approximation explains the overarching disparity between the NO and IO sensitivities, its assumption of exact $\mu-\tau$ symmetry ($|U_{\mu i}|^2 = |U_{\tau i}|^2$) predicts identical limits for $|U_{\mu N}|^2$ and $|U_{\tau N}|^2$. To explain the stark divergence of these limits under IO seen in \figu{limits_Usq_vs_M} in the main text, we must examine how the physical mass pull of the HNL interacts with the true, asymmetric flavor content of the mass eigenstates.

The mass threshold correction introduces a negative shift to the diagonal elements of the active neutrino mass matrix, $\delta m_{\alpha\alpha} \propto -|U_{\alpha N}|^2$. This correction forces the flavor $\alpha$ to align more strongly with the lightest available mass eigenstate. (This is because reducing the diagonal mass term associated with a specific flavor lowers its contribution to the overall mass, and by  definition, the lightest mass eigenstate must shift its composition toward that flavor to remain the eigenstate with the lowest possible mass.) The strength of the resulting exclusion limit depends on how much the standard mixing matrix must reorganize its native flavor distribution to accommodate this forced alignment.

Global fits reveal that the atmospheric mixing angle, $\theta_{23}$, breaks exact $\mu-\tau$ symmetry, but its preferred value depends strictly on the assumed mass ordering. 

In the IO, fits strongly prefer the second octant ($\sin^2\theta_{23} \approx 0.55$). This asymmetry ensures that $\nu_3$ is electron-depleted and muon-heavy, $\boldsymbol{w}_3 \approx (0.02, 0.55, 0.43)^{\rm T}$. Conversely, the bulk of the tau flavor is trapped in the $\nu_1$ and $\nu_2$ states. Because $\nu_3$ is the lightest state in IO, introducing the $|U_{\mu N}|^2$ coupling forces the muon flavor to align with it. However, because $\boldsymbol{w}_3$ is already natively 55\% muon flavor, the standard mixing matrix is pre-aligned with this perturbation. The diagonal mass correction merely reinforces an existing eigenvector configuration, requiring minimal structural rotation. Consequently, $\delta \boldsymbol{f}_{\rm HNL}$ remains small and highly collinear with the standard band ($\sin^2 \gamma \to 0$), allowing the statistical profiling to  absorb the shift via $f_{e, {\rm S}}$.

Conversely, when $|U_{\tau N}|^2$ is active in IO, the mass correction attempts to drag the tau flavor into $\nu_3$. Because the vast majority of the tau fraction resides in the heavier eigenstates ($\nu_1$ and $\nu_2$), satisfying this pull requires a significant reorganization of the mixing matrix. The tau flavor must be drastically rotated out of the heavy eigenstates and forced into $\nu_3$. This large rotation shatters the near-collinearity with the standard-mixing region established in the TBM limit earlier, generating an appreciable, orthogonal shift $\delta \vec{f}_{\rm HNL}$ ($\sin^2 \gamma \gg 0$) that cannot be spoofed by astrophysical uncertainties, yielding a stronger limit on $|U_{\tau N}|^2$.

In the NO, global fits prefer the first octant ($\sin^2\theta_{23} \approx 0.47$). While this flips the asymmetry in $\nu_3$ (making it slightly tau-rich), the HNL-modified dynamics under NO are governed entirely by the lightest eigenstate, now $\nu_1$. Crucially, $\nu_1$ remains electron-rich ($\approx 68\%$). Regardless of the exact value of $\theta_{23}$, the muon and tau flavors are severely depleted, yielding roughly symmetric, low amounts: $\boldsymbol{w}_1 \approx (0.68, 0.15, 0.17)^{\rm T}$.

Therefore, activating either $|U_{\mu N}|^2$ or $|U_{\tau N}|^2$ in NO attempts to drag a muon or tau flavor into this isolated, electron-rich state. Because neither the muon nor the tau flavor natively populates $\nu_1$ in significant amounts, satisfying either mass pull demands an equally drastic reorganization of the mixing matrix. The targeted flavor must be completely extracted from the heavy $\nu_2$ and $\nu_3$ states. As a result, both couplings generate massive, orthogonal displacements from the standard allowed band, yielding comparably stringent limits differentiated only by the inherently lower experimental resolution for identifying $\nu_\tau$ in neutrino telescopes (Sec.~\ref{sec:astro_production-flavor_measurement}).


\subsection{Absence of limits on $|U_{e N}|^2$}

In the main text, we reported the absence of limits on $|U_{e N}|^2$.  Below, we explain its origin.

In the NO, the vanishing sensitivity to $|U_{eN}|^2$ is a consequence of severe structural redundancy. The lightest mass eigenstate is $\nu_1$, which global fits dictate is overwhelmingly electron-dominated, containing roughly $68\%$ electron flavor [$\boldsymbol{w}_1 \approx (0.68, 0.15, 0.17)^{\rm T}$]. The mass threshold correction acts as a negative diagonal perturbation to the electron sector, $\delta m_{ee} \propto -|U_{eN}|^2$. Because reducing this diagonal mass term lowers its contribution to the overall mass, $\nu_1$ is forced to shift its composition toward the electron flavor. However, because $\nu_1$ already consists almost entirely of electron flavor, standard physics has already placed the system very close to the flavor configuration the perturbation is trying to enforce. Consequently, the mixing matrix requires only minimal structural rotation to accommodate the mass pull. The resulting shift in the Earth flavor composition is small ($\delta \boldsymbol{f}_{\rm HNL} \to 0$), insufficient to be detected against the uncertainty on $f_{e, {\rm S}}$, leading to a correspondingly weak limit.

Conversely, in the IO, the failure to constrain $|U_{eN}|^2$ stems from a combination of mixing suppression and geometric degeneracy. Here, the lightest state is $\nu_3$, which is electron-depleted [$\boldsymbol{w}_3 \approx (0.02, 0.55, 0.43)^{\rm T}$]. While the perturbation $\delta m_{ee} \propto -|U_{eN}|^2$ attempts to drag the electron flavor into $\nu_3$, this rotation is suppressed by the smallness of the reactor mixing angle ($\sin^2\theta_{13} \approx 0.022$). More importantly, any rotation that does occur generates a flavor shift $\delta \boldsymbol{f}_{\rm HNL}$ that the likelihood profiling cannot penalize, as we explain next. 

By the definition of normalized flavor fractions at Earth, \equ{renormalized_fraction} in the main text, any shift in the electron fraction must be balanced by the muon and tau flavors: $\delta f_{\mu} + \delta f_{\tau} = -\delta f_{e}$. Because the mass perturbation acts exclusively on the electron row of an approximately unitary $3\times3$ active mixing matrix (non-unitarity is small because the mixing with the HNL is assumed to be small), the flavor shift is distributed between $\nu_\mu$ and $\nu_\tau$ according to the standard atmospheric mixing angle, which governs the relative muon and tau couplings to $\nu_3$ ($|U_{\mu 3}|^2 = c_{13}^2 s_{23}^2$, $|U_{\tau 3}|^2 = c_{13}^2 c_{23}^2$).

Thus, due to near-maximal atmospheric mixing ($\theta_{23} \approx 45^\circ$), the distribution between the muon and tau sectors is nearly even, forcing $\delta f_{\mu} \approx \delta f_{\tau} \approx -\frac{1}{2} \delta f_{e}$. Consequently, the HNL-induced flavor shift takes the  form $\delta \boldsymbol{f}_{\rm HNL} \propto (1, -1/2, -1/2)^{\rm T}$. This has the same direction as the standard-mixing allowed band, $\boldsymbol{v}_{\rm std}$ in \equ{flavor_direction_vector_std}. Because $\delta \boldsymbol{f}_{\rm HNL}$ and $\boldsymbol{v}_{\rm std}$ are parallel ($\sin^2 \gamma \to 0$), the statistical fit absorbs the HNL flavor shift by sliding the value of $f_{e, {\rm S}}$, resulting in a flat $\Delta \chi^2$ and the absence of a limit on $|U_{e N}|^2$.


\section{Impact of the neutrino mass ordering}
\label{app:mass_ordering}

\renewcommand{\theequation}{E\arabic{equation}}
\renewcommand{\thefigure}{E\arabic{figure}}
\renewcommand{\thetable}{E\arabic{table}}
\setcounter{figure}{0} 
\setcounter{table}{0}

The quantitative predictions for the astrophysical flavor shifts depend profoundly on the assumed neutrino mass ordering. Because the Majorana threshold correction is fundamentally driven by the mass differences between the high-energy eigenstates [as isolated in \equ{Delta_alpha_i}], flipping the ordering from normal (NO, $m_1 < m_2 \ll m_3$) to inverted (IO, $m_3 \ll m_1 < m_2$) rewires the kinematic dynamics of the mixing matrix. So far, we have shown quantitative results only for the NO case.  If we apply the exact same HNL parameters (mass and couplings) to the IO case, three distinct physical mechanisms emerge that alter the expected flavor shifts at Earth.

\medskip

\textbf{\textit{Kinematic amplification.---}}The most significant difference between the two orderings is the intrinsic separation between the first and second mass eigenstates, $\nu_1$ and $\nu_2$. In the absolute hierarchical limit ($m_{\rm lightest} \approx 0$), the NO yields $m_1 \approx 0$ and $m_2 \approx \sqrt{\Delta m_{21}^2} \approx 0.0087$~eV. The smallest denominator governing the mixing corrections is therefore $(m_2 - m_1)_{\rm NO} \approx 0.0087$~eV.

In contrast, under IO, the third state is the lightest ($m_3 \approx 0$) and the two heavier states are pushed up by the large atmospheric mass splitting, yielding $m_1 \approx \sqrt{|\Delta m_{31}^2|} \approx 0.05$~eV, and $m_2 = \sqrt{m_1^2 + \Delta m_{21}^2}$. Because both $m_1$ and $m_2$ are anchored to the large absolute scale of the atmospheric splitting, the solar splitting that separates them is fractionally tiny. We can explicitly expand their mass difference as:
\begin{equation}
 (m_2 - m_1)_{\rm IO} = \frac{\Delta m_{21}^2}{m_1 + m_2} \approx \frac{\Delta m_{21}^2}{2\sqrt{|\Delta m_{31}^2|}} \,.
\end{equation}
Evaluating this using best-fit NuFIT 6.1 parameters yields $(m_2 - m_1)_{\rm IO} \approx (7.5 \times 10^{-5}) / 0.1 \approx 7.5 \times 10^{-4}$~eV. 

This reveals that the 1-2 mass difference in IO is roughly an order of magnitude smaller than in NO, \ie, $(m_2 - m_1)_{\rm NO} / (m_2 - m_1)_{\rm IO} \approx 11$. Consequently, even in the absolute hierarchical limit, IO is intrinsically quasi-degenerate in the $e$--$\mu$ sector. This compression acts as a massive kinematic amplifier. The perturbative correction to the 1-2 mixing sector---which scales as $(m_2 - m_1)^{-1}$---is amplified by a factor of $\sim 10$, generating significantly larger observable flavor displacements for the exact same HNL coupling strength.

\medskip

\textbf{\textit{Swapping the inertial states.---}}As established in the derivation of inertial suppression [\equ{Majorana_decoupling}], heavier choices of the lightest mass, $m_\text{lightest}$, resist HNL-induced structural reorganization because the perturbative correction is suppressed by $(m_i + m_j)^{-1}$. Switching the mass ordering swaps which flavors experience this damping.

In NO, $m_3 \approx 0.05$~eV is the heaviest mass (assuming $m_1 = 0$). Because of the standard mixing angles ($\theta_{13} \approx 8.6^\circ$, $\theta_{23} \approx 45^\circ$), this state is composed almost entirely of muon and tau flavors. Consequently, any new-physics perturbation attempting to displace the muon or tau fractions must alter the composition of $\nu_3$. However, its large mass endows it with strong kinematic inertia: the large $m_i + m_j$ denominator suppresses the required mixing corrections, making $\nu_3$ resistant to structural reorganization. This effectively anchors the $\mu$--$\tau$ flavor sector in place, capping the overall flavor shifts (which nevertheless remain appreciable, as shown in Figs.~\ref{fig:flavor_vs_M}--\ref{fig:flavor_vs_phases}).

In IO, this geometry is inverted: $m_3 \approx 0$ becomes the lightest state, possessing zero kinematic inertia. Now this specific mass state, containing the bulk of the muon and tau flavors, becomes exceptionally easy to perturb. For the flavor ratios, this implies that muon- or tau-philic HNLs will induce more extreme flavor shifts in IO than in NO. Conversely, $\nu_1$ and $\nu_2$ (which contain the bulk of the electron flavor via $\theta_{12}$ and $\theta_{13}$) become the heavy, inertial states in IO, subjecting the electron sector to stronger damping than it experiences in NO.

\medskip

\textbf{\textit{Enhanced sensitivity to Majorana phases.---}}Because the $(m_2 - m_1)$ denominator is shrunk in IO, the mass threshold correction is overwhelmingly dominated by the mixing perturbation between $\nu_1$ and $\nu_2$. Looking back at the analytical expansion for the probability shift in \equ{Delta_alpha_i}, the magnitude of this specific correction is driven by the cross-term of the HNL projections, $\xi_1 \xi_2$. 

Because $\xi_1$ and $\xi_2$ [\equ{xi_explicit}] isolate the two Majorana CP-violation phases ($\rho$ and $\sigma$), this forces the flavor shifts to be governed by their interference. While NO produces moderate phase-dependent oscillations (\figu{flavor_vs_phases}), the amplified kinematic correction in the 1-2 sector in IO causes the flavor fractions to swing more sharply when varying the Majorana phases. Therefore, if the true neutrino mass ordering is inverted, the flavor ratios become a sensitive interferometer for the Majorana phases, even for highly suppressed HNL couplings.


\section{Flavor regions: impact of standard mixing uncertainties and mass ordering}
\label{app:flavor_extra}

\renewcommand{\theequation}{F\arabic{equation}}
\renewcommand{\thefigure}{F\arabic{figure}}
\renewcommand{\thetable}{F\arabic{table}}
\setcounter{figure}{0} 
\setcounter{table}{0}

\begin{figure*}[t]
 \centering
 \includegraphics[width=0.49\textwidth]{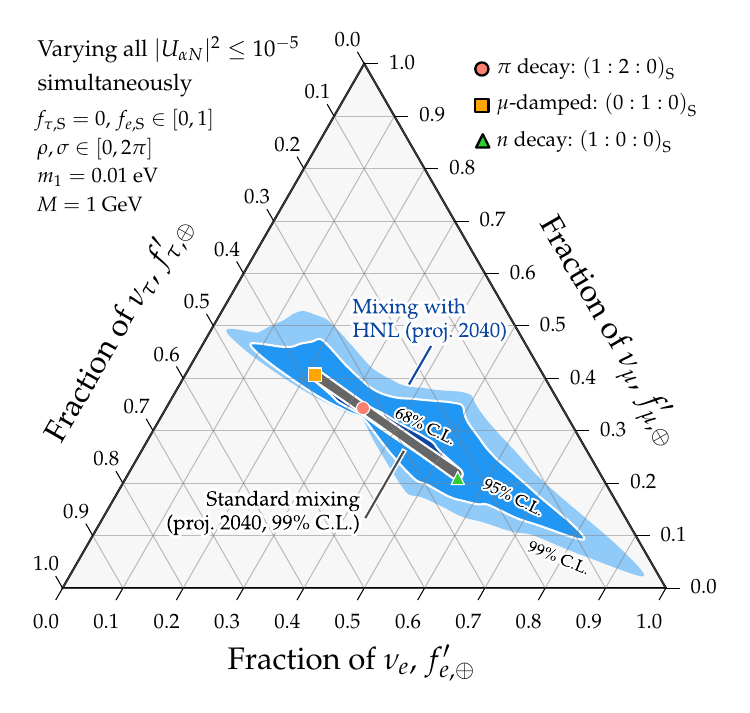}
 \hfill
 \includegraphics[width=0.49\textwidth]{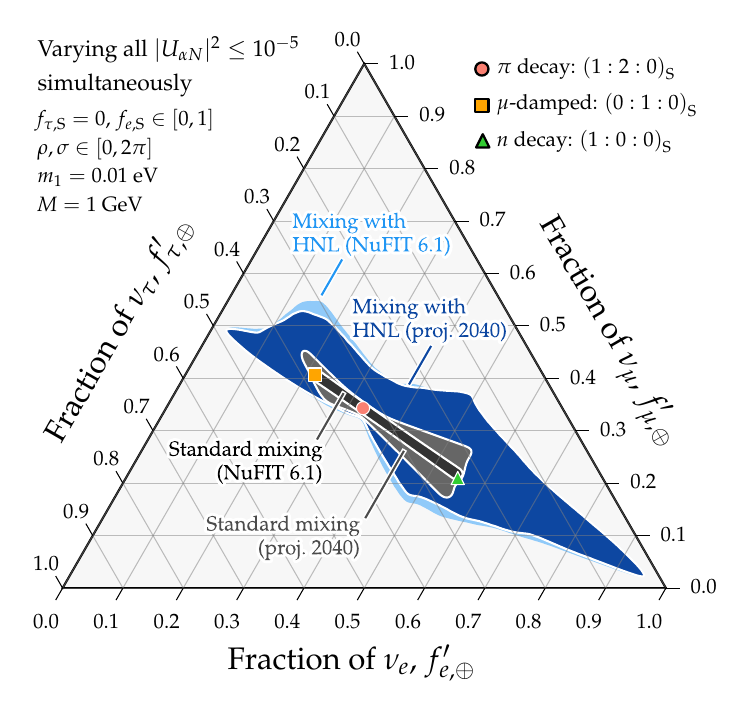}
 \vspace*{-0.5cm}
 \caption{\textbf{Dependence of the HNL-induced flavor composition at Earth on confidence levels and standard-mixing uncertainties.} In both panels, all three active-sterile mixing parameters $|U_{\alpha N}|^2$ are varied simultaneously up to $10^{-5}$ for a generic source composition devoid of initial $\nu_\tau$ ($f_{\tau, {\rm S}} = 0$, $f_{e, {\rm S}} \in [0,1]$), fixed $M = 1$~GeV and $m_1=0.01$~eV, and varying the CP-violation Majorana phases, $\rho, \sigma \in [0, 2\pi]$. \textit{Left:} The accessible region under mixing with an HNL, contoured at the 68\%, 95\%, and 99\% confidence levels (C.L.), and using the 2040-projected allowed ranges of the standard mixing parameters (Table~\ref{tab:free_parameters}). \textit{Right:} A comparison of the 99\% C.L.~allowed regions derived using present allowed ranges of the standard mixing parameters, from NuFIT 6.1, versus projected ones.}
 \label{fig:hnl_flavor_uncertainties}
 \vspace*{-0.5cm}
\end{figure*}

Figure~\ref{fig:hnl_flavor_uncertainties} demonstrates the robustness of the HNL threshold effect against uncertainties in the standard mixing parameters. The left panel isolates the statistical expansion of the accessible region under projected 2040 sensitivities. The allowed footprint at 95\% C.L.~already occupies a substantial fraction of the ternary space. The right panel contextualizes this by comparing the 99\% C.L.~regions under current NuFIT 6.1 uncertainties against the precision expectations for 2040. Future improvements in the standard oscillation parameters will predominantly compress the standard-mixing flavor region, and marginally compress the HNL-modified region. 

\begin{figure}[b!]
 \centering
 \includegraphics[width=\columnwidth]{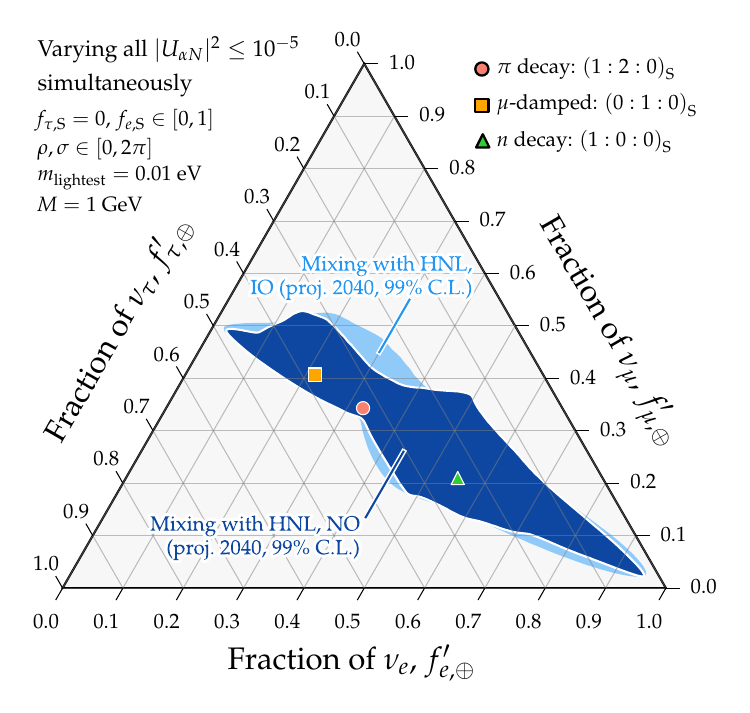}
 \caption{\textbf{Flavor composition at Earth for normal vs.~inverted neutrino mass orderings under HNL mixing.} Similar to \figu{hnl_flavor_uncertainties}, but showing the difference in the allowed flavor regions between normal ordering (NO, $m_1=0.01$~eV) and inverted ordering (IO, $m_3=0.01$~eV), adopting 2040 projections of the standard mixing parameters.}
 \label{fig:hnl_flavor_orderings_compare}
\end{figure}

Figure~\ref{fig:hnl_flavor_orderings_compare} shows the impact of the mass ordering on the HNL-modified flavor composition at Earth. Under NO, the mass state most closely associated with the muon and tau flavors is the heaviest ($\nu_3$). This large baseline mass suppresses the HNL-induced mixing corrections. IO flips this structure, making the muon- and tau-rich eigenstate the lightest. Without this kinematic inertia, the active-sterile couplings drive larger flavor shifts. Consequently, the allowed flavor region under IO is larger than under NO. Because the standard mixing angles are nearly identical for both orderings, the allowed regions expand in the same directions in flavor space; IO simply extends further outward due to its reduced mass suppression.

\section{Example testable scenarios}
\label{app:testable_scenarios}

\renewcommand{\theequation}{G\arabic{equation}}
\renewcommand{\thefigure}{G\arabic{figure}}
\renewcommand{\thetable}{G\arabic{table}}
\setcounter{figure}{0} 
\setcounter{table}{0}

\begin{figure*}[t!]
 \centering
 \includegraphics[width=\textwidth]{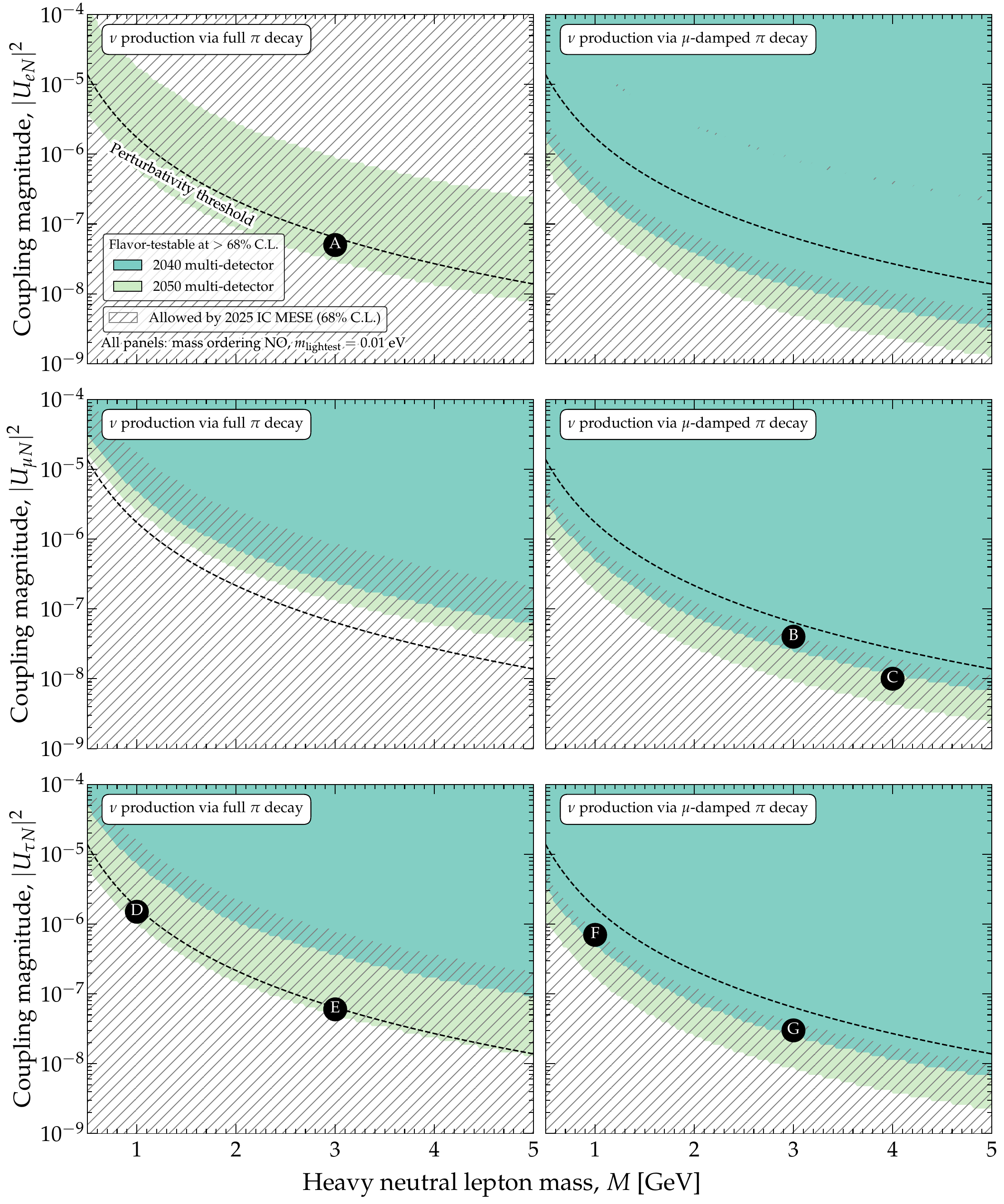}
 \caption{\textbf{Projected flavor-testability of heavy neutral leptons (HNLs).} We show this as a function of mass $M$ and active-sterile mixing $|U_{\alpha N}|^2$ for $\alpha \in \{e, \mu, \tau\}$ (rows), evaluated across different astrophysical neutrino production models (columns). Filled contours denote regions where HNL-induced flavor deviations exceed the projected sensitivities of future multi-detector networks enough to be resolvable. The hatched overlays bound the complementary parameter space currently compatible with present 11.4-year IceCube MESE flavor measurements~\cite{Abbasi:2025fjc}. The mass-threshold calculation remains perturbative below the ``perturbativity threshold''. The Majorana phases $(\rho, \sigma)$ are optimized at each grid point to maximize the Euclidean flavor distance from the standard unperturbed expectation. Labeled benchmark points (A--G) indicate specific flavor ratios selected to test our methods.}
 \label{fig:testability_regions}
\end{figure*}

Figure~\ref{fig:testability_regions} shows the ``sweet spot'' in parameter space where HNLs can be discovered by future neutrino telescopes. For any combination of HNL mass and coupling to be considered both valid and testable, the flavor composition at Earth, $(f_e, f_\mu, f_\tau)_\oplus$, that it predicts must simultaneously satisfy three criteria: 
\begin{enumerate}
 \item[(i)]
  \textbf{Compatibility with current flavor measurements:} To avoid contradicting existing observations, the predicted flavor composition must lie within the current 68\% C.L. flavor contour inferred from the 11.4-year IceCube MESE sample~\cite{Abbasi:2025fjc}. 
 \item[(ii)]
  \textbf{Perturbativity:} The point must lie below the perturbativity threshold, defined by $|U_{\alpha N}|^2 < 10^{-6} (1.2~\text{GeV} / M)^3$, to ensure that the computation of the neutrino mass threshold correction remains perturbative. 
 \item[(iii)]
  \textbf{Testability:} The induced flavor shift must be sufficiently distant from the standard-mixing expectation to be resolved by the projected flavor-measurement capabilities of future multi-detector networks (in \figu{testability_regions}, this is either in 2040 or 2050).
\end{enumerate}
At each grid point $(M, |U_{\alpha N}|^2)$, we optimize the values of the Majorana phases, $\rho$ and $\sigma$, in order to maximize the resulting flavor shift from standard-mixing expectations.

Within this valid and testable parameter space, the maximum Euclidean flavor-distance deviations are restricted to only 5--15\% relative to their standard-mixing expectations. This ceiling is primarily enforced by criterion (i). Because any viable HNL model must remain compatible with the existing IceCube MESE limits, the magnitude of the allowed active-sterile perturbation is bounded; if the deviations were significantly larger, they would have already been excluded by current data. Consequently, the discovery of HNLs in this mass-coupling regime relies entirely on the capability of the 2040 and 2050 detector generations to shrink the experimental uncertainties just enough to resolve these subtle, yet distinct, 5--15\% shifts in the flavor composition.

Table~\ref{tab:benchmark_points} in the main text shows the definitions of the seven benchmark scenarios, A--G, marked in \figu{testability_regions} that satisfy these three criteria across different production models and coupling sectors.


\section{Approximate SHiP forecasts}
\label{app:ship_forecasts}

\renewcommand{\theequation}{H\arabic{equation}}
\renewcommand{\thefigure}{H\arabic{figure}}
\renewcommand{\thetable}{H\arabic{table}}
\setcounter{figure}{0} 
\setcounter{table}{0}

In \figu{param_estimation} in the main text, to visually evaluate the complementarity between future high-energy astrophysical neutrino observations and terrestrial beam-dump experiments, we superimpose the projected constraints of SHiP~\cite{SHiP:2018xqz} onto the allowed parameter space derived from our astrophysical flavor analysis. We adopt a semi-analytic, approximate procedure to generate these SHiP contours without requiring a full Monte Carlo detector simulation, which lies beyond the scope of this paper. We model the SHiP measurement as a two-dimensional confidence region based on independent uncertainties in the HNL mass, $M$, and coupling, $|U_{\alpha N}|^2$.


\subsection{HNL mass resolution}

At a beam-dump facility like SHiP, the HNL mass is reconstructed kinematically from the invariant mass of its visible decay products (\eg, $N \to \mu + \pi$ or $N \to \mu + \mu + \nu$). Given the design specifications of the SHiP tracking spectrometer, the uncertainty on this invariant mass is tightly constrained. We approximate this measurement's error as a Gaussian standard deviation of $\sigma_M = 15\text{ MeV}$. This value is representative of the typical kinematic resolution for $\mathcal{O}(\text{GeV})$ long-lived particles decaying in the vacuum vessel at SPS energies~\cite{SHiP:2018xqz}.


\subsection{Mixing angle uncertainty and event scaling}

Unlike the mass, the mixing parameter $|U_{\alpha N}|^2$ is inferred from the total number of observed signal events, $N_{\text{obs}}$. To estimate $N_{\text{obs}}$ for a given benchmark point $(M_{\text{true}}, |U_{\alpha N}|^2_{\text{true}})$, we exploit the parametric scaling of HNL production and decay in the long-lived regime. 

At beam-dump facilities, the probability of producing an HNL and its subsequent probability to decay within the fiducial volume both scale linearly with $|U_{\alpha N}|^2$. Consequently, the total event yield scales as $N_{\text{obs}} \propto |U_{\alpha N}|^4$~\cite{Gorbunov:2007ak}. Using this scaling relation, we extrapolate the expected number of events from the official 90\% C.L.~exclusion contours published by the SHiP Collaboration~\cite{SHiP:2018xqz}, shown in Figs.~\ref{fig:discovery_window_var_angles_UmuN}, \ref{fig:discovery_window}, \ref{fig:discovery_window_var_angles_UtauN}, and \ref{fig:discovery_window_var_angles_UeN}.

Assuming a zero-background experimental environment, the 90\% C.L. lower boundary corresponds to an expected yield of $N_{\text{obs}} = 2.3$ events, following Poisson statistics. Thus, the number of observed events for our chosen true benchmark is approximated by
\begin{equation}
 N_{\text{obs}} = 2.3 \times \left( \frac{|U_{\alpha N}|^2_{\text{true}}}{|U_{\alpha N}|^2_{\text{limit}}(M_{\text{true}})} \right)^2 \;,
\end{equation}
where $|U_{\alpha N}|^2_{\text{limit}}(M_{\text{true}})$ is the specific mixing value evaluated at the lower boundary of the SHiP 90\% C.L. contour for the corresponding flavor $\alpha$.

The total fractional uncertainty on the $|U_{\alpha N}|^2$ is then computed by adding the statistical Poisson error and a systematic uncertainty floor in quadrature, \ie,
\begin{equation}
 \frac{\sigma_{|U|^2}}{|U_{\alpha N}|^2_{\text{true}}} = \sqrt{\frac{1}{N_{\text{obs}}} + \sigma_{\text{sys}}^2} \;.
\end{equation}
For GeV-scale HNLs produced at SPS energies, the systematic uncertainty is dominated by the heavy-flavor ($D$- and $B$-meson) production cross-sections and secondary cascade enhancement factors in the thick target. Based on evaluations by the SHiP Collaboration, the $c\bar{c}$ production cross-section carries an uncertainty of roughly 10\%, while the $b\bar{b}$ cross-section uncertainty is approximately 19\%~\cite{SHiP:2018xqz}. For simplicity in our projected contours, we adopt a representative systematic uncertainty floor of $\sigma_{\text{sys}} = 0.15$ (15\%) across the parameter space, noting that in reality the exact systematic penalty depends on the dominant production meson at a given HNL mass.


\subsection{Measurement contours}

Because the HNL mass is derived from kinematic tracking and the mixing from total event counting, the two measurements are statistically independent, yielding a covariance of zero (in our simplified treatment). We map this projected constraint onto the $(M, \log_{10}|U_{\alpha N}|^2)$ parameter space used in \figu{param_estimation}. The constant fractional uncertainty on $|U_{\alpha N}|^2$ translates into a constant absolute uncertainty in logarithmic space via error propagation:
\begin{equation}
\sigma_{\log_{10}|U|^2} = \frac{1}{\ln 10} \frac{\sigma_{|U|^2}}{|U_{\alpha N}|^2_{\text{true}}} \;.
\end{equation}
The SHiP projection is therefore implemented as an axis-aligned elliptical contour representing the $1\sigma$, $2\sigma$, and $3\sigma$ confidence intervals, overlaid on the astrophysical contours, separately for benchmark Points B, C, F, and G. 

This procedure is an approximation designed specifically to estimate the complementarity of terrestrial and astrophysical probes. A rigorous projection would require simulating the detailed acceptance of  SHiP for the exact benchmark models under consideration.


\end{document}